\documentclass[numberedappendix,apj]{emulateapj}            
\pdfoutput=1

\usepackage{natbib}
\usepackage{mathtime}  
\usepackage{lscape}

\usepackage[
debug=true,
hyperindex=true,
colorlinks=false,
bookmarksnumbered=true,
baseurl={http://www.cfa.harvard.edu},
hyperfootnotes=false, 
pdfkeywords={infrared, stars, circumstellar matter, open clusters, Spitzer Space Telescope},
pdftitle={A Spitzer census of the IC 348 nebula},
pdfauthor={August Muench}]{hyperref} 

	\newcommand{\eprint}[1]{\href{http://arxiv.org/abs/#1}{#1}} 
	 
	\newcommand{\adsurl}[1]{\href{#1}{ADS}} 
	\providecommand{\url}[1]{\href{#1}{#1}}
\journalinfo{To appear in the Astronomical Journal}
\submitted{Submitted November 20, 2006; Accepted March 30, 2007; Version \today}
\shorttitle{\sst\ census of IC~348}
\shortauthors{Muench et al.}

\newcommand{\solarmass}{\ensuremath{ \mathcal{M}_{\Sun} }}
\newcommand{\jupmass}{\ensuremath{ \mathcal{M}_{Jup} }}
\newcommand{\mass}{\ensuremath{ \mathcal{M} }}

\newcommand{\halpha}{\ensuremath{\mbox{H}\alpha}}

\newcommand{\ntwohplus}{\ensuremath{\mbox{N}_{2}\mbox{H}^{+}}}

\newcommand{\comap}{\ensuremath{{\ ^{13}}{\mbox{CO}}}}
\newcommand{\av}{\ensuremath{\mathnormal{A}_{V}}}

\newcommand{\lbol}{\ensuremath{{L}/{L}_{\Sun}}}
\newcommand{\teff}{\ensuremath{{T}_{eff}}}

\newcommand{\sst}{{\it Spitzer}}
\newcommand{\cxo}{{\it Chandra}}
\newcommand{\irac}{{IRAC}} 
\newcommand{\mips}{{MIPS}} 

\newcommand{\htwo}{H\,{\small II}}

\newcommand{\iracalpha}{\ensuremath{\alpha_{3-8{\tiny \micron}}}}
\newcommand{\omicron}{\ensuremath{o}}

\newcommand{\SIa}{\ensuremath{3.6}}
\newcommand{\SIb}{\ensuremath{4.5}}
\newcommand{\SIc}{\ensuremath{5.8}}
\newcommand{\SId}{\ensuremath{8.0}}
\newcommand{\SMa}{\ensuremath{24}}
\newcommand{\SMb}{\ensuremath{70}}
\begin{document}

\title{A Spitzer census of the IC~348 nebula}

\author{August A. Muench\altaffilmark{1},
        Charles J. Lada\altaffilmark{1},
        K. L. Luhman\altaffilmark{2,3},
        James Muzerolle\altaffilmark{4} \&
        Erick Young\altaffilmark{4}}

\altaffiltext{1}{Smithsonian Astrophysical Observatory.
60 Garden Street, Mail Stop 72. Cambridge, MA. 02138  USA;
~gmuench@cfa.harvard.edu,~clada@cfa.harvard.edu}
\altaffiltext{2}
{Visiting Astronomer at the Infrared Telescope Facility, which is operated
by the University of Hawaii under Cooperative Agreement no. NCC 5-538 with
the National Aeronautics and Space Administration, Office of Space Science,
Planetary Astronomy Program.}
\altaffiltext{3}{Department of Astronomy and Astrophysics,
The Pennsylvania State University, University Park, PA 16802, USA;
~kluhman@astro.psu.edu.}

\altaffiltext{4}{Steward Observatory, University of Arizona
Tucson, AZ 85712;~jamesm@as.arizona.edu,~eyoung@as.arizona.edu}


%
\begin{abstract}

\sst\ mid-infrared surveys enable accurate census of young stellar objects by sampling 
large spatial scales, revealing very embedded protostars and detecting low luminosity objects.  
Taking advantage of these capabilities, we present a \sst\ based census of the IC~348 nebula and 
embedded star cluster, covering a~2.5~pc region and comparable in extent to the Orion nebula.  
Our \sst\ census supplemented with ground based spectra has added 42 class II T-Tauri sources 
to the cluster membership and identified $\sim20$ class 0/I protostars. The population of IC~348 likely 
exceeds 400 sources after accounting statistically for unidentified diskless members.    
Our \sst\ census of IC~348 reveals a population of class I protostars that is anti-correlated spatially 
with the class II/III T-Tauri members, which comprise the centrally condensed cluster around a B star.  
The protostars are instead found mostly at the cluster periphery about $\sim1$~pc from the B star
and spread out along a filamentary ridge.  We further find that the star formation rate in this 
protostellar ridge is consistent with that rate which built the older exposed cluster while the 
presence of fifteen cold, starless,  millimeter cores intermingled with this protostellar population 
indicates that the IC~348 nebula has yet to finish forming stars.  Moreover, we show that the 
IC~348 cluster is of order 3-5 crossing times old, and, as evidenced by its smooth radial 
profile and confirmed mass segregation, is likely relaxed. 
While it seems apparent that the current cluster configuration is the result of dynamical evolution
and its primordial structure has been erased, our finding of a filamentary ridge of class I protostars
supports a model where embedded clusters are built up from numerous smaller sub-clusters. 
Finally, the results of our \sst\ census indicate that the supposition that star formation 
must progress rapidly in a dark cloud should not preclude these observations that show it 
can be relatively long lived.

\end{abstract}


\section{Introduction}
\label{sec:intro}

The \objectname[IC 348]{IC~348} nebula on the northeastern corner of the Perseus Molecular Cloud 
\citep{1915ApJ....41..253B} has been known to harbor pre-main sequence T-Tauri stars since they were revealed 
through a slitless \halpha\ grism survey by \citet{1954PASP...66...19H}.  Slitless \halpha\ grism 
surveys were once the most powerful tool for searching for young  stars 
\citep[c.f.,][]{1988cels.book.....H}, while the subsequent development of infrared bolometers permitted 
better census of the darker regions of molecular clouds, including very young protostars which
are young stars that still retain infalling envelopes.   Such infrared observations in  IC~348 by 
\citet{1974PASP...86..798S} led,  for example, to the discovery of an optically invisible bright 
$2\micron$ source about 1pc from the clustering of  \halpha\ members.  
Strom's \objectname[Name IC 348 IR]{IR} source was the first such
hint that the  stars forming in the IC~348 nebula might not all have the same age.
Modern tools for identifying young stars include X-ray surveys, which parse young stellar objects
(YSOs) using energetic emissions from their rotationally enhanced, magnetic activity,  
and wide-field infrared imaging surveys,  which identify YSOs using the signature in the star's  broadband spectral 
energy distribution (SED) of thermal reprocessing of the star's light by an optically thick circumstellar disk.  
To date roughly 300 young stars have been identified in the IC~348 nebula from X-ray 
\citep[e.g.,][]{2001AJ....122..866P,2004A&A...422.1001P}, optical \citep[e.g.,][]{1997A&A...324..549T,1998ApJ...497..736H}, near-infrared
\citep[][hereafter, LL95 and M03 respectively]{1995AJ....109.1682L,2003AJ....125.2029M} 
and spectroscopic surveys 
\citep{1998ApJ...508..347L,1999ApJ...525..466L,2003ApJ...593.1093L,2005ApJ...618..810L}.
These known members are clustered at the center of the nebula and have a median age of $\sim2-3$~My
\citep{2003ApJ...593.1093L}; we examined the disk properties of these members in  \citet[][hereafter, Paper 1]{2006AJ....131.1574L},

For this paper we undertook a mid-infrared survey of the IC~348 nebula with the \sst\ Space Telescope 
\citep{2004ApJS..154....1W} to make a more complete membership census over a large cluster area.
Statistical studies of the surface density of stars around IC~348 \citep{2002ApJ...578..523T,2006A&A...445..999C}
anticipated the discovery of more cluster members, but they could not identify individual members and could give no 
information about their evolutionary status. The classification of a young star as protostellar (class I)
or more evolved class II sources with optically thick disks \citep[see][etc]{1987ApJ...312..788A}
is best accomplished using its broadband spectral energy distribution.  Thus, we have identified and classified 
approximately 60 new cluster members, including $\sim20$ protostellar objects, by constructing each source's broad 
band $(0.5-70\micron)$ SED and through spectroscopic follow up.  Our census has expanded the confirmed 
boundaries of IC~348  to a physical size comparable to that well studied portion of 
the \objectname[Name Orion Nebula Cluster]{Orion Nebula Cluster} \citep{1998ApJ...492..540H}.

\begin{figure*}
     \centering \includegraphics[angle=90,width=0.8\textwidth]{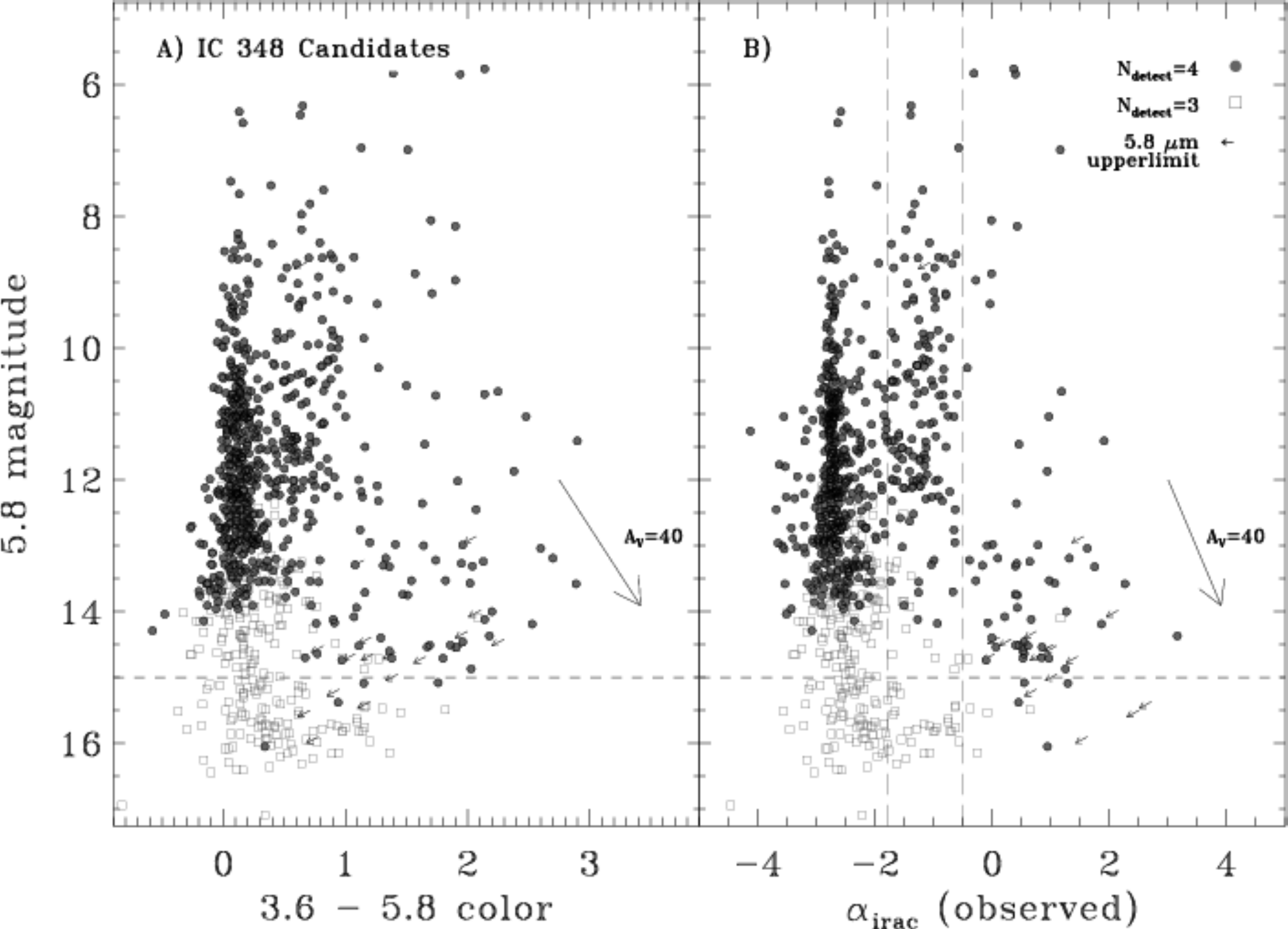}
\caption{ 
A)~\sst\ $m_{\SIa}\,-\,m_{\SIc}$~color vs $m_{\SIc}$ magnitude diagram;~
B)~\sst\ $\iracalpha$ spectral index vs $m_{\SIc}$ magnitude diagram.
Symbol types differentiate four band (filled circles) and three band (open squares) \irac\ detections.
Upperlimits  on $m_{\SIc}$ are shown with arrows (19 sources).  
Subsequent sorting of these sources into respective YSO classes was restricted 
to that sample with $m_{\SIc}<15$ (or fainter sources which were detected 
in the  $\SIa, \SIb, \mbox{and } \SId\micron$~bands).
Vertical dashed lines in panel (B) correspond to $\alpha$ values used to 
segregate different YSO classes (e.g., class I, II; see \S\ref{sec:select:sed} and Paper I).
The reddening law is from \protect\citet{2005ApJ...619..931I} or as derived in \S\ref{app:av}.
\label{fig:cmds}}
\end{figure*}

Paper I contains all details of the  data processing except for the far-infrared Multiband Imaging Photometer for
\sst\ \citep[\mips;][]{2004ApJS..154...25R} observations  (see \S\ref{sec:select:mips})\footnote{The \sst\ data obtained for 
this paper were taken from AORs~\dataset[ADS/Sa.Spitzer#3955200]{3955200}, \dataset[ADS/Sa.Spitzer#3651584]{3651584},
\dataset[ADS/Sa.Spitzer#4315904]{4315904}.}.  Candidate members were  selected initially using spectral
indices to identify the presence of infrared excess in their composite SEDs (\S\ref{sec:select:sed}).
Ground-based spectra, including new observations presented in this paper, support the membership status for 
nearly all of the class II candidates and many of the protostars.  Our census of very low luminosity protostars 
required the removal of an overwhelming population of extragalactic sources that masquerade 
as young stars (\S\ref{sec:select:protostars}).
In \S \ref{sec:visual} we examine the nature of the IC 348 protostellar and class II members 
by comparing their positions to  recent dense gas and dust maps from the COMPLETE\footnote{The COordinated 
Molecular Probe Line Extinction Thermal Emission Survey of Star Forming Regions, \url{http://cfa-www.harvard.edu/COMPLETE/}.}
project \citep{2006AJ....131.2921R}, by analyzing their physical separations, and by placing them on the
Hertzsprung-Russell (HR) diagram.  We discuss briefly the implications of the cluster's inferred structure and star 
forming history and examine the timescales for dynamical evolution of the central star cluster,  pointing out
their relevance for the timescale for dark cloud and circumstellar disk evolution (\S\ref{sec:discuss}).  
Appendices include a discussion of the effects of reddening on the $3-8\micron$ portion of a stellar or star+disk SED 
(\S \ref{app:av}), spectra of $\sim40$ new members (\S \ref{app:spex}) and a photometric catalog of candidate 
apparently diskless (class III)  members selected from  X-ray surveys of this region (\S \ref{app:class3}).

\begin{figure*}
    \centering \includegraphics[angle=90,width=0.6\textwidth]{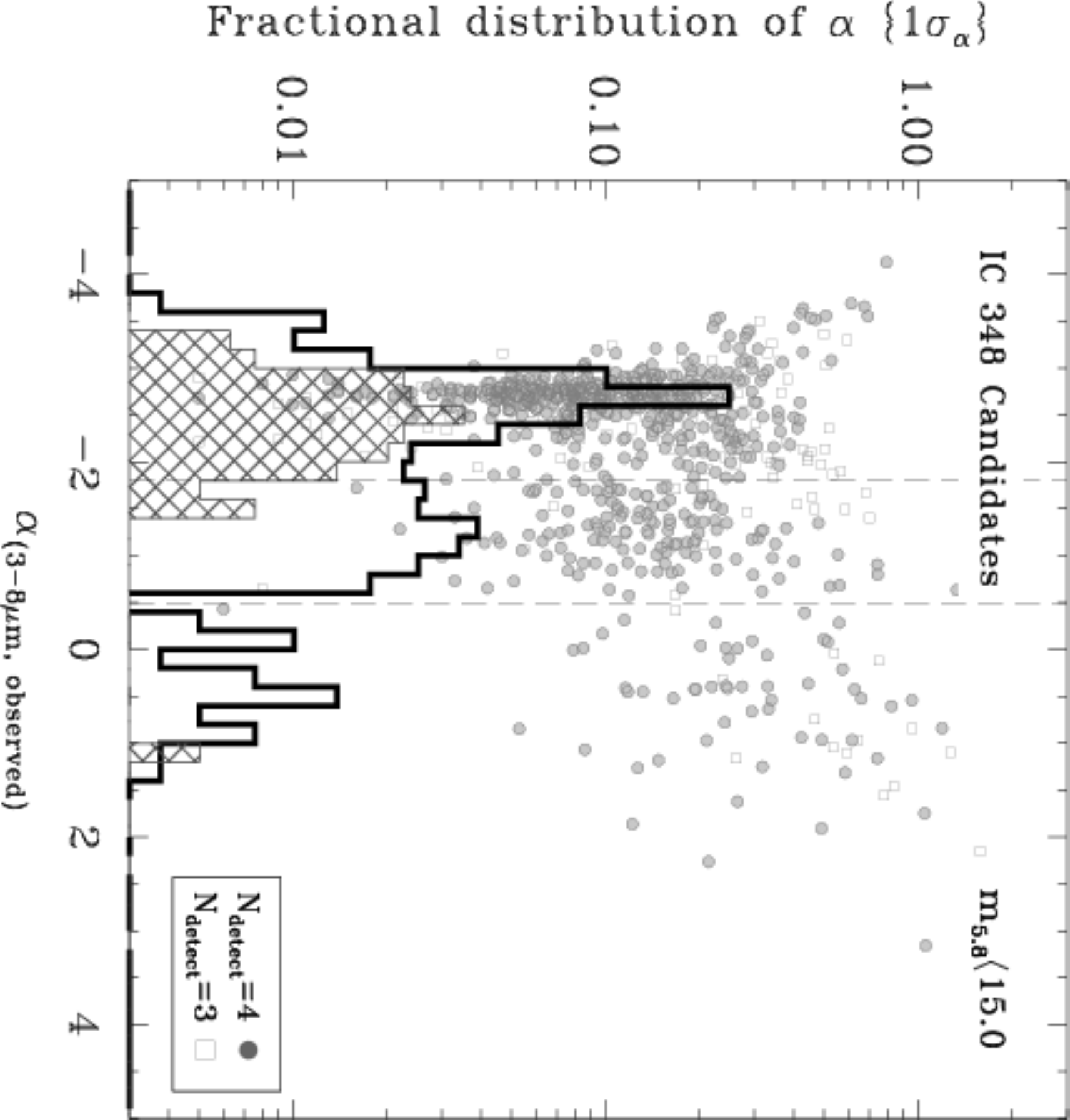}
\caption{Distribution function of $\iracalpha$ for sources with $m_{\SIc}<15.0$. 
The open histogram corresponds to \sst\ sources detected in the four \irac\ bands; 
the hatched histogram corresponds to the distribution function of three \irac\ band detections. 
The distribution functions were normalized by the total  number of candidates. 
Two vertical lines ($\alpha=-1.8\mbox{ and }\alpha=-0.5$) separate the candidates into three YSO classes.  
For each source we overplotted the $1\sigma$ uncertainty in the $\iracalpha$ fit versus its $\iracalpha$ 
on the x-axis; again, open and filled symbols differentiate 3 and 4 band detections, respectively.   
\label{fig:iracalpha}}
\end{figure*}

\section{Spitzer census}
\label{sec:select}

\subsection{SED selected young stellar objects}
\label{sec:select:sed}

Studying the previously known members of IC~348 in Paper I, we showed that the power-law fit of 
the  $3-8\micron$ portion of the young stars' SEDs as observed by \sst, provided a good diagnostic of these 
members' disk properties.  We were able to empirically separate members with optically thick T-Tauri disks 
(hereafter class II sources)  from those with little (termed anemic) or no apparent disk excess at these 
wavelengths\footnote{Hereafter,  we use the term ``class III'' to describe all members having SEDs
indicative of ``anemic'' inner disks or simple photospheres; see Paper I.}. 
In this section we describe how we used this SED parameter  to identify new young stellar objects (YSOs) from 
our entire \sst\ catalog of IC~348, including new embedded protostars that were not studied in Paper I.
In searching for new members using disk excess it is also important to avoid selecting reddened background stars; 
as we discuss in Appendix \ref{app:av}, the $3-8\micron$ SED slope is fairly insensitive to extinction, which
allows us to be confident in the quality of our initial member selection. 

To fit a power law to the $3-8\micron$ SED  we required sources in our catalog to be detected in at least three 
of the four \sst\ \irac\ \citep[InfraRed Array Camera;][]{2004ApJS..154...10F} bands; 
this restricted our search to a $26.8\arcmin$ by $28.5\arcmin$  region of 
the GTO~\irac\ maps centered at 03:44:20.518, +32:10:34.87 with a PA of $81\degr$. 
Note, this entire region was also surveyed with \mips.
The resulting $\sim2.5$~pc region enclosed both the $\av$-limited 
completeness census of \citet{2003ApJ...593.1093L} and the $20\arcmin$ FLAMINGOS\footnote{The 
FLoridA Multi-object Imaging Near-IR Grism Observational Spectrometer. See \url{http://flamingos.astro.ufl.edu/}.}
region studied by \citet{2003AJ....125.2029M}\footnote{These survey regions are compared in Figure \ref{fig:map1}.}.
In this region there are 906  sources detected in three \irac\ bands, including 282 of the 300 known members studied in 
Paper~I.  Of these 906 candidates, 648 were detected in all four \irac\ bands. Only 19 of the 906 sources lacked 
$\SIc\micron$~detections while 238 sources detected from $\SIa\;\mbox{to}\;\SIc \micron$ lacked 
$\SId\micron$~detections\footnote{One spectroscopically confirmed member, \#396~(M5.25), did not have photometry
at $\SIa\micron$ due to a nearby bright star; $\omicron$\ Persi was saturated at $\SIa$\ and\ $\SIb\micron$; there
were nine sources detected only at $\SIc\micron$.  Otherwise all the sources in our field were detected 
in bands $\SIa$ and $\SIb\micron$.  We found no sources detected only at $\SId\micron$.}.
To better constrain the candidates' SEDs we derived $95\%$ upper limits for 
all sources lacking either $\SIc\mbox{ or }\SId\micron$ detections.

We constructed the $m_{\SIa}\,-\,m_{\SIc}$ versus $m_{\SIc}$~color-magnitude diagram (CMD, Figure \ref{fig:cmds}a) 
for these 906 candidates to further refine our selection criteria.  Upper-limits for the $19$ sources lacking $\SIc\micron$ flux
measurements are displayed with arrows.  Two loci are clearly evident in the CMD: one of nearly colorless stars
and the second redder locus we expect to consist primarily of T-Tauri stars with disks.   A strong $\SId\micron$
magnitude cutoff for colorless stars is evident at $m_{\SIc}\sim~14$; most the $\SIc\micron$
upperlimits are $m_{\SIc}>14.5$.  In Figure 1b we have replaced the $m_{\SIa}\,-\,m_{\SIc}$ color with
the power-law fit to the observed slope of the \irac\ portion of these sources' SEDs 
($\iracalpha$)\footnote{Upper limits were not used in these calculations, although they were useful 
for filtering sources; see \S\ref{sec:select:sed}.}; this clearly reinforces the existence of two
intrinsic loci in the CMD.  The two loci are more distinct when plotting $m_{\SIc}$ as a function of $\iracalpha$  
because we have assumed the correct underlying shape of the objects' SEDs; i.e., whether it is the Raleigh Jeans portion of
a star's photospheric SED or the thermal infrared SED of a passive, re-radiating optically thick disk.%
\begin{figure}
    \centering \includegraphics[angle=90,width=0.475\textwidth]{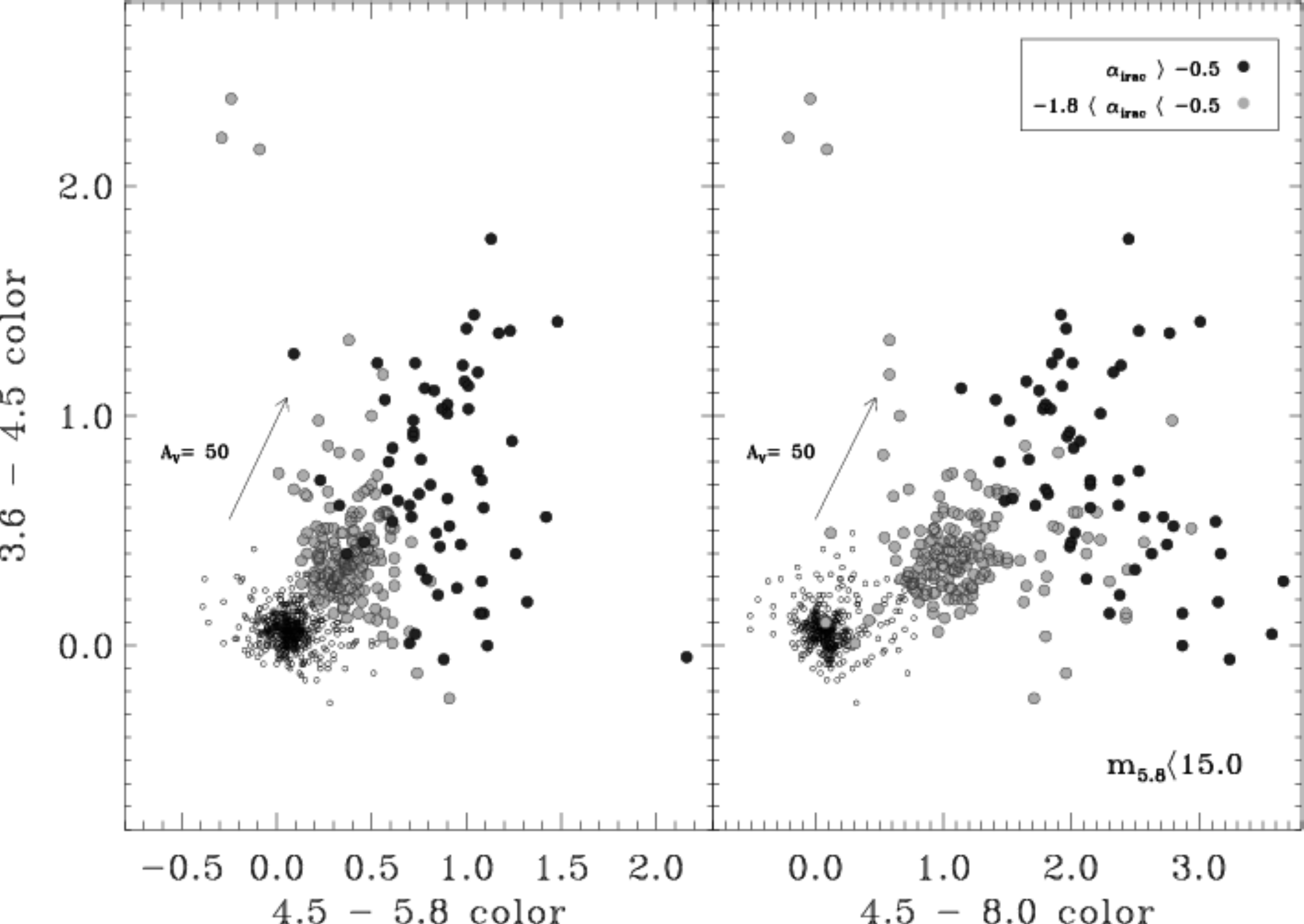}
\caption{\sst\ \irac\ color-color diagrams for all 3 band \irac\ detections with $m_{\SIc} < 15$. 
Sources are color coded by candidate YSO class defined by $\iracalpha$: class I $(\alpha>-0.5)$, 
solid grey filled circles; class II $(-0.5>\alpha>-1.8)$, light grey filled circles; other (class III, ``anemic'' 
disk or non-member)  objects are open circles.  A)  $m_{\SIb} - m_{\SIc}$ vs $m_{\SIa} - m_{\SIb}$.
This pane includes sources not detected in $\SId\micron$ band;
B) $m_{\SIb} - m_{\SId}$ vs $m_{\SIa} - m_{\SIb}$.  This panel includes sources lacking $\SIc\micron$ detections.  
Note, we found that IC~348 sources in the far upper left are contaminated by shocked emission 
from Herbig-Haro objects. The reddening law is from \protect\citet{2005ApJ...619..931I}.
\label{fig:ccs}}
\end{figure}%
Further, these power-law fits are less sensitive to uncorrelated photometric
uncertainties than colors, which are of course ratios between only two wavelengths.
Nonetheless, we further filtered our sample of candidate members based on photometric quality.  
We imposed an empirical flux limit of $m_{\SIc}<15$ based on the increased spread in the 
value of $\iracalpha$ for fainter colorless stars; we did, however include fainter $\SIc\micron$ \irac\ sources 
if they were also detected at $\SId\micron$\footnote{We will use the $m_{\SIc}$ magnitude to parse 
the sources in our subsequent analysis for four reasons: 
1) it is clearly more sensitive than the $\SId\micron$ channel; 
2) when combined with a standard extinction law it will be the primary \irac\ bandpass for emergent flux from
heavily reddened sources (\citet{2004ApJ...617.1177W}; Appendix \S\ref{app:av} and Figure \ref{fig:app})
due at least in part to silicate absorption in the $\SId\micron$ bandpass; 3) it is somewhat less contaminated by 
PAH emission than the $\SId\micron$ bandpass frequently evident in the SEDs of non-cluster sources.}.  
We also required the detections to have photometric errors of less than 0.2 magnitudes.  

After applying these photometric constraints we had 657 candidates in our IC~348 \sst\ region.
Figure \ref{fig:iracalpha} displays the distribution function of $\iracalpha$ as a histogram
for  these  candidates.  The first narrow peak in the $\iracalpha$ distribution function at
-2.8 reflects the narrowly constrained value of $\iracalpha$ for stellar photospheres;
photospheric $\iracalpha$ has very little spectral type dependence (Paper I). 
A second peak at $\iracalpha=-1.3$ corresponds to class II T-Tauri stars with optically thick disks and 
a third peak corresponds to sources with flat or rising mid-IR SEDs. 
Using our empirical boundary between anemic and class II disks ($\iracalpha >-1.8$; Paper I; shown in
Figure \ref{fig:iracalpha} ) we identified $192$ candidate YSOs in our IC~348 region.  
Our tally of IC~348 YSOs is $20\%$ larger than the total number of IC~348 YSOs (158) 
identified by \citet{2006ApJ...645.1246J}.
While we are using slightly lower luminosity limits than \citeauthor{2006ApJ...645.1246J}, 
the statistics of their Legacy survey come from a different and $70\%$ larger cluster area,
correspond to a different definition of the spectral index and include class III 
(by their definition) sources; thus, we do not further discuss the statistics of this Legacy project.
Finally, we did not search for new members with ``anemic'' type disks ($-2.6 < \iracalpha < -1.8$; Paper I)
since a search for sources with very small excesses can be hampered by poor photometry, in this case due to the
nebula (see the scatter in the power-law fit sigma overplotted Figure \ref{fig:iracalpha}).

We subdivided the $\iracalpha >-1.8$ YSO sample into two classes based on the shape of the $\iracalpha$ 
distribution function:  thick disk class II sources in the peak, $-1.8 < \iracalpha < -0.5$ 
and class I ``protostellar'' candidates with $\iracalpha > -0.5$.  
Flat spectrum sources, considered to be  protostars in a later stage of envelope dispersal or with highly flared disks, 
can have slightly falling mid-infrared SED slopes, $0.3 >\alpha >-0.3$ \citep{1987IAUS..115....1L}.
A distinction between highly flared class II disks and emission from disk +remnant envelope 
may require data at wavelengths longer than $10\micron$.
There are a total of $136$ candidate class II sources in our IC~348~\sst\ region  and $56$ red class I candidates.   
For comparison to other \sst\ studies of YSOs in clusters that 
use color-color classification techniques \citep[e.g.,][]{2004ApJS..154..367M}, 
we plot two such diagrams in Figure \ref{fig:ccs}.  Together these encompassed all $3$ band \irac\ detected
sources; protostellar and class II sources are color-coded on these plots.   Sources parsed by $\iracalpha$ are
well segregated in the color-color diagram except where photometric errors in a single color yield some scattering.   

\begin{figure}
    \centering  \includegraphics[angle=90,totalheight=0.425\textwidth]{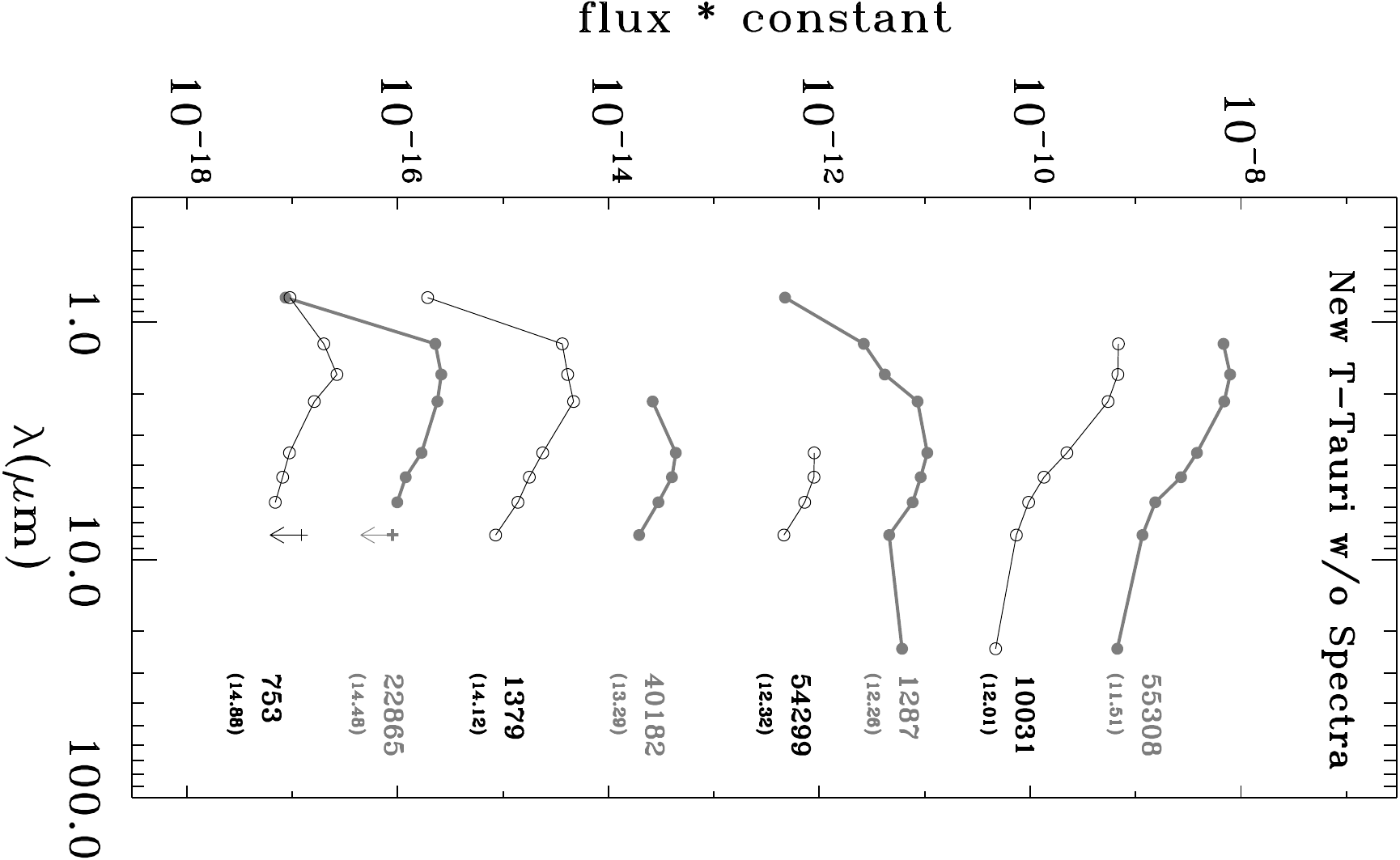}
\caption{Additional class II candidates considered to be IC~348 members but lacking spectral types.
Again, sources are sorted according to their $\SIc\micron$ magnitude, which is listed
in parenthesis below each source ID.  Plotting symbols, line thickness and line color alternate 
from SED to SED for clarity.
\label{fig:class2nos}}
\end{figure}%

\begin{deluxetable*}{lccccrr}
\tablewidth{0pt}
\tablecaption{Spectroscopy of IC~348 \sst~excess sources\label{tab:spex}}
\tablehead{
\colhead{ID\tablenotemark{a}} &
\colhead{$\alpha$(J2000)} &
\colhead{$\delta$(J2000)} &
\colhead{$f_{p}$\tablenotemark{b}} &
\colhead{Spectral Type} &
\colhead{Membership\tablenotemark{c}} &
\colhead{Class}}
\startdata
   70 & 03 43 58.55 & 32 17 27.7 &cfht&   M3.5(IR),M3.75(op) &  $A_V$,H$_2$O,ex,e,Li,NaK & II  \\
  117 & 03 43 59.08 & 32 14 21.3 &2m  &   M3-M4(IR)          &  $A_V$,H$_2$O,e,ex        & II   \\
  132 & 03 44 27.25 & 32 14 21.0 &cfht&   M3.5(IR,op)        &  $A_V$,H$_2$O,ex,NaK      & II  \\
  162 & 03 43 48.81 & 32 15 51.7 &cfht&   M4.5(IR)           &  $A_V$,H$_2$O,ex          & II  \\
  179 & 03 44 34.99 & 32 15 31.1 &cfht&   M3.5(IR,op)        &  $A_V$,H$_2$O,ex,NaK      & II    \\
  199 & 03 43 57.22 & 32 01 33.9 &wfpc&   M6.5(IR)           &  $A_V$,H$_2$O,ex          & II   \\
  215 & 03 44 28.95 & 32 01 37.9 &cfht&   M3.25(IR)          &  $A_V$,H$_2$O,ex          & II   \\
  231 & 03 44 31.12 & 32 18 48.5 &cfht&   M3.25(IR)          &  $A_V$,H$_2$O,ex          & II \\
  234 & 03 44 45.22 & 32 01 20.0 &cfht&   M5.75(IR)          &  $A_V$,H$_2$O,ex          & I  \\
  245 & 03 43 45.17 & 32 03 58.7 &cfht&      ?(IR)           &  ex                       & I \\
  265 & 03 44 34.69 & 32 16 00.0 &cfht&   M3.5(IR)           &  $A_V$,ex                 & II \\
  280 & 03 44 15.23 & 32 19 42.1 &cfht&   M4.75(IR,op)       &  $A_V$,H$_2$O,ex,NaK      & II \\
  321 & 03 44 22.94 & 32 14 40.5 &cfht&   M5.5(IR)           &  $A_V$,H$_2$O,ex          & II  \\
  327 & 03 44 06.00 & 32 15 32.3 &cfht&   M6.5(IR)           &  H$_2$O,ex                & II \\
  364 & 03 44 43.03 & 32 15 59.8 &cfht&   M4.75(IR,op)       &  $A_V$,H$_2$O,ex,NaK      & II   \\
  368 & 03 44 25.70 & 32 15 49.3 &cfht&   M5.5(IR)           &  $A_V$,H$_2$O,ex          & II  \\
  406 & 03 43 46.44 & 32 11 06.1 &cfht&   M6.5(IR),M5.75(op) &  $A_V$,H$_2$O,ex,NaK      & II   \\
  643 & 03 44 58.55 & 31 58 27.3 &cfht&   M6.5(IR)           &  $A_V$,H$_2$O,ex          & II  \\
  723 & 03 43 28.47 & 32 05 05.9 &cfht&   M4(IR)             &  $A_V$,H$_2$O,e,ex        & II   \\
  904 & 03 45 13.81 & 32 12 10.1 &cfht&   M3.5(IR)           &  $A_V$,H$_2$O,ex          & I   \\
 1679 & 03 44 52.07 & 31 58 25.5 &cfht&   M3.5(IR)           &  $A_V$,H$_2$O,ex          & II   \\
 1683 & 03 44 15.84 & 31 59 36.9 &cfht&   M5.5(IR),M5.25(op) &  $A_V$,H$_2$O,ex,e,NaK    & II   \\
 1707 & 03 43 47.63 & 32 09 02.7 &cfht&   M7(IR)             &  H$_2$O,ex                & II \\
 1761 & 03 45 13.07 & 32 20 05.3 &2m  &   M5(IR)             &  $A_V$,H$_2$O,ex          & II   \\
 1833 & 03 44 27.21 & 32 20 28.7 &cfht&   M5.25(IR),M5(op)   &  $A_V$,H$_2$O,ex,NaK      & II   \\
 1843 & 03 43 50.57 & 32 03 17.7 &cfht&   M8.75(IR)          &  $A_V$,H$_2$O,ex          & II   \\
 1872 & 03 44 43.31 & 32 01 31.6 &2m  &    ?(IR)             &  e,ex                     & I   \\
 1881 & 03 44 33.79 & 31 58 30.3 &cfht&   M4.5(IR),M3.75(op) &  $A_V$,H$_2$O,ex,e,NaK    & II \\
 1889 & 03 44 21.35 & 31 59 32.7 &2m  &    ?(IR)             &  e,ex                     & I \\
 1890 & 03 43 23.57 & 32 12 25.9 &cfht&   M4.5(op)           &  $A_V$,NaK                & II   \\
 1905 & 03 43 28.22 & 32 01 59.2 &cfht&$>$M0(IR),M1.75(op)   &  $A_V$,H$_2$O,ex,e,Li     & II \\
 1916 & 03 44 05.78 & 32 00 28.5 &2m &     ?(IR)             &  ex                       & I \\
 1923 & 03 44 00.47 & 32 04 32.7 &2m &    M5(IR)             &  $A_V$,H$_2$O,ex          & II \\
 1925 & 03 44 05.78 & 32 00 01.3 &cfht&   M5.5(IR)           &  $A_V$,H$_2$O,ex          & II   \\
 1933 & 03 45 16.35 & 32 06 19.9 &cfht&   ?(IR),K5(op)       &  $A_V$,ex,e               & II   \\
10120 & 03 45 17.83 & 32 12 05.9 &cfht&   M3.75(op)          &  e,NaK,$A_V$,ex           & II   \\
10176 & 03 43 15.82 & 32 10 45.6 &cfht&   M4.5(IR)           &  $A_V$,H$_2$O,ex          & II   \\
10219 & 03 45 35.63 & 31 59 54.4 &cfht&   M4.5(IR,op)        &  $A_V$,H$_2$O,ex,NaK,e    & II   \\
10305 & 03 45 22.15 & 32 05 45.1 &cfht&   M8(IR)             &  $A_V$,H$_2$O,ex          & II \\
22232 & 03 44 21.86 & 32 17 27.3 &cfht&   M5(IR),M4.75(op)   &  $A_V$,H$_2$O,ex,e,NaK    & II  \\
30003 & 03 43 59.17 & 32 02 51.3 &wfpc&   M6(IR)             &  $A_V$,H$_2$O,ex          & I  \\   
\enddata
\tablenotetext{a}{The running number identifiers used in this work corresponds to and extends that system used in \protect\citet{1998ApJ...508..347L,1999ApJ...525..466L,2003ApJ...593.1093L,
2005ApJ...623.1141L,2005ApJ...618..810L,2006ApJ...643.1003M,2006AJ....131.1574L}.}
\tablenotetext{b}{$f_{p}$ is a flag on the source's position indicating the origin of that astrometry: 
\protect\citet[2m: ][; FLAMINGOS]{2003AJ....125.2029M}; 
\protect\citet[cfht: ][]{2003ApJ...593.1093L};
\protect\citet[wfpc: ][]{2005ApJ...623.1141L};
irac: \irac\ mosaics, this paper.}
\tablenotetext{c}{Membership in IC~348 is indicated by $A_V\gtrsim1$ and
a position above the main sequence for the distance of IC~348 (``$A_V$"),
excess emission in the \irac\/\mips\ data (``ex"), the shape of the gravity-sensitive
steam bands (``H$_2$O"), Na~I and K~I strengths intermediate
between those of dwarfs and giants (``NaK"), strong Li absorption (``Li")
or emission in the Balmer, Paschen or Brackett lines of hydrogen (``e").}
\end{deluxetable*}

%

%
\subsection{Class II census results}
\label{sec:select:class2}

\subsubsection{Membership}
\label{sec:select:class2:members}

In this section we explore the membership status of the 136 class II candidates identified in \S\ref{sec:select:sed}, 
revealing that the vast majority of them are confirmed spectroscopicallly as members.
Seventy six of our 136 $\iracalpha$ selected class II objects were cataloged previously as members of IC~348 
\citep{1998ApJ...497..736H, 1998ApJ...508..347L, 1999ApJ...525..466L, 2003ApJ...593.1093L, 2005ApJ...618..810L}.
For this paper we obtained optical and near-infrared spectroscopy of  34 more class II sources; these
observations are detailed in  Appendix \ref{app:spex} and in Table \ref{tab:spex} we list new members with spectral types.
From the remaining 26 class II candidates, we  identified an additional eight sources whose SEDs suggest they are
high quality candidates (55308, 10031, 1287, 1379, 22865, 753; see Figure \ref{fig:class2nos} and Table \ref{tab:nospex}).
The four latter objects are very faint $(H>16$; see also Figure \ref{fig:complete}), and 
if they are cluster members rather than background sources (e.g. galaxies) 
then they are almost certainly brown dwarfs given their low luminosities. 
Three sources classified initially as class II sources using \irac\ data 
were reclassified as protostellar (\S\ref{sec:select:protostars}) based on their \mips\ SEDs.
The remaining sources were class II contaminants, consisting of either HH knots (2) or false excesses
sources detected in only 3 bands and contaminated by nebular emission (13).
We conclude that the technique of using $\iracalpha$ as a discriminator of class II YSOs is
successful for roughly 90\% of the initial class II sample (118 members from 136 candidates).

\begin{figure}
    \includegraphics[angle=90,width=0.475\textwidth]{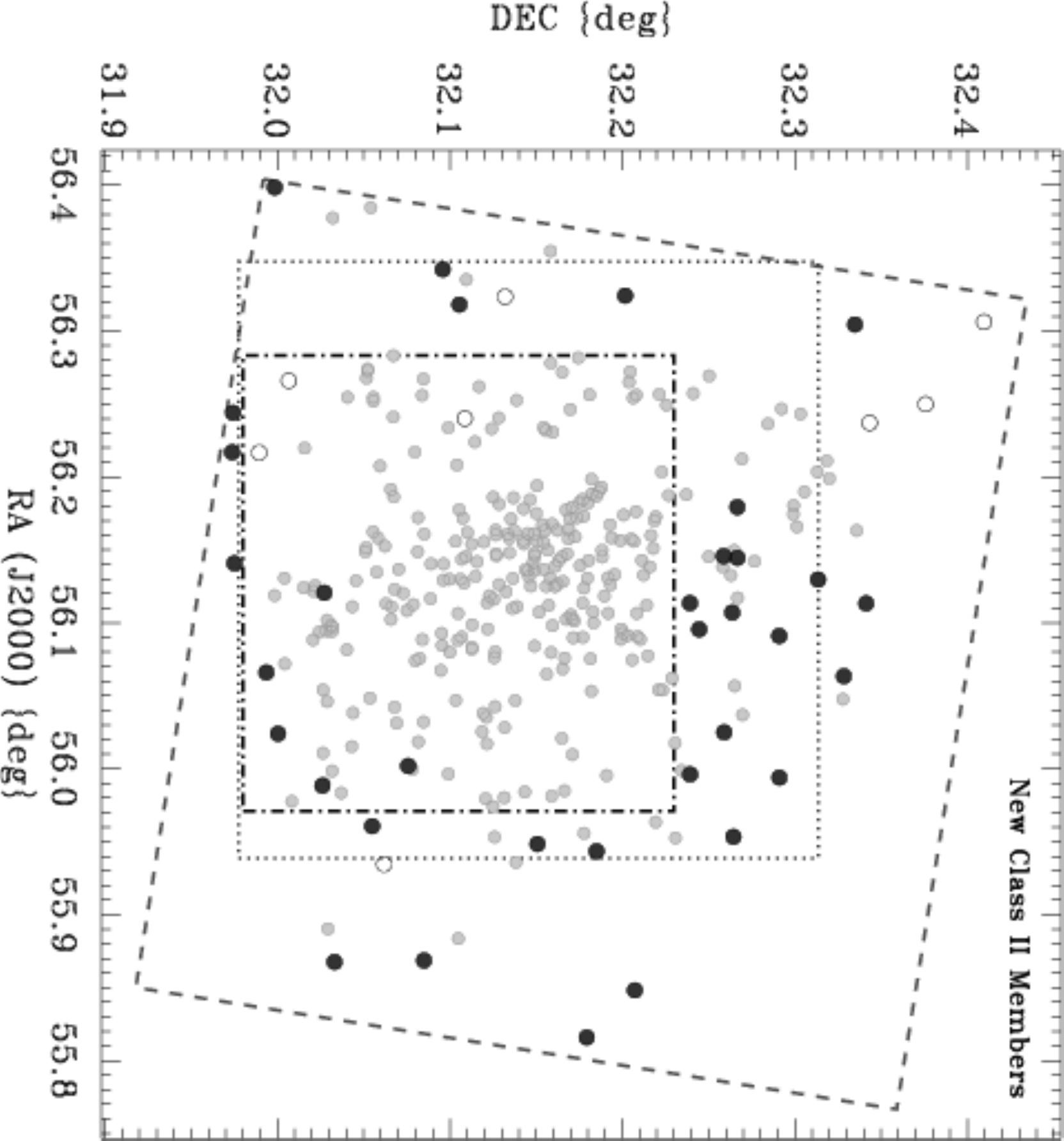}
\caption{Comparison of our \sst\ survey region (dashed box) to the \protect\citet{2003ApJ...593.1093L}
$\av<4,\;\mass>0.03\solarmass$ complete census region (dot-dashed box) and the 
\protect\citet{2003AJ....125.2029M}~FLAMINGOS near-IR survey (dotted box). 
Filled dark circles are new class II sources with spectral types; unfilled circles are
class II sources identified from their SEDs (\S\ref{sec:select:class2}) but 
lack optical/near-IR spectra; light filled circles are previously known members. 
\label{fig:map1}}
\end{figure}

Figure \ref{fig:map1}  compares the locations of our new class II sources to previous deep IR/spectroscopic census. 
New class II sources roughly correlate spatially with previously known members although all but 8 lie outside the 
\citet{2003ApJ...593.1093L} $\av<4;\;\mass>0.03\solarmass$ completeness region. Inside that survey region 
five new class II members are deeply embedded $(\av>4)$ in a dark molecular gas cloud at the cluster's
southwestern boundary,  while three are very faint, likely lying below that survey's $0.03\solarmass$ limit 
(e.g. \# 1379) and lack spectroscopic followup.  Most (27) of the new class II members fall within the boundaries of the 
\citet{2003AJ....125.2029M} near-infrared survey and confirmed cluster members can now be found as far 
as 2~pc from the cluster core.  Compared to $40\:\pm\:6$ unidentified $K<13$ members predicted
by \citet{2006A&A...445..999C} we found 23 new class II and 4 class I with $K<13$ while our disk based SED selection
criteria could not have revealed new class III members, 
which outnumber class II members by a factor of two.  If we consider that the surface density excess seen 
in the \citeauthor{2006A&A...445..999C} 2MASS map of IC~348 extends well beyond the borders of our
\sst\ survey then we would conclude that \citeauthor{2006A&A...445..999C} has underestimated somewhat
the true population size at larger radii.  A simple ratio of 2MASS excess to \sst\ survey areas suggests a
correction factor of 3-4.  Section \ref{sec:discuss:extent}
includes further discussion of the total cluster population size inferred from our \sst\ survey statistics.

\subsubsection{Completeness}
\label{sec:select:class2:complete}

We explored the completeness of our class II membership as affected by the selection 
requirements we used when identifying new candidates and by the depth of our spectroscopic observations.  
Intrinsically, our \sst\ census is  very sensitive to  faint sources while insensitive to the effects of dust extinction.  
For example, the $m_{\SIc}$ magnitude limit in the \sst\ color-magnitude diagram of Figure \ref{fig:cmds}a 
corresponds to the ability to detect a diskless 10 My $20\:\jupmass$\ brown dwarf 
\citep[$K\sim\:15.6; K-\SIc\sim\:0.6$;][; see their Figure 12]{2003AJ....125.2029M}
at multiple \sst\ wavelengths.; further, we could easily detect a 3 My brown dwarf seen through through $\sim\:40$ 
visual  magnitudes of extinction.  Our first selection requirement, requiring detection at  three bands 
short-ward of $8\micron$, would have included $80\%$ of the known $H\:<\:16$ IC~348 members examined in Paper I;  
those missing were primarily class III members (those whose SEDs  lack disk excess signatures).  
We were more concerned the application of two photometric constraints, $m_{\SIc}\:<\:15$ and $m_{err}\:<\:0.2$~mag 
and how these filters might affect the completeness of our census.

\begin{figure*}
    \centering \includegraphics[angle=90,totalheight=0.5\textwidth]{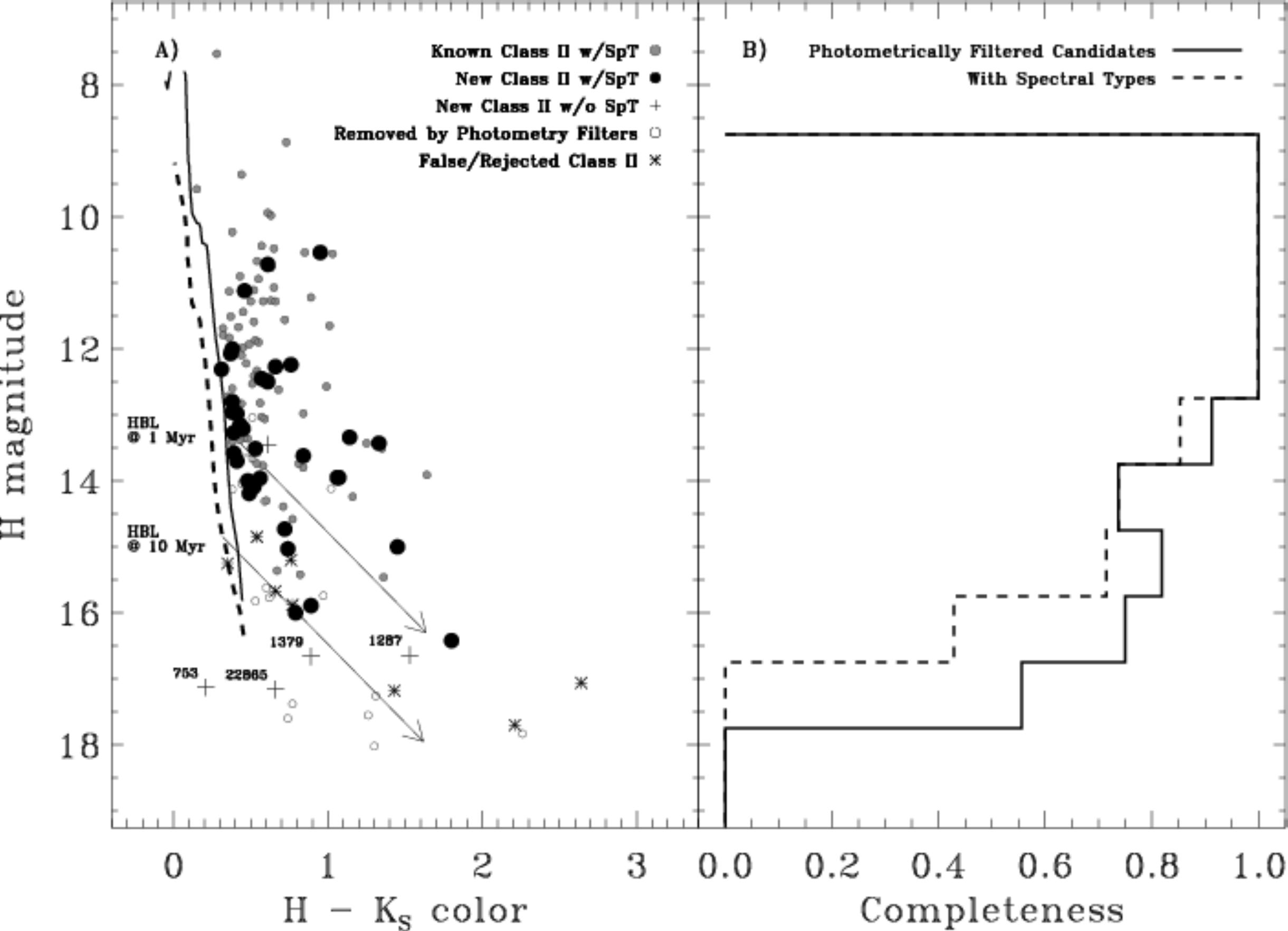}
\caption{Completeness of the IC~348 class II census.
A) $H-K$ vs $K$ color-magnitude diagram  plotted for all class II candidates independent
of \sst\ photometric quality. 
Symbols correspond to: 
class II $(-0.5>\iracalpha>-1.8)$ members with previously known spectral types (light filled circles); 
class II members with new spectra (dark filled circles); 
candidate class II members w/o spectral types (crosses);
potential class II candidates filtered from this study due faint or poor \sst\ photometry (open circles);  
and those class II candidates rejected as members based upon further SED analysis (asterisks).  
Isochrones and the hydrogen burning limit are shown for 1 and 10 Myr from \protect\citet{1998A&A...337..403B};  
B) The completeness fraction from the ratio of $H$ band luminosity functions.
Solid line: the ratio of photometrically filtered class II candidates to all class II candidates;
Dashed line: the ratio of class II sources with spectral types to all class II candidates.  See text.
\label{fig:complete}}
\end{figure*}

Figure \ref{fig:complete}a is an $H-K/H$ CMD for all potential class II candidates. 
This includes the 136 class II candidates and those sources excluded by our photometric constraints; 
each category is plotted with different symbols.  
Using the $H$ band magnitude as our proxy for the mass+age+extinction limits of this study,
the ratio of the photometrically filtered to the unfiltered Class  II $H$ band LFs gives an 
estimate  of our incompleteness due to these quality filters  (Figure \ref{fig:complete}b).  
Sources photometrically filtered from our catalog correspond to about $10-20\%$ of the sample over a
range of $H$ magnitude, due probably to the variable intensity of the nebular background.   
Our class II census is probably more complete than suggested ($>\:80\%$ complete for $H\:<\:16$) 
because some fraction of the photometrically filtered class II  candidates 
would have been rejected as non-members.  
Most of our high quality candidate members lacking spectra are faint. 
Dividing the $H$ LF of our spectroscopic  sample by the $H$ LF of the unfiltered class II 
candidates yields a similar completeness limit: $>\:80\%$ complete for $H\:<\:16$.

\begin{deluxetable}{lcccr}
\tablewidth{0pt}
\tablecaption{\sst~excess sources without spectral types\tablenotemark{a}\label{tab:nospex}}
\tablehead{
\colhead{ID} &
\colhead{$\alpha$(J2000)} &
\colhead{$\delta$(J2000)} &
\colhead{$f_{p}$} &
\colhead{Class}}
\startdata
     753 &   03 44 57.617 &    32 06 31.25 &    cfht  &    II   \\
    1287 &   03 44 56.904 &    32 20 35.86 &   cfht  &    II  \\
    1379 &   03 44 52.010 &    31 59 21.92 &   cfht  &    II  \\
    1401 &   03 44 54.690 &    32 04 40.28 &   cfht  &    I   \\
    1517 &   03 43 20.029 &    32 12 19.38 &    cfht  &    I\tablenotemark{b}   \\
    1898 &   03 44 43.893 &    32 01 37.37 &    2m  &     0/I   \\
    4011 &   03 44 06.914 &    32 01 55.35 &    cfht  &    I   \\
   10031 &   03 44 59.979 &    32 22 32.83 &    2m  &    II   \\
   21799 &   03 43 51.586 &    32 12 39.92 &   cfht &     I\tablenotemark{b}   \\
   22865 &   03 45 17.647 &    32 07 55.33 &    cfht  &    II   \\
   22903 &   03 45 19.053 &    32 13 54.85 &    cfht  &    I\tablenotemark{b}   \\
   40150 &   03 43 56.162 &    32 03 06.11 &    irac  &    I   \\
   40182 &   03 45 03.838 &    32 00 23.54 &    irac &    II   \\
   52590 &   03 44 20.384 &    32 01 58.45 &    irac  &    I   \\
   52648 &   03 44 34.487 &    31 57 59.60 &    irac  &    I   \\
   54299 &   03 43 44.284 &    32 03 42.41 &    irac  &    II   \\
   54361 &   03 43 51.026 &    32 03 07.74 &    irac  &    I   \\
   54362 &   03 43 50.948 &    32 03 26.24 &    irac   &    I   \\
   54419 &   03 43 59.400 &    32 00 35.40 &    irac  &    I   \\
   54459 &   03 44 02.415 &    32 02 04.46 &    irac  &    I   \\
   54460 &   03 44 02.622 &    32 01 59.58 &    irac  &    I   \\
   55308 &   03 45 13.497 &    32 24 34.68 &    irac  &    II   \\
   55400 &   03 44 02.376 &    32 01 40.01 &    irac  &    I   \\
   57025 &   03 43 56.890 &    32 03 03.40 &    m24m  &    0   \\
  HH-211 &   03 43 56.770 &    32 00 49.90 &    m70m   &    0 \\ 
\enddata
\tablenotetext{a}{Column descriptions same as in Table \ref{tab:spex}.}
\tablenotetext{b}{These class I sources are located away from any
molecular material and may be background sources with SEDs that 
mimic circumstellar disks.}
\end{deluxetable}
\begin{deluxetable}{lcccc}
\tablewidth{0pt}
\tablecaption{Non-members\tablenotemark{a}\label{tab:nonmembers}}
\tablehead{
\colhead{ID\tablenotemark{a}} &
\colhead{$\alpha$(J2000)} &
\colhead{$\delta$(J2000)} &
\colhead{$f_{p}$} &
\colhead{Spectra} }
\startdata
398      & 03 43 43.28  &  32 13 47.3 & cfht     & op   \\
424      & 03 43 43.11  &  32 17 47.7 & cfht     & op   \\
1920     & 03 43 23.55  &  32 09 07.8 & cfht    & op  \\
22898    & 03 45 18.713 &  32 05 31.0 & cfht  & IR  \\
40163    & 03 44 39.994 &  32 01 33.5 & irac    & IR \\
52827    & 03 45 14.012 &  32 06 53.0 & irac   & IR  \\      
52839    & 03 45 13.199 &  32 10 01.9 & irac    & IR  \\
\enddata
\tablenotetext{a}{The optical/near-IR spectra of these sources indicate they are 
galaxies (\S\ref{sec:select:lowlum}) or field stars (\S\ref{sec:imacsobs}). 
The wavelength regime of the spectral observation is given (op/IR). 
Column descriptions same as in Table \ref{tab:spex}.}
\end{deluxetable}

\subsection{Protostellar census}
\label{sec:select:protostars}

\begin{figure*}[!t]
    \centering   \includegraphics[angle=90,totalheight=0.45\textwidth]{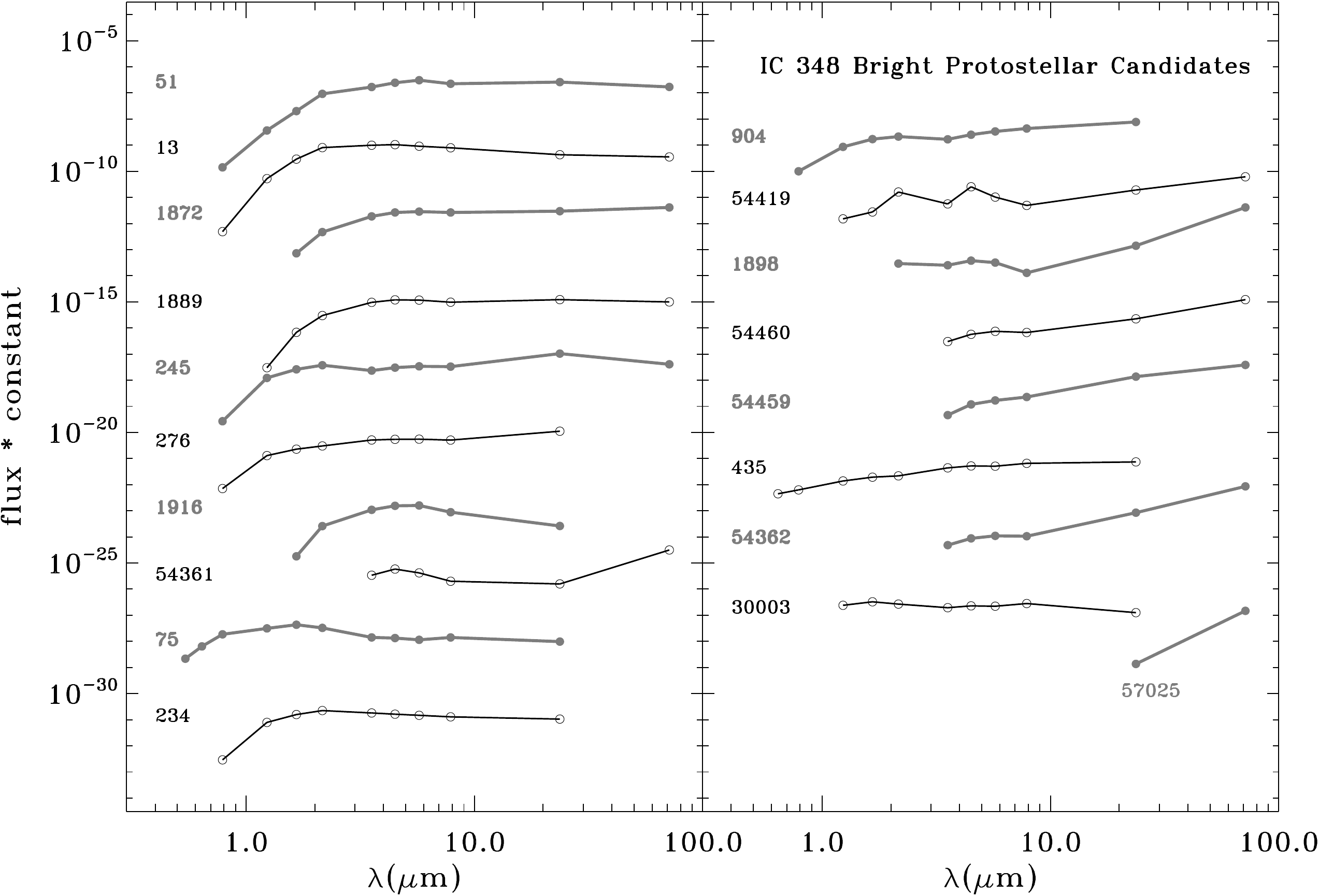}
\caption{Bright IC~348 protostars.  Most (15) of these sources were initially selected using
$\iracalpha>-0.5$ and three were added where the \irac\ SED is contaminated by shock or scattered
light emission (e.g., 54419~or~1898) and/or the \mips\ SED appears protostellar (\#54361).
The steep $24-70\micron$ SED of source \#57025, the apparent class 0 driving source for HH-797, 
is shown for comparison. Sources are ordered by decreasing $\SIc\micron$ flux.
Plotting symbols, line thickness and line color alternate from SED to SED for clarity.
\label{fig:bclass1}}
\end{figure*}

The results of scrutinizing the 56 \iracalpha\ selected class I candidates are given in this section.
Figure \ref{fig:bclass1} displays the spectral energy distributions of 
the 15 brightest class I protostellar ($\iracalpha>-0.5$)  candidates,
including \mips\ photometry out to $70\micron$. Sources are sorted on 
decreasing $\SIc\micron$ flux; all have $m_{\SIc}<12.5$, 
which should reduce the chance that they might be galaxies \citep{2006ApJ...645.1246J}.  
All of these sources are clearly protostars 
from their SEDs; previous speculation on the nature of some of these objects based 
on the association of such red \sst\ point sources 
with HH objects \citep{2006AJ....132..467W} appears to be confirmed.
There is an interesting apparent correlation of SED shape with $\SIc\micron$ flux.
As has been shown for protostars in Taurus \citep{1995ApJS..101..117K}, 
the most luminous IC~348 protostars are exclusively 
flat spectrum sources, while source SEDs longward of $10\micron$ become progressively 
steeper with decreasing source luminosity.
Moreover, the location of the flat spectrum sources in Figure \ref{fig:cmds}b mirrors 
another fact shown by the \citeauthor{1995ApJS..101..117K} 
Taurus study, namely, that  flat spectrum protostars are {\it intrinsically} more 
luminous than class II sources.  
If we were to ``deredden'' our flat spectrum protostars along the reddening vector in Figure \ref{fig:cmds}b 
then we would find them to be 2-3+ magnitudes brighter than essentially all other IC~348 members.  
This indicates to us that flat spectrum protostars have a star+disk+envelope 
structure distinct from class II sources and likely correspond to a different evolutionary phase.  
For comparison to the fainter steeper class I protostars we plot the steep 
slope of the $\SMa\rightarrow\SMb\micron$ \mips\ SED of \#57025,  
which lacks detection in \irac\ bands (it is placed on this plot using the $\SIc \micron$ upperlimit) and which 
corresponds to a previously known class 0 source that drives the HH-797 jet (\S\ref{sec:select:mips}).   

\begin{figure}
	\centering \includegraphics[angle=0,width=0.4\textwidth]{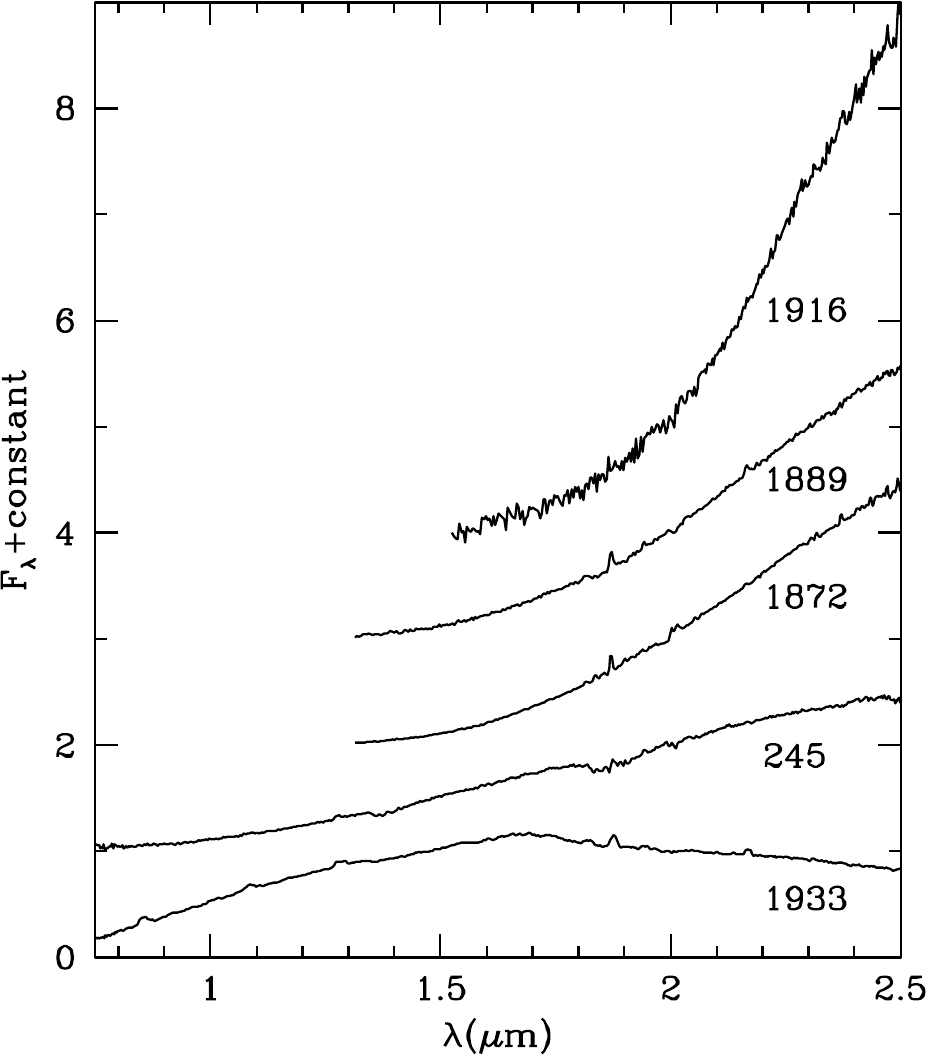}
\caption{SpeX near-IR spectra of candidate  members of IC~348 that show no detectable photospheric 
absorption features; these data have a resolution of $R=100$ and are normalized at $2\,\micron$.
We consider sources 245, 1872, 1889 and 1916, to be protostellar candidates based on their
broadband SEDs.    Source \#1933, while featureless in the IR, is an accreting (H$\alpha\sim$55\AA)~
K5 class II $(\iracalpha=-0.57)$ member, which we were able to type using optical spectra (\S \ref{sec:imacsobs};
see Figures \ref{fig:class2s} and \ref{fig:op-spec1})).
\label{fig:spec5}}
\end{figure}

Five of these bright protostars had existing spectroscopy to which we have added seven new spectra (see also 
Appendix \ref{app:spex}).  Seven of these twelve IC~348 protostars have M type spectra, 
ranging from M0 for the luminous IR source first identified by \citet[][(our source \#13)]{1974PASP...86..798S} 
to the newly typed faint M6 source \#30003, which is enshrouded in a scattered light cavity that can been seen in  
HST/optical \citep{2005ApJ...623.1141L}, near-IR (M03) and \sst\ $\SIb\micron$ images.    
Spectral types were not measurable for the other five sources because no absorption 
features were detected in the infrared.  New featureless infrared spectra of four of these sources are shown in 
Figure~\ref{fig:spec5}; a spectrum of the  fifth, \#51, appeared in  \citet{1998ApJ...508..347L}.  
Based on their mid-IR SEDs, the featureless nature of their near-IR spectra is probably due to veiling by 
continuum emission from circumstellar material \citep{1992MNRAS.258..399C,1996AJ....112.2184G} 
though these spectra do not exclude the possibility that they are embedded, early type (thus hotter) YSOs.
Much hotter YSOs (corresponding to A or B type) are excluded because we do not observed the characteristics
of massive protostars, namely, very large bolometric luminosities $(>100\,\lbol)$, hydrogen absorption lines and/or 
evidence of embedded \htwo\ regions.  Given the  presence of hydrogen emission lines in a number of the 
objects and their proximity to other class I objects and mm cores, 
it is very likely they are low mass members of IC~348 rather than massive
members or background sources. Three additional bright class I candidates are physically associated with 
molecular cloud cores (54460, 54459, 54362) but lack spectra. All three of these sources have \mips\ SEDs consistent with 
significant reprocessing of their emergent flux by cold  envelopes, and \citet{2006A&A...456..179T} recently identified a 
molecular outflow associated with \#54362. We tabulated all of these bright sources as protostellar members of IC~348; 
again, sources with spectra are listed in Table \ref{tab:spex}; those without are in Table \ref{tab:nospex}.

Additionally, we reclassified three \iracalpha\ selected class II sources as protostellar based upon their $24-70\,\micron$ SEDs.  
Sources \#1898, 54361 and \#55419, which all appeared as nebulous blobs in the near-IR images of M03, 
appear as point sources in \sst\ data, have class II \irac\ SED slopes yet have sharply rising \mips\ SEDs 
(See Figure \ref{fig:bclass1}).  Source \#1898 is infact the brightest far-IR source  in IC~348 
(fluxes of $\sim10\mbox{ Jy }@ 70\micron$ and $\sim60\mbox{ Jy }@ 160\micron$) and is almost 
certainly a newly identified protostellar member (see also \S\ref{sec:visual:swridge:3}; Figure \ref{fig:color3}a).  
The slope of its 70/160\micron\ SED is 1.2 compared to 1.1 and 2.4, respectively for the HH-797 and 
HH-211 class 0 sources. Were it not for its detection in scattered light in the near-IR and that a 
strong molecular outflow has not (yet) been found, its far-IR SED would suggest that it is
also a class 0 source.  Source \#54361 is a point source from  $\SIa\mbox{ to }\SMa\micron$ 
but is blended with \#54362 at $\SMb\micron$; the $\SMb\micron$ emission is elongated N-S,
peaks right between \#54361 and \#54362 and cannot be ascribed confidently to either.
This source also  appears to lie along the axis of the \#54362 molecular outflow found recently by 
\citet{2006A&A...456..179T},  who suggested that \#54361 may be a bright $24\micron$ but
unresolved knot of shocked gas instead of a young embedded star.  
Source \#55419 also appears to be blended with parts of the HH-211 outflow and is detected 
at $\SMb\micron$.  Additional spectroscopic data may clarify these latter 2 candidates' 
true nature;  in this work we have included them as candidate class I sources. 

\subsubsection{Low luminosity protostellar candidates}
\label{sec:select:lowlum}

\begin{figure*}
    \centering       \includegraphics[angle=90,totalheight=0.6\textwidth]{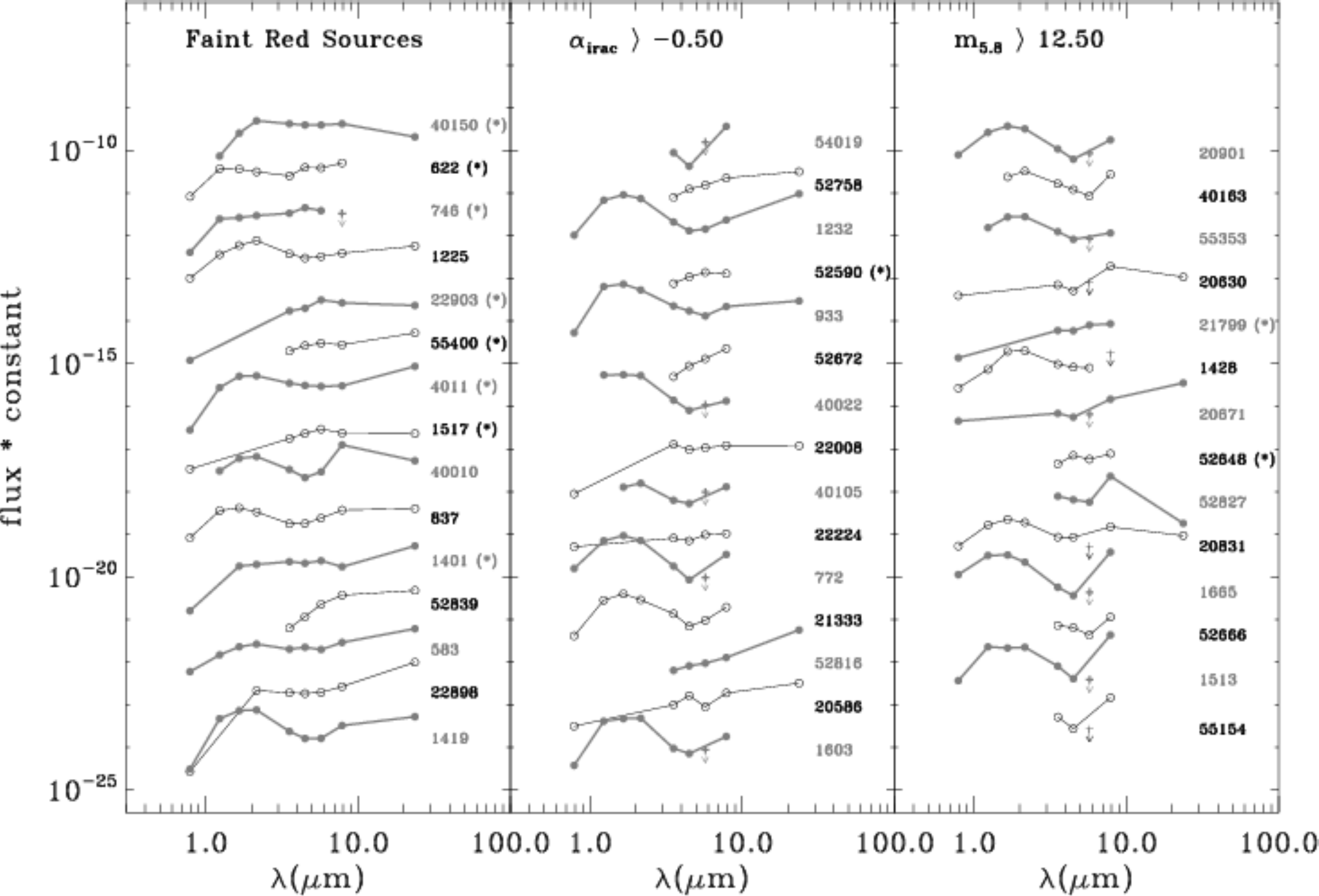}
\caption{Observed spectral energy distributions of faint $(m_{\SIc}>12.5)$ candidate 
protostellar $(\iracalpha>-0.5)$~objects ordered by decreasing $\SIc\micron$ flux.  
Clearly a mix of source types is present at these faint magnitudes and PAH rich sources, 
which are identified by the strongly inflected $8\micron$ SED point (e.g., \#40010),  
are excluded from our study (\S \ref{sec:select:lowlum}).  The near-IR spectra of 
monotonically increasing  SEDs like \#52839 indicate these are also likely extra-galactic 
interlopers  (Figure \ref{fig:faintspectra}). 
Plotting symbols, line thickness and line color alternate from SED to SED for clarity.
Sources distinguished as good candidate low-luminosity protostars are marked with (*) 
next to their id (see text).
\label{fig:fclass1}}
\end{figure*}

\begin{figure}
    \centering  \includegraphics[angle=0,width=0.25\textwidth]{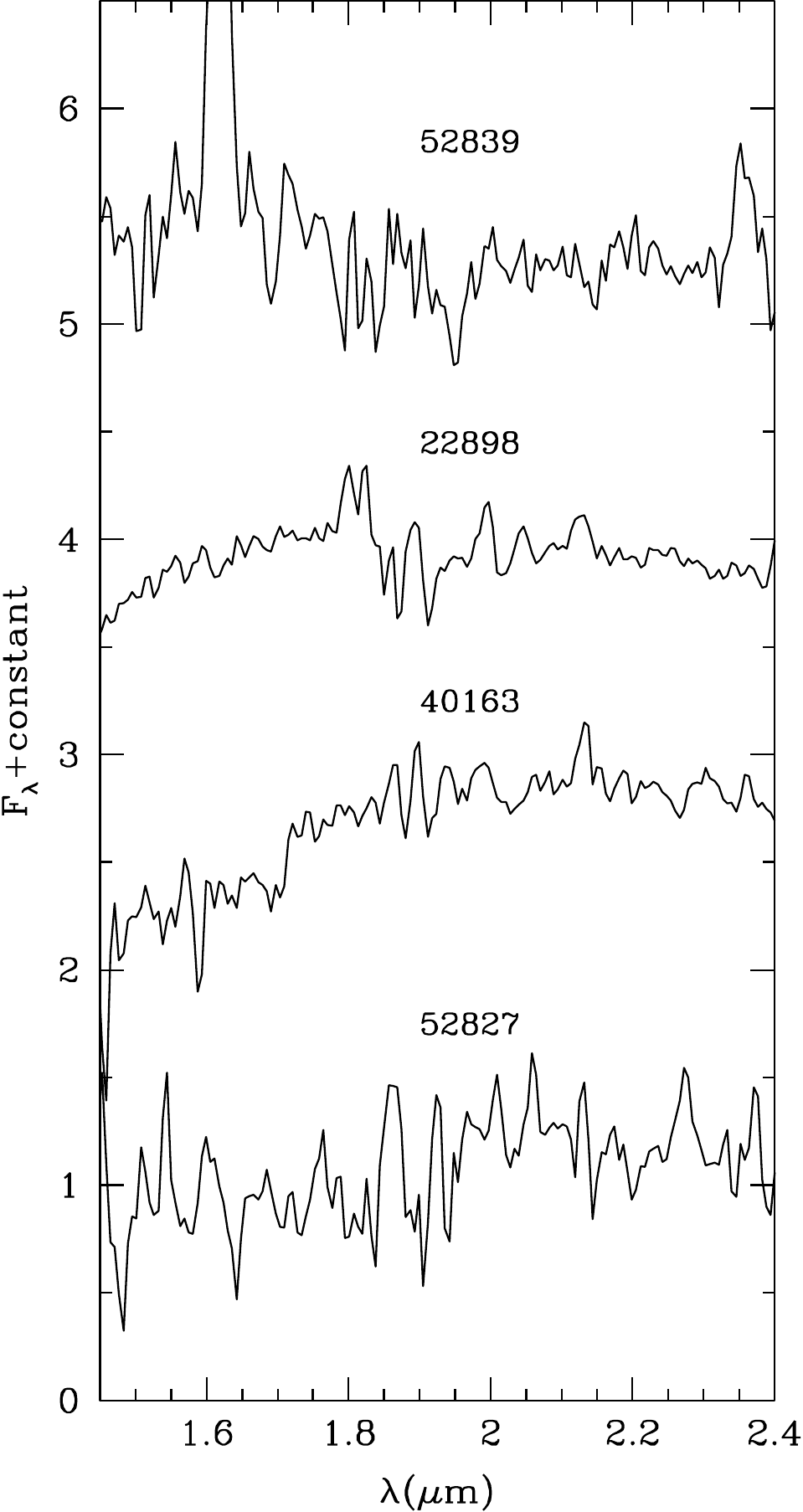}
\caption{Keck NIRC spectroscopy of candidate low-luminosity protostars $(\iracalpha>-0.5)$.
Despite their rising mid-IR SEDs (Figure \ref{fig:fclass1}), 
these sources are not intrinsically red nor do they show absorption features typical of brown dwarfs.  
They are likely galaxies; source \#52839 displays emission lines that do not 
correspond to typical transitions observed in young stars.
\label{fig:faintspectra}}
\end{figure}

Finally, we noted an interesting trend in Figure \ref{fig:cmds}b: most of the class II candidates are 
bright, while most of the class I candidates are very faint.  
Although this low luminosity range has a high likelihood of galaxy contamination,
it is important to investigate these faint candidates to search for low luminosity young
stars that would be missed by the flux limits suggested by \citet{2006ApJ...645.1246J} and co-workers.  
We began our exploration of these sources by plotting in Figure \ref{fig:fclass1} the SEDs of the 
41~faint $\iracalpha>-0.5$ protostellar candidates, sorting them by $\SIc\micron$ magnitude (or its upper limit).%
The ensemble population is clearly dominated by a class of objects with non-power law SEDs,
which was a fact previously evident in the poor quality of many of the class I SED power-law fits (Figure \ref{fig:iracalpha}). 
Many have stellar-like continuum out to  5 microns with sharply inflected and rising SEDs beyond.  
Such a SED feature can be ascribed to PAH emission at 6 and 8 microns, which appear in galaxies and 
evolved stars \citep{2006ApJ...637L..45J}. 
To substantiate this point,  we obtained Keck NIRC \citep{1994iaan.conf..239M} 
$HK$  spectra of 4 of these red low luminosity sources.
Two of these targets have monotonically rising \sst\ SEDs, while 2 have sharp $8\micron$ inflections.  
These spectra, which were  obtained on 23 November 2004, are shown in Figure \ref{fig:faintspectra}.
The fact that these sources are not very red (especially compared to those spectra in Figure \ref{fig:spec5}) 
indicates they are not class I objects, and the lack of steam indicates they are not brown dwarfs.  
They are probably all galaxies. Source \#52839  is almost certainly a galaxy based on its emission lines, 
which do not correspond to rest-frame wavelengths of any lines typical of young stars.  

\begin{figure}
   \centering  \includegraphics[angle=90,width=0.45\textwidth]{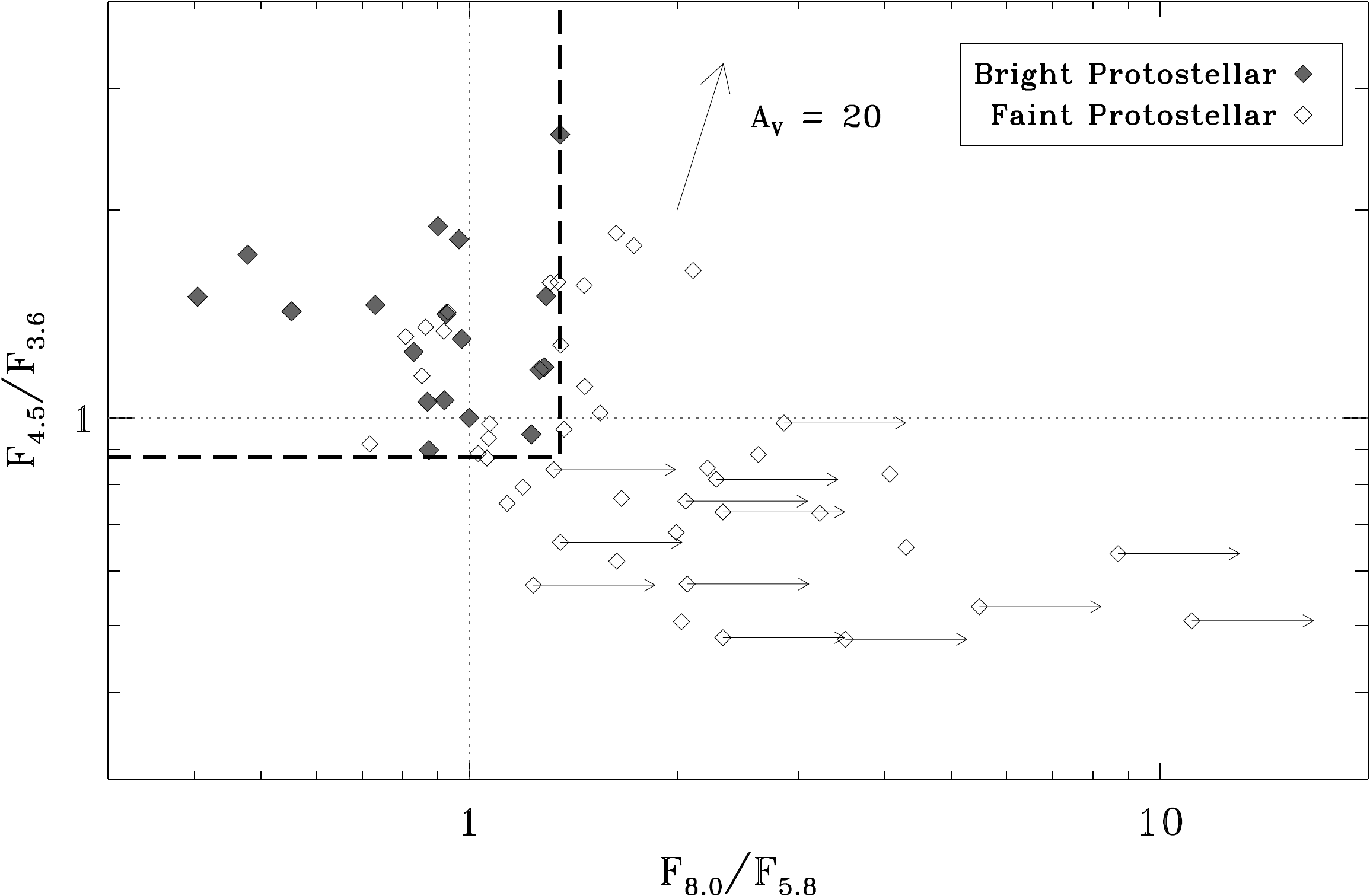}
\caption{Parsing faint YSOs from $6-8\micron$ PAH emission sources using \irac\ 
flux ratios. Flat spectrum flux ratios are marked with light dashed lines.
The locations of bright protostars in this diagram were used to select the best candidate
low luminosity protostars. The selection box is traced by heavy dashed lines and includes
sources in the upper left quadrant.  This acts to exclude sources with obvious PAH emission and
include sources with silicate absorption. The SEDs of these 11 candidates are 
tagged with (*)  in Figure \ref{fig:fclass1}.
Sources with $\SIc\micron$  upperlimits are marked with arrows.
\label{fig:fluxdiagram}}
\end{figure}

We chose to exclude {\it all} sources with PAH or similar features from our census of faint YSOs. 
To identify the best YSO candidates out of these faint sources and exclude PAH rich sources, we compared 
the monochromatic flux ratios $\SIb/\SIa\micron$ and $\SId/\SIc\micron$ of these faint candidate YSOs to these 
flux ratios for the brighter protostars none of which show obvious $6-8\micron$ PAH emission (Figure \ref{fig:fluxdiagram}).
Flat spectrum sources are located at (1,1) and sources with strong silicate absorption fall into the upper left quadrant. 
We traced a box around the locations of the brighter protostellar candidates in this diagram and chose the 
11 faint candidates within it as additional protostellar candidates.  
This box excluded the four Keck sources whose spectra are clearly not those of YSOs.

Some of these 11 low-luminosity class I candidates are more likely to be young stars than others. 
Two sources in particular have stellar (or sub-stellar) spectral features (\#622, M6; \#746, M5) and
two sources are close companions to bright class I sources (\#55400 and 40150).
Source \#4011 lies in the center of narrow dark lane/shadow clearly seen in the infrared images of \citet{2003AJ....125.2029M}. 
This strongly suggests it is a young star-disk system seen nearly edge-on, which is reinforced by the
presence of a jet (HH799) that was observed and associated with this source by \citet{2006AJ....132..467W}. 
However, the presence of an edge-on disk could cause a class II member to appear as a class I source \citep{1999ApJ...519..279C};
thus, the exact evolutionary stage of these young stars is unclear.
Edge-on geometries also cause sources to appear subluminous on the HR diagram due to the fact the optical/near-IR 
flux is likely scattered light, which leads to low measured values of extinction, while the mid-IR flux is still 
quenched by the disk extinction.  Both \#622 and \#746 are, for example, subluminous on the HR diagram.
Note, if the dust in the disk is grey, the reddening vector(s) in Figure 1 are vertical
and the basic IRAC SED classification remains nearly unchanged \footnote{See also the SED dependence
of nearly-edge on disks and dust settling in \citet{1999ApJ...527..893D}.}.

Although these faint class I candidates are spatially correlated near dark cores,
others lie far from the molecular cloud, including candidates \#1517 and 21799 to the NW.
These are likely extragalactic contaminants despite their convincing SEDs; for completeness,
all of the protostellar candidates lacking spectral confirmation are given Table \ref{tab:nospex}.
In total we find the accuracy of $\iracalpha$ for uniquely selecting class I sources is quite low 
(less than $\sim50\%$). While the  application of additional selection criteria (flux limits) like 
those used by \citet{2006ApJ...645.1246J} and we used in Figure \ref{fig:fluxdiagram} can improve the accuracy of 
a class I census, the reality is that galaxies and PAH sources  masquerading as protostars 
dominate the statistics even for this nearby young cluster and follow up spectroscopy is clearly 
needed to confirm low luminosity protostellar candidates. 

\begin{deluxetable*}{lrrrrrrrrrr}
\tablewidth{0pt}
\tablecaption{\sst\ $3-24\micron$ data for new IC~348 members \label{tab:fclass2}}
\tablehead{
\colhead{ID} &
\multicolumn{5}{c}{Magnitudes} &
\multicolumn{5}{c}{Uncertainties\tablenotemark{a}} \\
\colhead{} &
\colhead{$3.6\micron$} &
\colhead{$4.5\micron$} &
\colhead{$5.8\micron$} &
\colhead{$8.0\micron$} &
\colhead{$24\micron$} &
\colhead{$3.6\micron$} &
\colhead{$4.5\micron$} &
\colhead{$5.8\micron$} &
\colhead{$8.0\micron$} &
\colhead{$24\micron$} 
}
\startdata
        70 &    9.93 &    9.58 &    9.28 &    8.59 &    5.40 &    0.02 &    0.01 &    0.02 &    0.03 &    0.03    \\
       117 &   10.87 &   10.27 &    9.99 &    9.27 &    6.40 &    0.01 &    0.02 &    0.03 &    0.06 &    0.03    \\
       132 &   11.06 &   10.64 &   10.16 &    9.61 &    6.25 &    0.01 &    0.05 &    0.01 &    0.04 &    0.03    \\
       162 &   11.26 &   10.86 &   10.32 &    9.51 &    6.81 &    0.02 &    0.02 &    0.03 &    0.03 &    0.03    \\
       179 &   11.35 &   10.98 &   10.47 &    9.60 &    7.19 &    0.02 &    0.05 &    0.03 &    0.05 &    0.03    \\
       199 &   11.96 &   11.59 &   11.16 &   10.54 &    7.79 &    0.03 &    0.05 &    0.06 &    0.03 &    0.05    \\
       215 &   11.08 &   10.73 &   10.28 &    9.62 &    5.91 &    0.01 &    0.02 &    0.03 &    0.02 &    0.03    \\
       231 &   11.82 &   11.28 &   10.89 &   10.02 &    6.66 &    0.01 &    0.06 &    0.05 &    0.02 &    0.03    \\
       234 &   11.57 &   10.94 &   10.30 &    9.46 &    6.13 &    0.02 &    0.03 &    0.05 &    0.03 &    0.03    \\
       245 &   10.05 &    9.02 &    8.15 &    7.18 &    2.40 &    0.01 &    0.03 &    0.03 &    0.01 &    0.03    \\
       265 &   11.00 &   10.43 &    9.85 &    9.15 &    4.59 &    0.02 &    0.06 &    0.04 &    0.02 &    0.03    \\
       280 &   12.22 &   11.99 &   11.69 &   11.07 &    5.87 &    0.01 &    0.01 &    0.05 &    0.04 &   -9.00    \\
       321 &   12.70 &   12.45 &   12.12 &   11.40 &    5.61 &    0.01 &    0.03 &    0.07 &    0.08 &   -9.00    \\
       327 &   12.58 &   12.24 &   11.79 &   11.11 &    8.60 &    0.01 &    0.02 &    0.05 &    0.05 &    0.04    \\
       364 &   12.04 &   11.66 &   11.04 &   10.16 &    6.46 &    0.01 &    0.04 &    0.03 &    0.09 &    0.03    \\
       368 &   12.74 &   12.32 &   12.01 &   11.35 &    7.11 &    0.01 &    0.02 &    0.08 &    0.08 &    0.06    \\
       406 &   13.15 &   12.81 &   12.58 &   12.00 &    9.09 &    0.01 &    0.02 &    0.06 &    0.14 &    0.04    \\
       643 &   13.19 &   12.87 &   12.25 &   11.54 &    8.86 &    0.01 &    0.02 &    0.03 &    0.04 &    0.04    \\
       723 &   12.18 &   11.67 &   11.42 &   10.62 &    7.44 &    0.01 &    0.03 &    0.04 &    0.05 &    0.03    \\
       753 &   16.03 &   15.45 &   14.88 &   13.25 &    6.83 &    0.12 &    0.12 &    0.15 &   -9.00 &   -9.00    \\
       904 &   12.91 &   11.72 &   10.66 &    9.39 &    5.22 &    0.05 &    0.01 &    0.05 &    0.02 &    0.03    \\
      1287 &   13.40 &   12.82 &   12.26 &   11.82 &    7.98 &    0.02 &    0.10 &    0.07 &    0.05 &    0.04    \\
      1379 &   15.02 &   14.60 &   14.12 &   13.65 &    8.35 &    0.03 &    0.04 &    0.04 &    0.10 &   -9.00    \\
      1401 &   15.07 &   14.43 &   13.53 &   12.89 &    8.13 &    0.02 &    0.04 &    0.06 &    0.13 &    0.13    \\
      1517 &   15.35 &   14.32 &   13.31 &   12.54 &    9.04 &    0.04 &    0.03 &    0.11 &    0.08 &    0.04    \\
      1679 &   11.30 &   11.07 &   10.74 &   10.28 &    6.47 &    0.01 &    0.02 &    0.03 &    0.03 &    0.03    \\
      1683 &   12.31 &   12.04 &   11.62 &   10.85 &    7.55 &    0.02 &    0.02 &    0.07 &    0.04 &    0.03    \\
      1707 &   13.04 &   12.59 &   12.29 &   11.50 &    8.68 &    0.02 &    0.02 &    0.06 &    0.04 &    0.04    \\
      1761 &   12.77 &   12.54 &   12.13 &   11.60 &    9.04 &    0.01 &    0.03 &    0.05 &    0.04 &    0.05    \\
      1833 &   12.10 &   11.79 &   11.52 &   10.92 &    8.03 &    0.01 &    0.02 &    0.04 &    0.04 &    0.03    \\
      1843 &   13.88 &   13.36 &   12.76 &   11.97 &    5.04 &    0.03 &    0.05 &    0.08 &    0.06 &   -9.00    \\
      1872 &    7.78 &    6.67 &    5.84 &    4.92 &    1.26 &    0.01 &    0.00 &    0.05 &    0.02 &    0.03    \\
      1881 &   10.99 &   10.75 &   10.50 &    9.97 &    6.44 &    0.02 &    0.03 &    0.06 &    0.04 &    0.03    \\
      1889 &    9.76 &    8.78 &    8.06 &    7.26 &    3.48 &    0.02 &    0.03 &    0.03 &    0.03 &    0.03    \\
      1890 &   12.11 &   11.86 &   11.55 &   11.10 &    8.63 &    0.01 &    0.03 &    0.04 &    0.04 &    0.04    \\
      1898 &   12.46 &   11.28 &   10.72 &   10.70 &    4.57 &    0.12 &    0.11 &    0.17 &    0.29 &    0.04    \\
      1905 &    9.30 &    8.98 &    8.78 &    8.25 &    5.75 &    0.02 &    0.03 &    0.04 &    0.03 &    0.03    \\
      1916 &   10.87 &    9.75 &    8.97 &    8.61 &    6.40 &    0.03 &    0.02 &    0.02 &    0.04 &    0.03    \\
      1923 &   13.11 &   12.54 &   12.00 &   11.42 &    7.62 &    0.02 &    0.02 &    0.06 &    0.05 &    0.04    \\
      1925 &   12.66 &   12.09 &   11.50 &   10.87 &    7.52 &    0.02 &    0.03 &    0.05 &    0.04 &    0.03    \\
      1933 &    8.09 &    7.42 &    6.96 &    6.06 &    3.26 &    0.01 &    0.01 &    0.03 &    0.02 &    0.03    \\
      4011 &   14.61 &   14.00 &   13.30 &   12.28 &    7.60 &    0.04 &    0.03 &    0.10 &    0.06 &    0.03    \\
     10031 &   12.59 &   12.39 &   12.01 &   11.30 &    8.26 &    0.01 &    0.04 &    0.03 &    0.03 &    0.04    \\
     10120 &   11.99 &   11.71 &   11.41 &   11.00 &    8.04 &    0.02 &    0.02 &    0.04 &    0.04 &    0.04    \\
     10176 &   13.36 &   13.01 &   12.63 &   12.10 &    7.59 &    0.02 &    0.03 &    0.06 &    0.04 &   -9.00    \\
     10219 &   11.28 &   10.96 &   10.69 &    9.87 &    6.98 &    0.01 &    0.01 &    0.05 &    0.01 &    0.03    \\
     10305 &   14.33 &   13.82 &   13.55 &   12.89 &    7.30 &    0.03 &    0.02 &    0.16 &    0.15 &   -9.00    \\
     21799 &   16.51 &   15.79 &   14.71 &   13.64 &    8.37 &    0.06 &    0.13 &    0.20 &    0.20 &   -9.00    \\
     22232 &   12.05 &   11.69 &   11.34 &   10.38 &    7.67 &    0.02 &    0.02 &    0.10 &    0.03 &    0.04    \\
     22865 &   15.40 &   15.03 &   14.48 &   13.60 &    9.86 &    0.02 &    0.05 &    0.05 &   -9.00 &   -9.00    \\
     22903 &   15.37 &   14.48 &   13.24 &   12.41 &    9.03 &    0.04 &    0.03 &    0.08 &    0.03 &    0.04    \\
     30003 &   13.99 &   13.08 &   12.36 &   11.11 &    8.44 &    0.06 &    0.03 &    0.10 &    0.07 &    0.12    \\
     40150 &   14.39 &   13.73 &   12.98 &   11.91 &    9.14 &    0.03 &    0.04 &    0.04 &    0.07 &    0.06    \\
     40182 &   14.37 &   13.72 &   13.29 &   12.75 &    9.34 &    0.12 &    0.03 &    0.04 &    0.13 &   -9.00    \\
     52590 &   16.26 &   15.13 &   14.12 &   13.20 &    6.61 &    0.05 &    0.04 &    0.05 &    0.18 &   -9.00    \\
     52648 &   16.84 &   15.61 &   15.08 &   13.76 &    8.47 &    0.04 &    0.07 &    0.09 &    0.17 &   -9.00    \\
     54299 &   13.59 &   12.85 &   12.32 &   11.82 &    5.66 &    0.02 &    0.02 &    0.05 &    0.10 &   -9.00    \\
     54361 &   10.88 &    9.55 &    9.17 &    8.97 &    5.69 &    0.03 &    0.02 &    0.06 &    0.03 &    0.03    \\
     54362 &   14.25 &   12.87 &   11.87 &   10.91 &    5.12 &    0.07 &    0.08 &    0.05 &    0.04 &    0.03    \\
     54419 &   12.84 &   10.46 &   10.70 &   10.50 &    5.49 &    0.09 &    0.05 &    0.08 &    0.06 &    0.04    \\
     54459 &   14.31 &   12.54 &   11.41 &   10.09 &    4.60 &    0.14 &    0.06 &    0.07 &    0.04 &    0.03    \\
     54460 &   13.52 &   12.08 &   11.04 &   10.16 &    5.32 &    0.04 &    0.06 &    0.07 &    0.04 &    0.03    \\
     55308 &   12.01 &   11.64 &   11.51 &   10.81 &    7.87 &    0.02 &    0.02 &    0.04 &    0.03 &    0.03    \\
     55400 &   15.22 &   14.17 &   13.27 &   12.37 &    8.13 &    0.10 &    0.04 &    0.13 &    0.07 &    0.06    \\
     57025 &  \nodata &   \nodata &  \nodata &\nodata &    7.10 &\nodata &   \nodata &   \nodata &   \nodata &    0.04    \\
    HH-211 & \nodata &   \nodata &   \nodata &   \nodata &   7.24  &   \nodata &  \nodata &   \nodata &  \nodata &    -9.00    \\
\enddata
\tablenotetext{a}{The listed magnitude is an upper limit if the listed uncertainty is given as -9. }
\end{deluxetable*}

\begin{deluxetable*}{lrrrrcccl}
\tablewidth{0pt}
\tablecaption{Far-IR \& submm flux densities for IC~348 protostars \label{tab:fproto}}
\tablehead{
\colhead{ID} &
\multicolumn{2}{c}{\mips\ $70\micron$} &
\multicolumn{2}{c}{\mips\ $160\micron$} &
\multicolumn{3}{c}{SCUBA $850\micron$\tablenotemark{c}} &
\colhead{Comments/} \\
\colhead{} &
\colhead{flux\tablenotemark{a}}  &
\colhead{unc\tablenotemark{b}}  &
\colhead{flux}  &
\colhead{unc}  &	
\colhead{$f_{20\arcsec}$}  &
\colhead{$f_{40\arcsec}$} &
\colhead{$S_{40\arcsec}$} &
\colhead{Blend~ID} 
}
\startdata
         13 &     2.601 & 0.368 &   14.492 &     -9.000 &     0.096 &    \nodata &     \nodata & MMP-10 \\
         51 &     3.944 & 0.453 &   19.126 &     -9.000 &     0.051 &     0.097 &        0.012 & \\
         75 &     5.078 &  -9.000 &  \nodata &    \nodata &     0.009 &   \nodata &      \nodata  & nebula \\
        234 &   \nodata &   \nodata &   \nodata &   \nodata &     0.110 &  \nodata &      \nodata  & \\
        245 &     0.945 & 0.262 &    \nodata &  \nodata &     0.049 &    \nodata &   \nodata & \\
        276 &     1.778 &  -9.000 & \nodata  &   \nodata &     0.031 &  \nodata &       \nodata &  \\
        435 &   \nodata &   \nodata &   \nodata &   \nodata &     0.006 & \nodata &   \nodata & nebula \\
        622 &   \nodata &   \nodata &   \nodata &   \nodata &     0.014 &   \nodata &     \nodata  & nebula \\
        746 &   \nodata &   \nodata &   \nodata &   \nodata &     0.012 &  \nodata &     \nodata  & \\
        904 &     1.144 &  -9.000 & \nodata &     \nodata &     0.037 &   \nodata &      \nodata  & \\
       1401 &   \nodata &   \nodata &   \nodata &   \nodata &  0.007 &   \nodata &      \nodata  & \\
       1517 &   \nodata &   \nodata &   \nodata &   \nodata &   \nodata &   \nodata &      \nodata  & Off SCUBA. \\
       1872 &     9.583 & 0.705 &   61.196 &      2.650 &     0.249 &     0.633 &        0.624  & 1898 \\
       1889 &     0.730 & 0.191 &       14.902 &  -9.000 &     0.058 &     0.193 &        0.205 &  \\
       1898 &     9.583 & 0.705 &      61.196 &   2.650 &     0.249 &     0.633 &        0.624 & 1872\\
       1916 &     0.248 &  -9.000 &   13.134 &    -9.000 &     0.034 &     0.116 &        0.055  & \\
       4011 &   \nodata &   \nodata &   \nodata &   \nodata &     0.110 &   \nodata &   \nodata&  \\
      21799 &   \nodata &   \nodata &   \nodata &   \nodata &   \nodata &   \nodata &      \nodata & Off SCUBA. \\
      22903 &   \nodata &   \nodata &   \nodata &   \nodata &   \nodata &   \nodata &      \nodata & Off SCUBA. \\
      30003 &     1.292 & -9.000 &   \nodata &    \nodata &     0.186 &   \nodata &    \nodata  & \\
      40150 &   \nodata &   \nodata &   \nodata &   \nodata &   0.327 &   \nodata &      \nodata & 57025 \\
      52590 &   \nodata &   \nodata &   \nodata &   \nodata &   0.005 &   \nodata &      \nodata &  \\
      52648 &   \nodata &   \nodata &   \nodata &   \nodata &  0.051 &   \nodata &      \nodata &  \\
      54361 &     2.320 & 0.337 &      26.142 &  -9.000 &     0.218 &     0.671 &        0.653  & 54362 \\
      54362 &     2.002 &  0.330 &    26.142 &     -9.000 &     0.218 &     0.671 &        0.653 & 54361 \\
      54419 &     0.452 &  0.207 &   \nodata &   \nodata &   0.042 &   \nodata &      \nodata \\
      54459 &     0.894 & 0.241 &   \nodata &    \nodata &     0.160 &   \nodata &    \nodata & 54460 \\
      54460 &     0.894 &  0.241 &  \nodata &      \nodata &     0.147 &  \nodata &    \nodata & 54459 \\
      55400 &   \nodata &   \nodata &   \nodata &   \nodata &   0.085 &   \nodata &      \nodata & \\
      57025 &     3.428 & 0.426 &      19.839 &   3.508 &     0.408 &     0.985 &        1.201 & 40150 \\
      HH211 &     2.854 &  0.388 &    48.655 &    4.803 &     0.618 &     1.695 &        1.617 & \\
\enddata
\tablenotetext{a}{All flux densities are in Janskys.}
\tablenotetext{b}{Sources with uncertainties equal to -9 correspond to $95\%$ upperlimits.}
\tablenotetext{c}{Aperture flux derivation same as Table \ref{tab:mmp}. Sources with SCUBA fluxes only in a $20\arcsec$
beam are 95\% upperlimits. }
\end{deluxetable*}

\begin{deluxetable*}{crrclll}
\tablecaption{Merged catalog of millimeter cores in IC~348\label{tab:mm}}
\tablehead{
\colhead{ID} &
\colhead{$\alpha$(J2000)} &
\colhead{$\delta$(J2000)} &
\colhead{$f_{p}$} &
\colhead{Other IDs} &
\colhead{Associated} &
\colhead{Comments} \\
\colhead{} &
\colhead{} &
\colhead{} &
\colhead{\tablenotemark{(a)}} &
\colhead{\tablenotemark{(b)}} &
\colhead{Protostars} &
\colhead{\tablenotemark{(c)}} 
}
\startdata%
MMS-01 & 3:44:43.7 & 32:01:32.3 & 3 & H05-14, Bolo116,    & 1898,  & Peak on       \\
\multicolumn{4}{c}{}                        & K034471+32015,    & 1872, & 1872/1898   \\
\multicolumn{4}{c}{}                        & SMM-07                &  234   &                    \\ 
MMS-02 & 3:44:21.4 & 31:59:20.3 & 4 & Bolo113,          & 1889      & \sst\ src. offset N. \\
\multicolumn{4}{c}{}                        & SMM-14          &              &                    \\ 
MMS-03 & 3:44:12.8 & 32:01:37.0 & 4 & SMM-17          & 51        & No Bolocam src.\\
MMS-04 & 3:44:05.0 & 32:00:27.7 & 2 & Bolo109          & 1916      & SCUBA peak $25\arcsec$ N.   \\
MMS-05 & 3:43:56.5 & 32:00:50.0 & 1 & H05-12, Bolo103,  & HH-211    &    \\
\multicolumn{4}{c}{}                         & K034393+32008   &              &    \\
\multicolumn{4}{c}{}                         & SMM-01              &              &                    \\ 
MMS-06 & 3:43:57.2 & 32:03:01.8 & 3 & H05-13, Bolo104, & 57025,    & \sst\ src.   \\
\multicolumn{4}{c}{}                        & K034395+32030,         & 40150     & offset SW. \\
\multicolumn{4}{c}{}                        & IC348-mm, HH-797      &              & \\
\multicolumn{4}{c}{}                        & SMM-02                      &              & \\
MMS-07 & 3:43:50.8 & 32:03:24.0 & 1 & H05-15, Bolo102,   & 54362,    & Peak on 54362. \\ 
            &               &                &    & K034383+32034    & 54361     &    \\ 
\multicolumn{4}{c}{}                         & SMM-03                &              & \\ \hline
MMP-01 & 3:45:16.8 & 32:04:46.4 & 4 & Bolo119          & Starless  & Near IRAS~03422+3156\\
MMP-02 & 3:44:56.0 & 32:00:31.3 & 2 & Bolo118          & Starless  & No SCUBA pt. src.\\
MMP-03 & 3:44:48.8 & 32:00:29.5 & 2 & H05-25, Bolo117  & Starless  & $24\micron$ abs. \\
\multicolumn{4}{c}{}                         & SMM-12                      &              & \\
MMP-04 & 3:44:36.8 & 31:58:49.0 & 1 & H05-19, Bolo115, & Starless  & $24\micron$ abs. \\
\multicolumn{4}{c}{}                         & K034460+31587    &           &    \\
\multicolumn{4}{c}{}                         & SMM-11                      &              & \\
MMP-05 & 3:44:14.1 & 31:57:57.0 & 4 & Bolo111          & Starless  & $24\micron$ abs.\\
\multicolumn{4}{c}{}                         & SMM-15                      &              & \\
MMP-06 & 3:44:06.0 & 32:02:14.0 & 4 & H05-22, Bolo110, & Starless  & SCUBA peak;\\
\multicolumn{4}{c}{}                         & K034410+32022    &           & $24\micron$ abs.   \\
\multicolumn{4}{c}{}                         & SMM-09                      &              & \\
MMP-07 & 3:44:05.4 & 32:01:50.0 & 4 & H05-20, Bolo110  & Starless  & SCUBA peak; $24\micron$ abs.\\
MMP-08 & 3:44:02.8 & 32:02:30.5 & 3 & H05-18, Bolo107,       & Starless  & SCUBA peak; \\
\multicolumn{4}{c}{}                         & K034405+32024        &           & $24\micron$ abs.   \\
\multicolumn{4}{c}{}                         & SMM-06                    &              & \\
MMP-09 & 3:44:02.3 & 32:02:48.0 & 1 & H05-21, Bolo107  & Starless  & SCUBA peak\\
MMP-10 & 3:44:01.3 & 32:02:00.8 & 3 & H05-16, Bolo106, & Starless  & $160\micron$ src?\\
\multicolumn{4}{c}{}                          & K034401+32019    &           &    \\
\multicolumn{4}{c}{}                         & SMM-05                    &              & \\
MMP-11 & 3:44:02.3 & 32:04:57.3 & 2 & Bolo108          & Starless  & No SCUBA pt. src.  \\
MMP-12 & 3:43:57.7 & 32:04:01.6 & 3 & H05-17, Bolo105, & Starless  &    \\
\multicolumn{4}{c}{}                         & K034395+32040    &           &    \\
\multicolumn{4}{c}{}                         & SMM-08                    &              & \\
MMP-13 & 3:43:45.6 & 32:01:45.1 & 2 & Bolo101          & Starless  & $24\micron$ abs.\\
MMP-14 & 3:43:43.7 & 32:02:53.0 & 4 & H05-26, Bolo100, & Starless  & SCUBA peak; \\
\multicolumn{4}{c}{}                         & K034373+32028    &           & $24\micron$ abs.   \\
\multicolumn{4}{c}{}                         & SMM-04                    &              & \\
MMP-15 & 3:43:42.5 & 32:03:23.0 & 1 & H05-24, Bolo100  & Starless  & SCUBA peak; $24\micron$ abs.\\
MMP-16 & 3:43:38.0 & 32:03:09.0 & 4 & H05-23, Bolo099, & Starless  & $24\micron$ abs. \\
\multicolumn{4}{c}{}                         & K034363+32031    &           &    \\
\multicolumn{4}{c}{}                         & SMM-10                    &              & \\
MMP-17 & 3:44:23.1 & 32:10:01.1 & 4 & Bolo114          & Starless  &    \\
MMP-18 & 3:44:15.5 & 32:09:13.1 & 4 & Bolo112          & Starless  &    \\
MMP-19 & 3:43:45.8 & 32:03:10.4 & 3 & Bolo100,         & Starless  & SCUBA Peak; \\
\multicolumn{4}{c}{}                         & K34346+32032     &           & $24\micron$ abs.    \\
\enddata%
\tablenotetext{(a)}{Origin of positions (\#):~~
(1) \protect\cite{2005A&A...440..151H}; 
(2) \protect\cite{2006ApJ...638..293E}; 
(3) \protect\cite{2006ApJ...646.1009K}; 
(4) Closed Contour SCUBA, this paper.}
\tablenotetext{(b)}{Origin of acronyms []:~~
[H]:~~ \protect\cite{2005A&A...440..151H}; 
[Bolo]:~~ \protect\cite{2006ApJ...638..293E};
[K]:~~\protect\cite{2006ApJ...646.1009K};
[SMM]:~~\protect\citet{2006AJ....132..467W}}
\tablenotetext{c}{Comments include the existence of a $24\micron$ absorption feature 
 and whether we agreed with \protect\citet{2005A&A...440..151H} 
 that a distinct SCUBA peak is present.   These two criteria frequently agreed.}
\end{deluxetable*}

\begin{deluxetable}{lrrrrrr}
\tablewidth{0pt}
\tablecaption{Far-IR/Submm flux densities and upperlimits for IC~348 starless cores\label{tab:mmp}}
\tablehead{
\colhead{ID} &
\multicolumn{3}{c}{\sst~\mips\tablenotemark{(a)}} &
\multicolumn{3}{c}{SCUBA $850\micron$\tablenotemark{(b)}} \\
\colhead{} &
\colhead{$24\micron$} &
\colhead{$70\micron$} &
\colhead{$160\micron$} &
\colhead{$f_{20\arcsec}$}  &
\colhead{$f_{40\arcsec}$} &
\colhead{$S_{40\arcsec}$}
}
\startdata%
     MMP-01 &    0.0007 &     0.209 &     6.234 &     0.072 &     0.227 &            0.198  \\
     MMP-02\tablenotemark{(c)} &    0.0008 &     0.393 &     9.285 &     0.034 &     0.115 &            0.119  \\
     MMP-03 &    0.0017 &     0.360 &    17.199 &     0.095 &     0.284 &            0.283  \\
     MMP-04 &    0.0009 &     0.162 &    18.773 &     0.107 &     0.287 &            0.303  \\
     MMP-05 &    0.0010 &     0.471 &    12.846 &     0.066 &     0.218 &            0.191  \\
     MMP-06 &    0.0016 &     1.726 &    23.118 &     0.150 &     0.486 &            0.402  \\
     MMP-07 &    0.0020 &     0.783 &    31.037 &     0.119 &     0.399 &            0.361  \\
     MMP-08 &    0.0012 &     0.631 &    22.191 &     0.137 &     0.462 &            0.510  \\
     MMP-09 &    0.0026 &     0.562 &    22.305 &     0.117 &     0.393 &            0.502  \\
     MMP-10 &    0.0190 &     2.326 &    14.492 &     0.164 &     0.536 &            0.592  \\
     MMP-11\tablenotemark{(c)} &    0.0328 &     2.251 &    20.301 &     0.042 &     0.158 &            0.131  \\
     MMP-12 &    0.0024 &     1.431 &    20.300 &     0.139 &     0.459 &            0.450  \\
     MMP-13 &    0.0005 &     0.521 &    17.214 &     0.087 &     0.298 &            0.156  \\
     MMP-14 &    0.0010 &     0.424 &    17.826 &     0.092 &     0.291 &            0.382  \\
     MMP-15 &    0.0036 &     0.350 &    12.689 &     0.112 &     0.393 &            0.368  \\
     MMP-16 &    0.0013 &     0.504 &    23.390 &     0.124 &     0.394 &            0.320  \\
     MMP-17 &    0.0300 &     9.158 &    66.411 &     0.050 &     0.164 &            0.165  \\
     MMP-18 &    0.0024 &     2.776 &    24.274 &     0.070 &     0.212 &            0.151  \\
     MMP-19 &    0.0060 &    \nodata &    \nodata  &     0.097 &     0.377 &            0.401  \\
\enddata%
\tablenotetext{(a)}{All \sst~flux density upperlimits are given in Jy. 
Central source fluxes for cores with protostars are given in Table \ref{tab:fproto}.}
\tablenotetext{(b)}{Aperture flux in a 20 or 40$\arcsec$ beam on 
the COMPLETE SCUBA $850\micron$  Perseus image. 
We corrected for the non-uniform nebular emission, which includes a pedestal flux contribution or
bowl-outs due to sky chopping, by subtracting a ``sky'' based on the mode of the pixel values in an
annulus from $120-140\arcsec$.  For comparison the last column labeled ``$S_{40\arcsec}$'' is the simple 
sum of the pixels in a $40\arcsec$ aperture without correction for the non-uniform background emission.
The conversion from Jy/beam to Jy was $0.0802$.}
\tablenotetext{(c)}{No SCUBA source is evident at the 1.1mm bolometer position; SCUBA flux given should be considered
an upperlimit in that aperture.}
\end{deluxetable}

\subsubsection{\mips\ survey of dark cloud cores near IC~348}
\label{sec:select:mips}

To identify the most embedded protostars we examined our $24$, $70$ and $160\micron$ \mips\ 
images of our \irac\ survey region and cross-correlated our \sst\ source list with a composite catalog (Table \ref{tab:mm}) 
of millimeter (mm) and sub-mm dark cloud cores  near to the IC~348 nebula. Our dark core list was 
cataloged from and contains cross-references to a number of recent mm-wave studies of the Perseus Molecular Cloud; 
it is similar to but encompasses a larger area than one presented in \citet{2006AJ....132..467W}.  
\citet{2005A&A...440..151H} performed a Submillimetre Common-User Bolometer Array \citep[SCUBA,][]{1999MNRAS.303..659H} 
survey that identified 15 unique dust continuum peaks near the IC~348 nebula, lying mostly in a 
molecular ridge south of the cluster center.   While also surveying the entire Perseus cloud at 1.1mm, 
\citet{2006ApJ...638..293E} found 21 compact sources within our \sst\ survey region. 
Finally, \citet{2006ApJ...646.1009K} produced a archival based SCUBA mosaic of the entire Perseus cloud  for the 
COMPLETE project and  these data are publicly available on their website.   
For source extraction \citeauthor{2006ApJ...646.1009K} used a single conservative 
threshold for identifying sources and recovered only some of the \citeauthor{2005A&A...440..151H} SCUBA sources.  
Yet all of the \citeauthor{2005A&A...440..151H}  sources and most of the 1.1mm bolometer objects are clearly detected in 
the COMPLETE SCUBA images. 

Our \mips\ observations of these cores were obtained with the camera in scan mode operating at medium scan rate 
and covering a total area of 30' by 30' common to all three detector arrays.  The map consisted of 12 scan legs;
half-array cross-scan offsets were employed to ensure full sky coverage at $160\micron$ and on 
side ``A" of the $70\micron$ array.  The total effective exposure time per pixel was 80 seconds at $24\micron$, 
40 seconds at $70\micron$, and 8 seconds at $160\micron$.  The data were reduced and mosaicked using the 
MIPS instrument team Data Analysis Tool \citep{2005PASP..117..503G}.    Coaddition and mosaicking of individual frames 
included applying distortion corrections and cosmic ray rejection.  The $70\mbox{ and }160\micron$~frames were 
further processed by applying a time filter on each scan leg in order to ameliorate time-dependent transient 
effects such as source and stimulator latency  and readout-dependent drifts.  We used IRAF and the DAOPHOT 
package to perform point-source photometry; specifically, at $24\micron$, we employed PSF fitting with an 
empirical PSF with a 5.6" fit radius  and 15-22.5" sky annulus.  For the  $70\mbox{ and }160\micron$ data
we used aperture photometry with beamsizes of 9" and 30" and sky annuli of 9-20" and 32-56", respectively. 
 We applied aperture corrections at all wavelengths as derived from STinyTim PSF models 
\citep{2006PASP.submittedE}.  No color corrections were applied.  Typical measurement uncertainties are 
$\sim 5-10\%$ at 24 $\mu$m and  10-20\% at 70 and 160 $\mu$m (though there may be somewhat 
larger  systematic uncertainties at 160 $\mu$m because of uncorrectable saturation effects).  The sensitivity 
at the latter two channels is limited by the very bright thermal emission from the molecular cloud
environs, and varies significantly with spatial position.  Only four sources are confidently detected at $160\micron$.

\begin{figure*}
    \centering     \includegraphics[angle=90,width=0.85\textwidth]{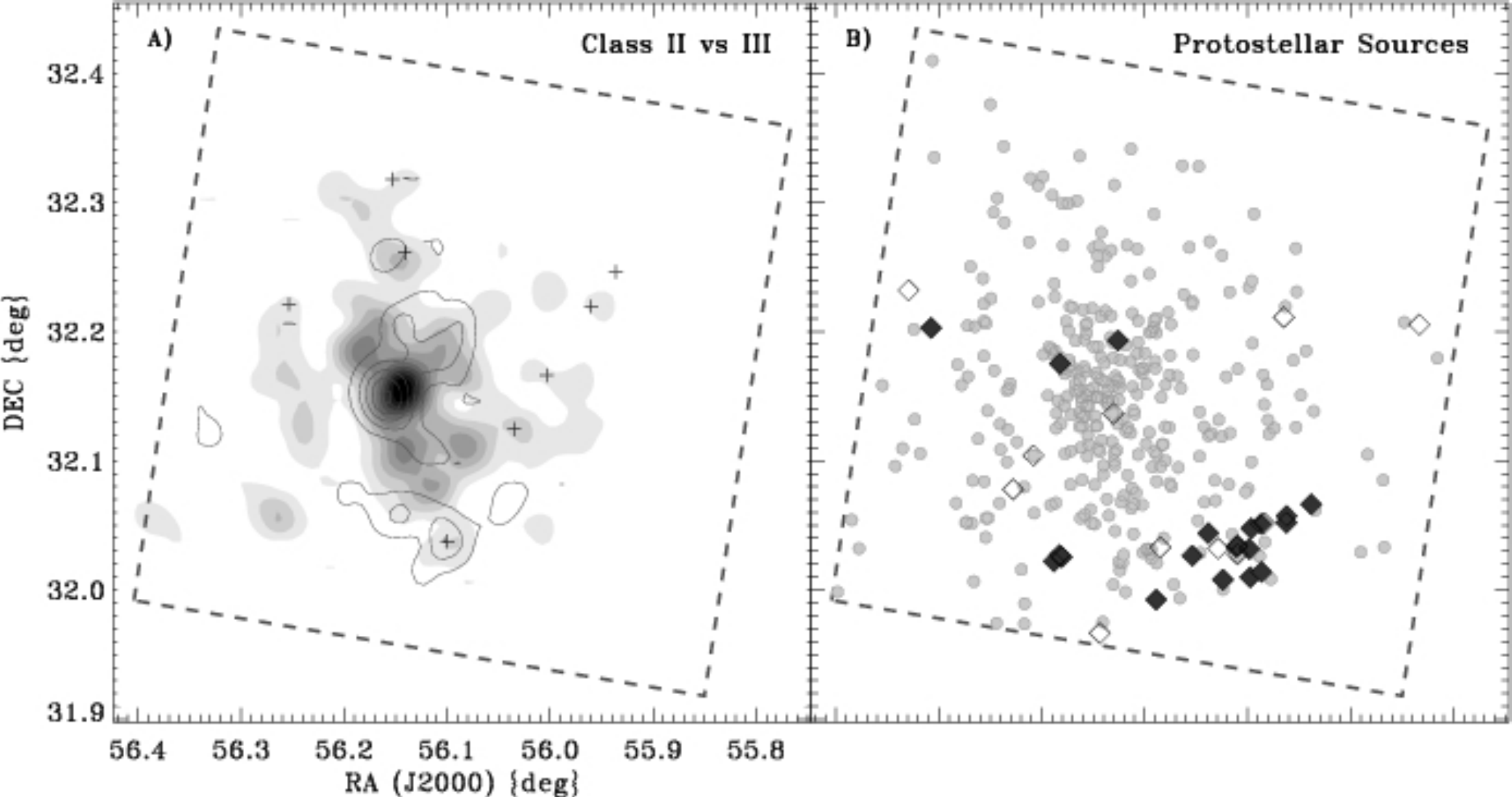}
\caption{Spatial distribution of young stars in the IC~348 nebula. 
Our \sst\ survey is marked in each panel with a dashed box.
A) class II vs III member surface density maps. 
The class III distribution is traced by filled grey contours, while the class II 
spatial distribution corresponds to red unfilled contours.  
In both cases contours start at 15 stars per square parsec and increase by 15 stars per square parsec. 
Subclusters reported by LL95 are plotted as crosses but do not correspond to actual clusterings of 
members except in 2-3 cases. 
B) class I protostars (\S\ref{sec:select:protostars})  compared to IC~348 ensemble membership 
(light filled symbols).   Bright protostars (filled diamonds), as well as many faint $(m_{\SIc}>12.5)$ 
filtered candidates (open diamonds),  are highly concentrated in a ridge $\sim1$~pc SW of the 
IC~348 core and anti-correlated with the central concentration of class II/III members.
See also Figure \ref{fig:nn}a.
\label{fig:map2}}
\end{figure*}

In these 26 IC~348 dark cores we found only two \mips\ sources which  lacked detections 
shortward of $8\micron$ and, thus, were not already identified as YSOs using  $\iracalpha$.  
These two~\mips-only sources corresponded to the previously identified driving 
sources of two outflows traced by Herbig-Haro objects: \objectname[HH 211]{HH-211} 
\citep{1994ApJ...436L.189M}  and \objectname[HH 797]{HH-797} \citep{2005AJ....129.2308W}.   
The HH-211 source appears only at 70 micron,  which is the position we recorded in Table \ref{tab:mm}. 
The HH-797 jet was originally detected in molecular hydrogen by \citet{1994ApJ...436L.189M} and 
\citet{2003ApJ...595..259E}. \citeauthor{2003ApJ...595..259E} discovered the 1.2~mm counterpart to 
the HH-797 driving source, naming it IC348-mm while \citet{2006A&A...456..179T} identified a strong molecular 
outflow correlating with the HH objects.  The apparent driving source appears first at $24\micron$ and 
corresponds to source \#57025 in our numbering system;  the position we tabulated corresponds to 
the $24\micron$ source.  Both of these sources have been previously  characterized as class 0 
sources \citep{2003ApJ...595..259E, 2005ApJS..156..169F}\footnote{The definition of embedded protostars 
was expanded by \citet{1993ApJ...406..122A} to include so called ``class 0'' sources, whose original definition 
included: 1) little or no flux shortward of $10\micron$, 2) a spectral energy distribution peaking in the sub-mm regime 
and characterized by a single black body temperature, and the somewhat less observable but more 
physical criteria 3) $\mass_{env} > \mass_{*}$. Their detection only at $\lambda>20\micron$ appear
to support this original definition.}.  Including these class 0 sources \#57025 and HH-211, we tally 20 
bright protostellar members of IC~348 as well as 11 fainter candidates.

Five other dark cloud cores contained sources we classified as protostellar based upon their $3-8\micron$ SEDs. 
These protostars are in systems of 1-3 bright members and we are confident of their association with these cores (also
\S\ref{sec:visual:swridge} and Table \ref{tab:mm}).  Thus, 19 of our composite list of 26 mm sources in 
our IC~348 \sst\ region appear to be starless.
To permit future SED analysis we tabulated all the relevant photometry for these starless cores.
Foremost we derived  $95\%$  \sst\ upperlimits  in the three \mips\ bandpasses (Table \ref{tab:mmp}).   
Since not all these sources were photometered in \citet{2006ApJ...646.1009K} we also derived aperture $850\micron$ fluxes
(or their upper limits) for all 26 cloud cores in the SCUBA mosaic\footnote{A sub-region SCUBA map of 
the IC~348 region was provided by J. Di Francesco, private communication; 
it had a pixel resolution of $3\arcsec$ compared to the $6\arcsec$ COMPLETE map.}.
Given the crowded nature of  these sources and the varying background emission, we tabulated SCUBA fluxes 
at different aperture beamsizes  and corrected for the varying nebular emission by subtracting a sky or pedestal value. 
In general the central positions of these aperture fluxes and upperlimits (and listed in Table \ref{tab:mm}) 
come from the better resolution SCUBA data.   In some cases the actual closed contour peaks in the SCUBA survey 
were much better correlated to what appear to be absorption features in the $24\micron$ nebulosity 
or to individual \sst\ sources than  those positions previously published.    If the published positions appeared to
us to be inaccurate then we used either the location of the closed contour SCUBA peak or the minima of 
the  $24\micron$ absorption features. 

\section{Analysis}
\label{sec:visual}

\begin{figure*}
     \centering \includegraphics[angle=0,width=0.95\textwidth]{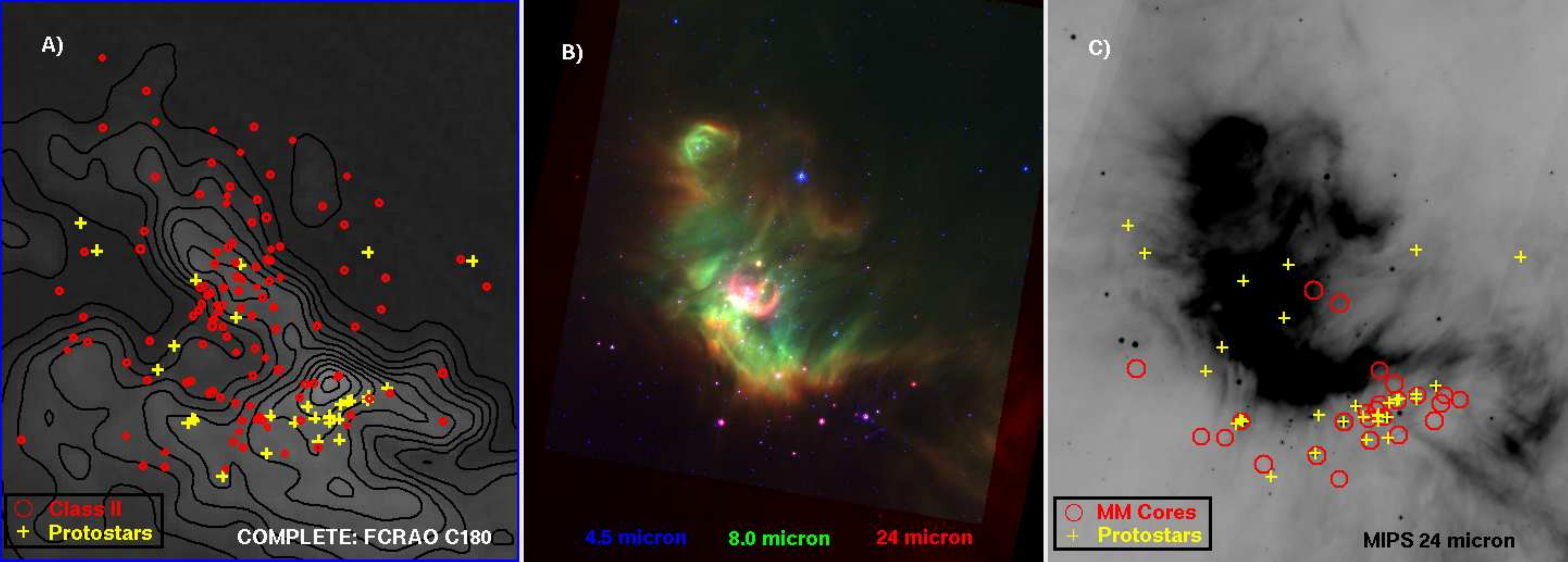}
\caption{Comparison of gas, dust and stars in the IC~348 nebula.~~~North is up and East is 
left in all three panels.~~~A) Locations of class II (red circles) and 
class I protostars (yellow crosses) compared to FCRAO \comap~data
\protect\citep[COMPLETE dataset][]{2006AJ....131.2921R};~~B) false color \irac\ \& \mips\ image of IC~348: 
red $(24\micron)$; green $(8\micron)$ and blue $(4.5\micron)$;~~~C) class 0/I protostars (yellow crosses) 
are compared to the merged list of millimeter cores (Table \ref{tab:mm}) against the \mips\ $24\micron$ 
image (inverse greyscale ).
\label{fig:color1}}
\end{figure*}%

\subsection{Spatial distribution of members}
\label{sec:visual:summary}

From our \sst\ census we have identified 42 new class II members of IC~348 and a population of $\sim30$ 
candidate class 0/I protostars of which we are confident in the membership and evolutionary status of $\sim20$.  
This section explores the spatial distribution of the cluster's class I, II and III members.  
Figures \ref{fig:map2}a compares the surface density maps of all (new and old) class II and III members; 
these maps were created by convolving the members' positions with a 0.2~pc box filter $(\sim2\arcmin)$. 
Note, class III source statistics are formally complete only in the region bounded by the \citet{2003ApJ...593.1093L} 
survey\footnote{Section \ref{sec:discuss:extent} discusses our incompleteness for class III members in more detail; 
as described in Appendix \ref{app:class3} we searched X-ray catalogs for additional class III sources, finding
27.  These candidates were included to the class III source list when creating  Figure \ref{fig:map2}a, 
although their addition or removal have little impact on our subsequent conclusions about the cluster's structure.}.   
Interestingly, the locations of the  class II and class III  surface density peaks are essentially identical;
we derive the same result when we directly calculated the median spatial centroids for each 
population\footnote{The median spatial centroids we derived are 03:44:30.053 +32:08:33.86 for
class II members and  03:44:32.809 32:09:6.00 for class III; both J2000.}.  This class II/III surface density 
peak corresponds approximately to the location of the B5 star  \objectname[HD~281159]{HD~281159} at the
center of the nebula and the concentration of members surrounding this peak in Figure \ref{fig:map2}a  
represents the centrally condensed IC~348 cluster core \citep{1998ApJ...497..736H,2003AJ....125.2029M}.  
Using a near-IR survey to derive the surface density distribution toward IC~348, LL95 found that the cluster
appeared to be constructed of this core, their IC~348{\it{a}}, and eight smaller sub-clusters.
At that time they did not have access to the refined cluster membership provided by subsequent surveys.  
Overplotting all nine of the LL95  sub-cluster centroids on our membership filtered map reveals that 
only two, or maybe three of them (a, b and possibly e)  represent significant cluster substructure; 
the rest are apparently  background surface density fluctuations likely due to counting statistics
and/or patchy line of sight extinction.

On the other hand, Figure \ref{fig:map2}b reveals that the class 0/I protostars have an entirely different 
spatial distribution.  While there are a few class I sources projected toward the cluster's class II/III center, 
most were found at the periphery, wrapping around the cluster from the east to the southwest.  
While many are widely spaced, a large concentration of IC~348 protostars lies $\sim1$~pc S.W. of the 
nominal cluster center and where there is no corresponding  surface density enhancement of class II or III members.  
Strom's IC~348 IR source lies near the center of this region, which is also the apex of most of the Herbig-Haro jets
found near IC~348, including HH-211, HH-797 and many new jets recently identified by \citet{2006AJ....132..467W}.

\subsection{Comparison of gas, mid-IR dust emission \\and young stars}
\label{sec:visual:visual}

We further examined the spatial distribution of disk bearing IC~348 members (class 0/I/II)  by 
comparing their locations to maps of the dust and gas emission from the associated Perseus molecular cloud.  
The most useful sets of such dust and gas maps come from the publicly available COMPLETE project, 
which were published by \citet{2006AJ....131.2921R}.  Figure \ref{fig:color1}a compares the locations of 
the young stars with disks to the Perseus COMPLETE integrated \comap~gas map.  

\begin{figure}
    \centering  \includegraphics[angle=0,width=0.425\textwidth]{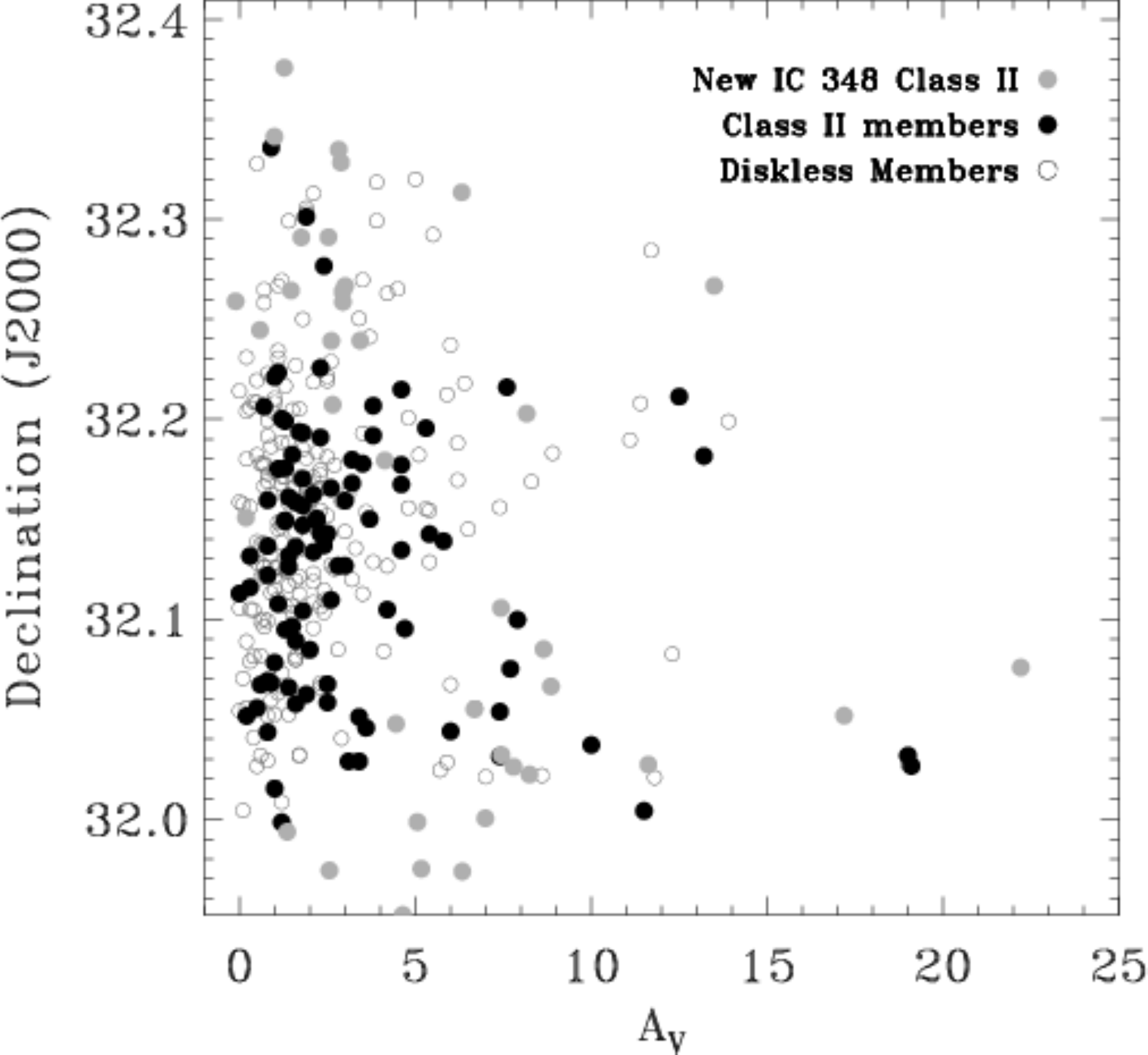}
\caption{Variation in reddening of members vs declination.  
There are increases in the dispersion of $\av$ at the declination of the cluster core 
(DEC.=$32.2\degr$) and along the southern protostellar ridge.  
There is a more significant segregation of reddened $(25\,>\,\av\,>\,4)$ class II/III members
other foreground cluster members. 
Values of $\av$ are from SED fitting (see  Paper I, \S\ref{sec:visual:ages} and Table \ref{tab:prop}).
\label{fig:dec_av}}
\end{figure}

Nearly all the disk bearing members of IC~348 are projected against dense molecular gas.
The class II members are more concentrated centrally near a gas filament that stretches 
from the southwest to the northeast. 
Even though our survey area is large class II sources are not distributed uniformly; there are
few class II members found to the northwest or southeast of this filament (hereafter termed
the central filament). 
Only those LL95 subclusters that are projected against the central filament are confirmed by 
our analysis of known members (IC~348{\it{a}}, {\it b} to the south  and perhaps {\it e} to the north).  
Unconfirmed LL95 sub-clusters are located off the central filament and where the reddening of
background stars is probably small and patchy.
Moreover, the stars in the IC~348{\it{a}} cluster core appear to be associated physically with the central filament.
Plotting the $\av$ of individual members versus declination in Figure \ref{fig:dec_av}, 
we see that the bulk of the cluster is infront of the central filament and they have fairly constant 
and low reddennings -- $\av<4$. Near the cluster core $(\delta\sim32.15\degr)$, however,  
$\av$ varies much more, reaching fairly large extinctions $(\av>10)$
and indicating that the cluster's core remains at least partially embedded in the central 
filament\footnote{Further evidence for the semi-embedded nature of cluster core is found in the 
reddened nebulosity surrounding it in the near-IR color image M03 (Figure 1; print edition).}.
Finally, there is no evidence that class II members avoid the cluster center or prefer the 
cluster halo as was suggested previously in $2\micron$ studies (LL95, M03).%
\begin{figure*}
     \centering \includegraphics[angle=0,width=0.725\textwidth]{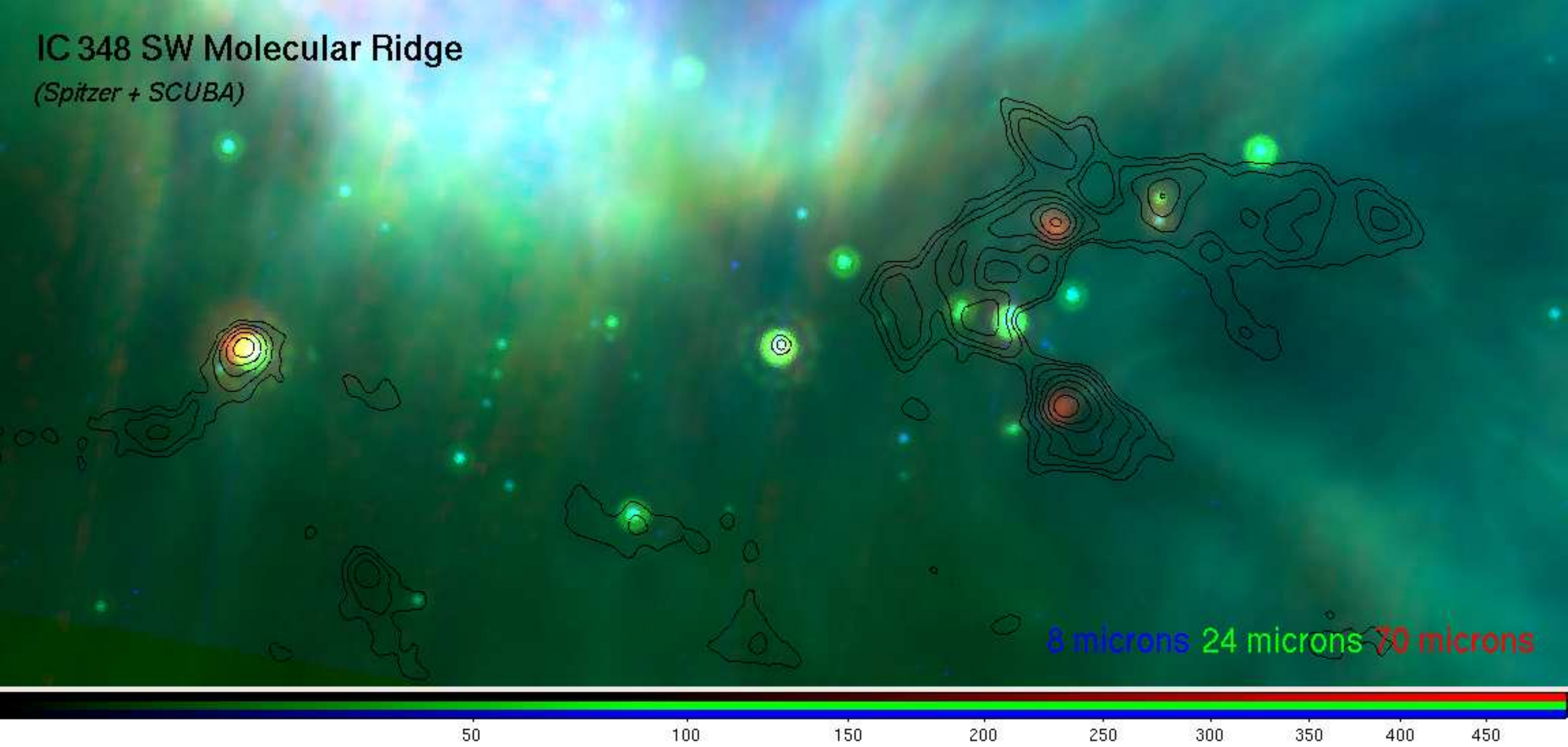}
\caption{Detailed view of the IC~348 SW protostellar ridge with \sst.
The false color image was created using $8$, $24$ and $70\micron$ \sst\ images.
Contours (Jy/beam) from the COMPLETE SCUBA $850\micron$ map (starting at 57 mJy/beam, $\sim10\sigma$)
are overlaid on this image.  
Note the five cases where the SCUBA data peak up on IC~348 protostars (also Figure \ref{fig:color3}).
Near the eastern edge of the ridge there is the $70\micron$ source \#1898 paired with the $24\micron$ 
source \#1872 both lying under a strong SCUBA core. 
The color bar scale corresponds to MJy/str in the $70\micron$ image.
\label{fig:color2}}
\end{figure*}%

Protostars are projected against, and embedded presumably within another molecular CO filament  
that stretches  west-to-east along the southern edge of the nebula. We term this the southern filament.
Note that the integrated CO emission is somewhat misleading in this respect: the apparent $^{13}$CO bridge 
(see Figure \ref{fig:color1}a) connecting the central and southern filaments is at a completely 
different (strongly blue-shifted) radial velocity compared to either  filament \citep{2005astro.ph..6604B}; 
therefo, the  central and southern filaments are infact {\it distinct}.
These distinct filaments, however, share a common radial velocity to within 0.45 km/sec 
\citep[$v_{lsr,cl}\sim8.15;\; v_{lsr,s}\sim8.6$;][]{2005astro.ph..6604B}; thus, 
they are physically {\it related}.
Further, there is significantly more contrast in the reddenings of foreground and embedded 
members along this southern filament than in the cluster core.   
In Figure \ref{fig:dec_av} we find that most members lie in front the southern filament
($\delta\,<\,32.1\degr$) with small reddenings,  $\av\sim1-2$ but the reddened members
very embedded with $\av>5$, ranging up to $25$ magnitudes.  
This segregation of members by $\av$ might be evidence of distinct cluster populations that orginate 
in the two distinct gas filaments.

Figure \ref{fig:color1}b reveals additional details about the IC~348 nebula, 
using the $\SIb, \SId, $ and $ 24\micron$ images to trace blue (scattered light and/or shocked 
hydrogen), green (scattered light and/or PAH) and red (24 micron dust emission). 
The optical portion of the nebula appears here as a blue-green cavity that surrounds the 
centrally condensed class II/III cluster core, providing more evidence that the central filament 
lies mostly but not far behind these stars.
At the cluster's periphery, on the other hand, the molecular gas contours in panel (a)  are
closely mirrored by emission or scattering traced by the \sst\ 8 and 24\micron\ data in panel (b).
Northwest of the cluster center, for example,  one can see the way the \sst\ dust emission traces around
the perimeter of a large low density CO clump facing the central B star.
We also find that the bright red nearly circular (r=0.13pc) $24\micron$ emission micron ring that surrounds 
the central B5 type binary has corresponding low level (30mJy/beam) SCUBA $850\micron$ emission that forms
small clumps around this ring.  This red ring also correlates with a ring of red near-IR $(K$ band) dust 
emission in the color Figure 1 of M03.  Unfortunately, SCUBA submm maps do not include another less 
red r=0.1pc cavity that surrounds a pair of A type stars (\#3 \& 14) to the NE. 
In this color image the class I protostars are found in strings of bright red 24 micron sources behind the 
southern emission wall and in the dark cloud core at the heart of the southern filament.

\begin{figure*}
    \centering \includegraphics[angle=0,width=0.725\textwidth]{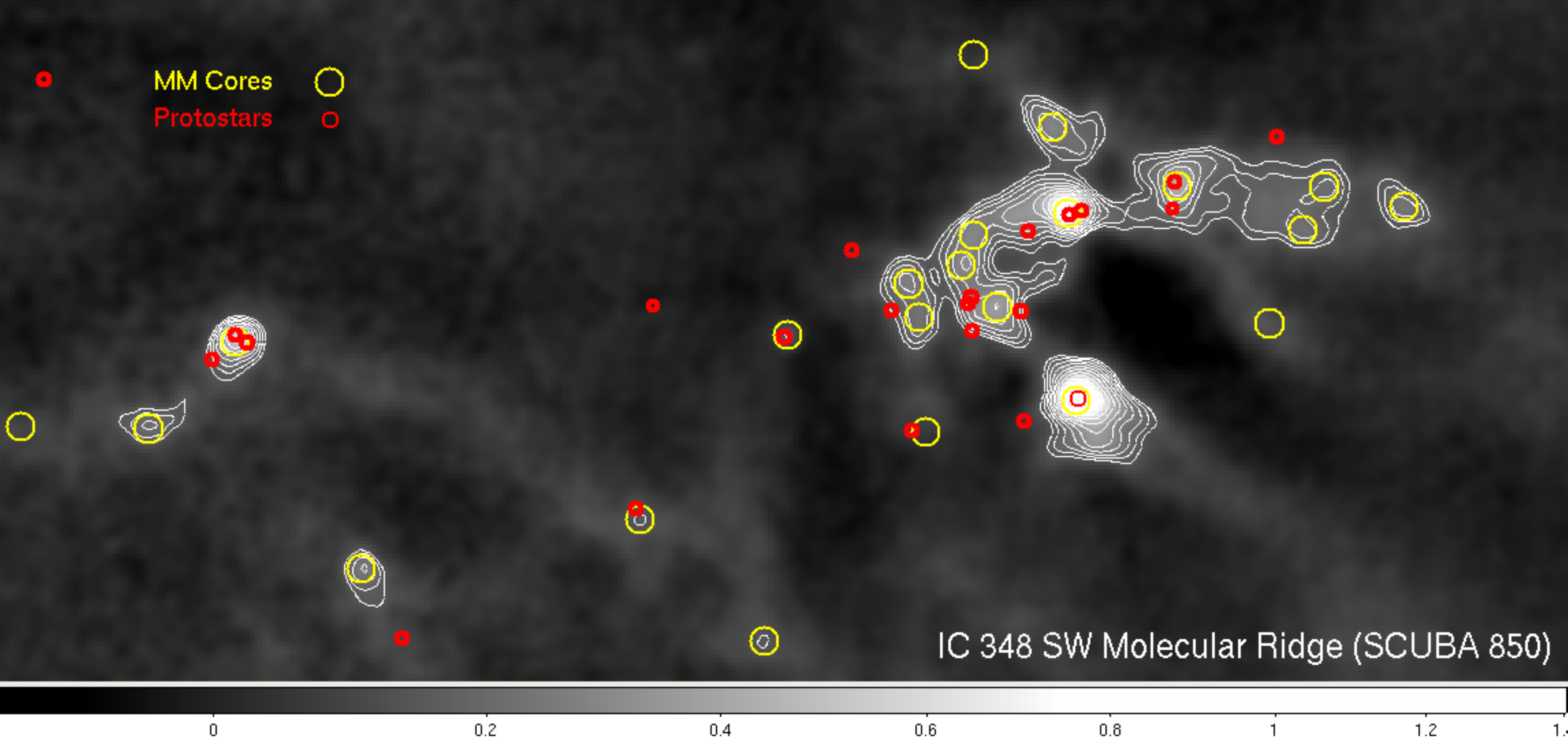}
\caption{Detailed view of the IC~348 SW protostellar ridge with COMPLETE SCUBA \protect\citep[][]{2006ApJ...646.1009K}. 
Ten logarithmically spaced contours from 100 to 1000 mJy/beam are overlaid for emphasis. 
MM cores from Table \ref{tab:mm} are shown as open (yellow) circles. 
\sst\ protostars (\S\ref{sec:select:protostars}) are plotted as small (red) circles. 
The color bar scale corresponds to Jy/beam in the SCUBA $850\micron$ image.
\label{fig:scuba}}
\end{figure*}%

Comparing Figure \ref{fig:color1}(a) to panel (c) one can better see how \mips\ $24\micron$ emission etches 
out the edges of the southern molecular filament,  wrapping around and into a $^{13}$CO cavity on the ridge's southern edge.
Many of the bumps and wiggles in the CO gas contours have counterparts in the $24\micron$ dust
emission and the dark molecular ridge does not appear sharp edged as it would were it a foreground cloud.
It is instead enveloped in and therefore immediately adjacent to (and we believe slightly behind) the nebula 
surrounding the central B star.  
Panel (c) includes symbols marking the locations of class I protostars and sub/millimeter dark cores (Table \ref{tab:mm}).  
While $24\micron$ dust emission closely follows the surface of the cloud as traced by molecular line data, 
the mid-IR dust emission does not closely follow the contours of the SCUBA dust emission. 
This could be due on the one hand to the spatially chopped nature of the SCUBA data, which acts to remove
larger scale and spatially smoother sub-mm emission structures.
On the other hand the SCUBA (starless) cores are frequently seen in silhouette as $24\micron$ absorption features 
(\S\ref{sec:visual:swridge}) against low level scattered mid-IR light that permeates this filament.
The source of that scattered light is not clear.  In projection, two starless cores are seen much closer to the  
central B star and its clustering, but have neither emission nor absorption features in \sst\ data;  
at least one (MMP-18)  is associated with an $1\solarmass$ \ntwohplus~core identified by \citet{2006A&A...456..179T} .
They appear to be hidden from the illuminating source of the mid-IR dust emission by the densest part
of the  central gas filament in which they reside and which reaches a peak reddening of  $\av\sim15$~magnitudes 
in the extinction maps presented in M03.  

\subsection{The protostars of the southern filament}
\label{sec:visual:swridge}

\subsubsection{\sst\ \& SCUBA correlations}
\label{sec:visual:swridge:1}

In this section  we focus on the southern molecular ridge, containing most of the class~0/I protostellar
objects; Figures \ref{fig:color2} and \ref{fig:scuba}  compare \sst\ and SCUBA images of this region.
We used a sub-region of the COMPLETE SCUBA Perseus map created by \citet{2006ApJ...646.1009K} 
to compare the cloud's dust continuum to our detected mid-IR point sources.  
In the IC~348 region of the Perseus SCUBA data we measured a mean and rms  of 8.6  and 4.8mJy/beam,
which were used to plot logarithmically spaced contours starting at 57 mJy/beam $(10\sigma)$. 
As discussed in the previous section there are a large number of starless SCUBA cores (Table \ref{tab:mm}) 
whose $850\micron$ contours correspond precisely to dark $24\micron$ absorption features.
These SCUBA emission/\sst\ dark cores appear across this protostellar ridge and indicate to us that 
the registration error for these SCUBA and \sst\ IC~348 comparisons is no more than 1-2 SCUBA pixels $(3-6\arcsec)$. 
We conclude similarly that the removal of large-scale $(>120\arcsec)$ structures from the SCUBA map 
by \citeauthor{2006ApJ...646.1009K} has had little affect on the spatial correlations 
of \sst\ and SCUBA point sources we discuss below.

In this southern molecular ridge there are 23 identified protostars and a comparable number (22) 
of MM cores of which 15 are starless.  Low level dusty filaments stretch across the region, 
threading  the various star forming sites; there are not, however, spatially distinct regions of star forming versus
starless cores.  Unlike the spatially anti-correlated distributions of class I and class II sources 
(Figure \ref{fig:color1}a), starless and star forming cores are intermingled and the empty SCUBA 
cores are typically no further  from the B star at the cluster center than are the protostars. 
While in Figure \ref{fig:color2} the three strongest 70 micron point sources shine through associated SCUBA core 
peaks and correspond to  class 0 sources, the fact is that most of the protostars are only peripherally associated 
with SCUBA cores (Figure \ref{fig:scuba}).
In  all but 6 cases, the closest SCUBA peak is more that 3000~AU $(10\arcsec)$ from a protostar and 
we conclude that these cores are neither the original ``infalling'' envelope nor  ``common'' envelopes 
encompassing a set of protostars \citep{2000ApJ...529..477L}.  
The intermingled SCUBA cores are instead probabe sites of future star formation.  
Moreover, six flat-spectrum protostars appear to be completely disassociated from 
the dust continuum, having neither SCUBA nor 1.1 mm detections.  
If the remnant envelopes are small ($\lesssim1000$~AU) or if we are seeing the envelope pole on, the 
integrated dust continuum might not have enough contrast  to be detected in larger beam size of the 
sub-mm observations\footnote{These sources are reminiscent of and may be similar to those nearby  
``peculiar'' class I Taurus sources detected by but unresolved with single dish 1.3mm data 
in \citet[][see also discussion in  \citet{2006astro.ph..4081W}]{2001A&A...365..440M}.   
Whether these non-detections  (or those unresolved Taurus detections) rule out the existence of an 
envelope (so removing the  protostellar moniker)  can only be firmly determined using observations of the 
silicate feature at $9.6\micron$ coupled with detailed SED modeling \citep[e.g.][]{2005ApJ...635..396E}.  
Indeed such SED modeling by \citeauthor{2005ApJ...635..396E} of one these \citet{2001A&A...365..440M} 
``peculiar'' class I objects, \objectname[LDN 1489]{L1489}, nonetheless prefers a disk+envelope structure; thus,
for now we retained the SED based protostellar classification.}. 
It is interesting to note that most of the protostars in or adjacent to SCUBA cores appear in systems of only 
1-3 members (at the \sst\ \mips\ resolution limit of $\sim1000$~AU).  Protostars distant from mm cores 
actually appear essentially solitary (down to separations of $\sim400$~AU  based on the near-IR data) and 
are typically separated from other protostars by $>20000$~AU. 

\subsubsection{Clustering}
\label{sec:visual:swridge:2}

\begin{figure}
    \begin{minipage}[c]{0.4\textwidth}
       \centering \includegraphics[angle=0,width=\textwidth]{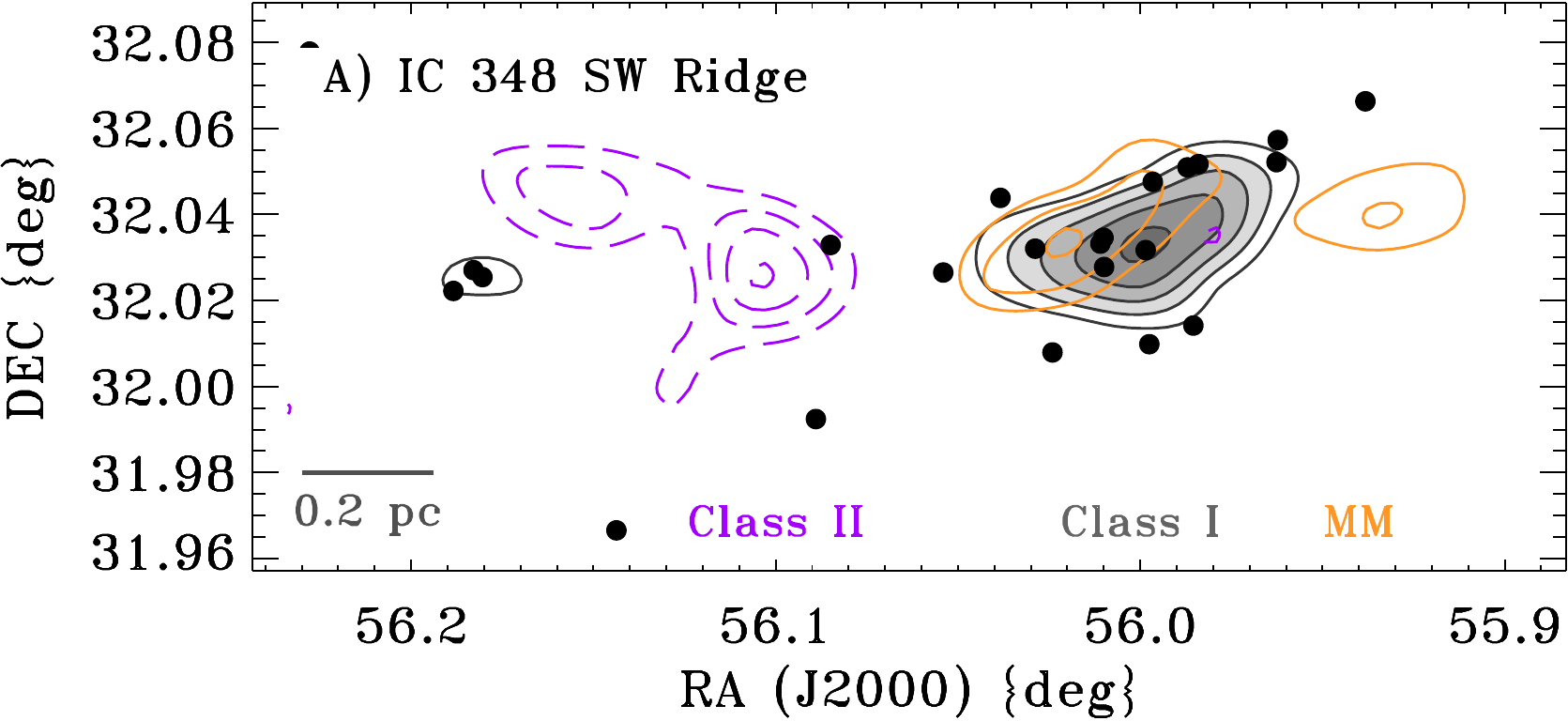}
    \end{minipage}%
	\hfill%
	\vspace{0.25in}
    \begin{minipage}[c]{0.4\textwidth}
       \centering \includegraphics[angle=0,width=\textwidth]{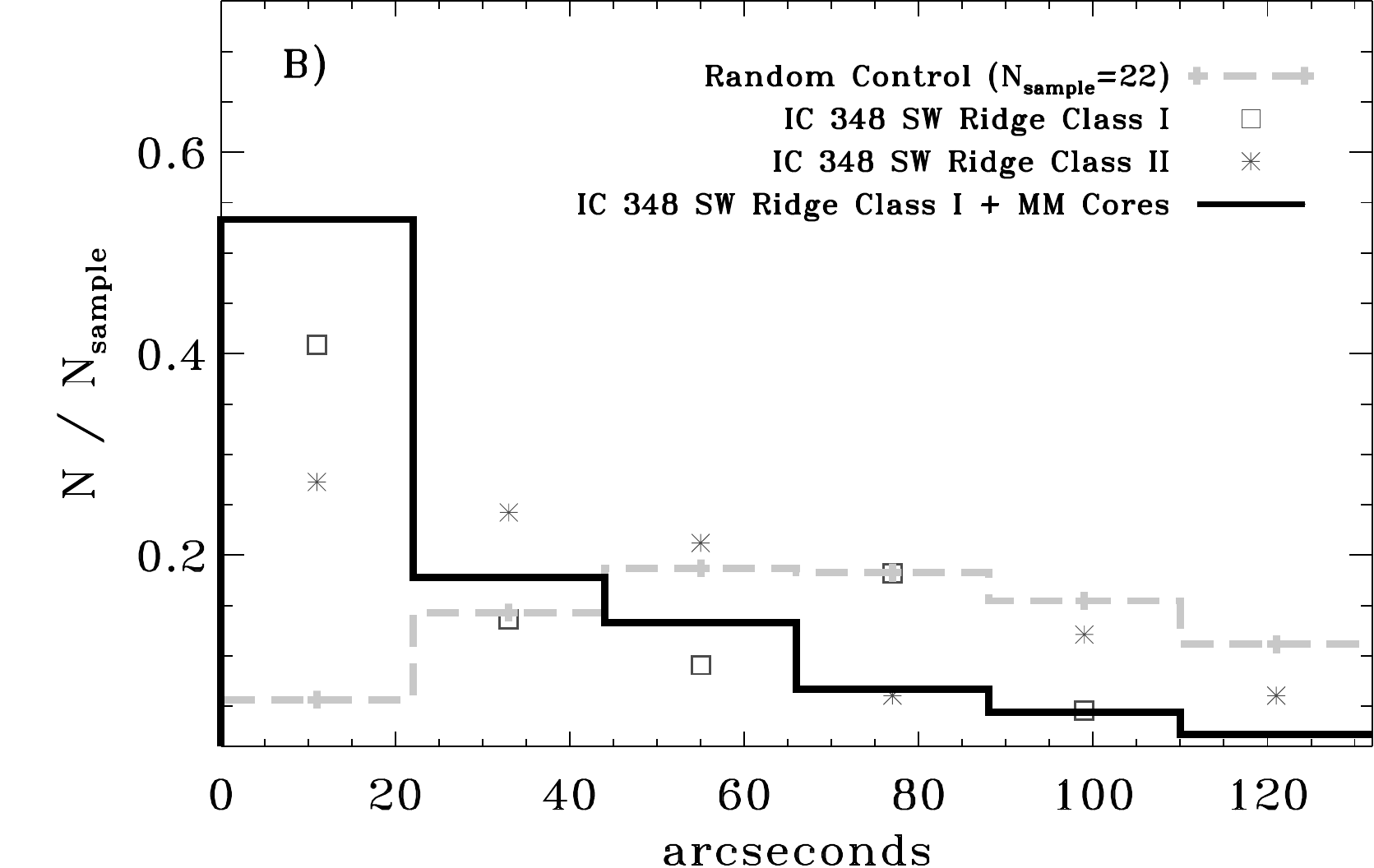}
    \end{minipage}%
\caption{Clustering in the IC~348 SW Ridge. A) Surface density map for class II,
class I and MM cores convolved with an $130\arcsec\:(\sim0.2\mbox{pc})$ square kernel.
Contours start at 3 objects/box $(\sim75\,\mbox{stars}\,\cdot\,\mbox{pc}^-2)$ and increase
in steps of 1 object per contour.  
The locations of the class I protostars are overplotted as filled circles.  
Note the correlated class I/MM core distribution and the anti-correlated class I/II distributions.
B) Nearest neighbor analysis for objects (22 mm cores; 23 protostars; 33 class II YSOs) in the SW ridge.  
All three distributions rise to the resolution limit of the surveys, which is 
smaller than the peak of the randomized distribution.  
The turnover in the MM core spacings appears to be due to the effective resolution $(\sim15\arcsec)$ 
of the SCUBA data.  
The unresolved class I peak for $r<20\arcsec$ corresponds to the small 2-3 member systems illustrated in Figure \ref{fig:color3}. 
\label{fig:nn}}
\end{figure}

Following our studies of protostars in Orion \citep{2000AJ....120.3162L,2004AJ....128.1254L},
we convolved the position of sources in the SW ridge with a box kernel to create surface density maps
and to identify and characterize any embedded sub-clusterings of protostars.   Unlike Orion where the embedded 
subclusters have $R<0.05\,\mbox{pc}$, convolution with kernels less than 0.2~pc~$(130\arcsec)$ 
produced no significant clumping in the protostellar ridge; the youngest IC~348 sources are much more spread out.  
Figure \ref{fig:nn}a displays the surface density maps for class II, class I and MM objects in the ridge 
convolved using an 0.2~pc kernel.  There is an apparent protostellar clustering whose peak coincidentally 
coincides with Strom's IR source and reaches $\sim200\,\mbox{stars}\,\cdot\,\mbox{pc}^{-2}$ , 
which is more than an order of  magnitude lower then we found for the embedded subclusters in the
molecular gas behind the  Orion nebula.
The class I and MM sources are correlated except for a small  group of starless cores 
on the western edge of the ridge. These two sets of sources are, however,
anti-correlated with the LL95 IC~348{\it{b}} subcluster of class II sources, which splits the ridge in half
and achieves a nearly identical peak surface density. This class II subgroup is obvious in 
Figure \ref{fig:color2}  as the central group of bright $24\micron$ sources lacking SCUBA emission.

\begin{figure*}
    \begin{minipage}[c]{0.3\textwidth}
        \centering \includegraphics[angle=0,width=\textwidth]{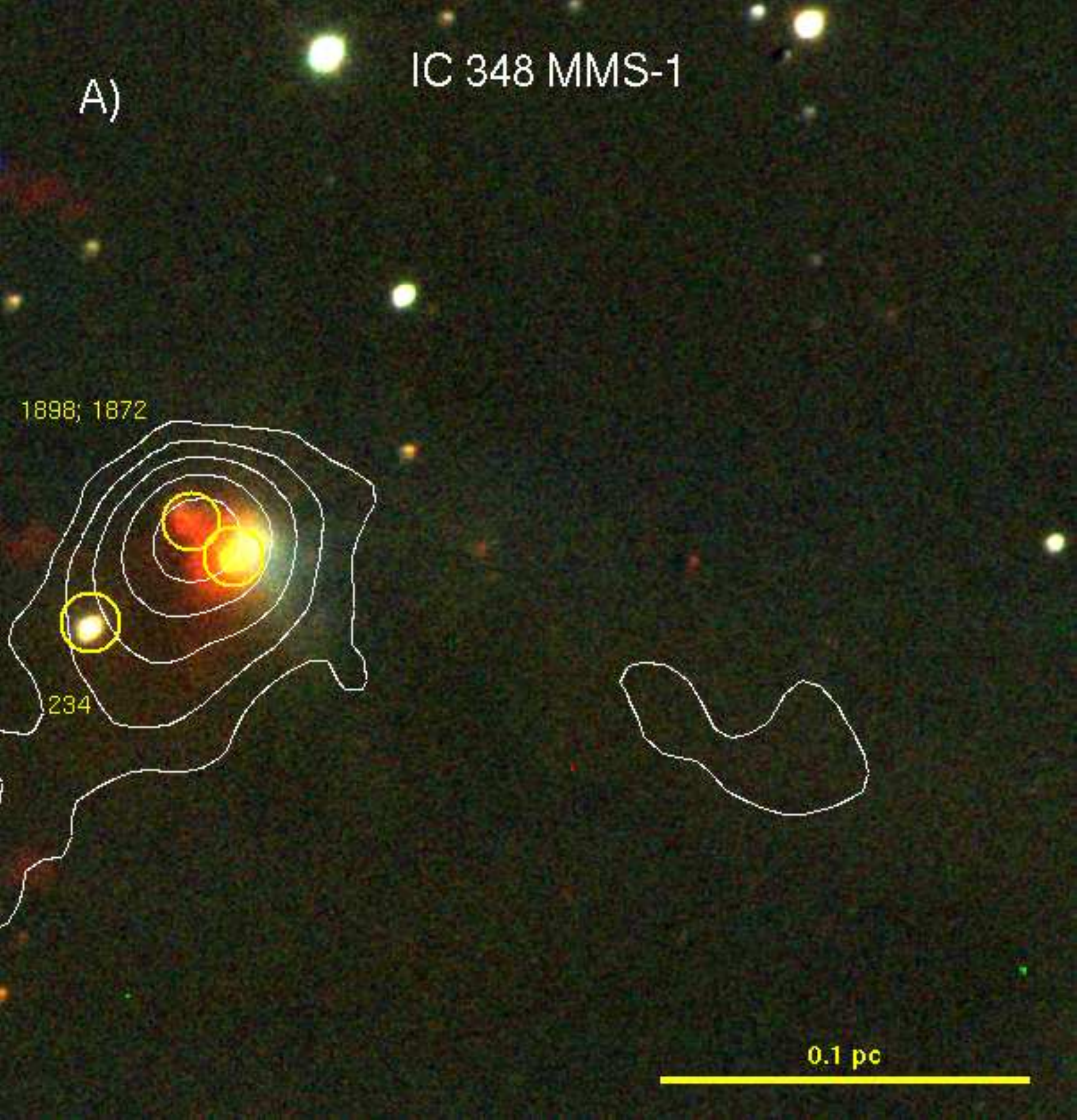}
    \end{minipage}%
    \hspace{0.04\textwidth}%
    \begin{minipage}[c]{0.3\textwidth}
        \centering \includegraphics[angle=0,width=\textwidth]{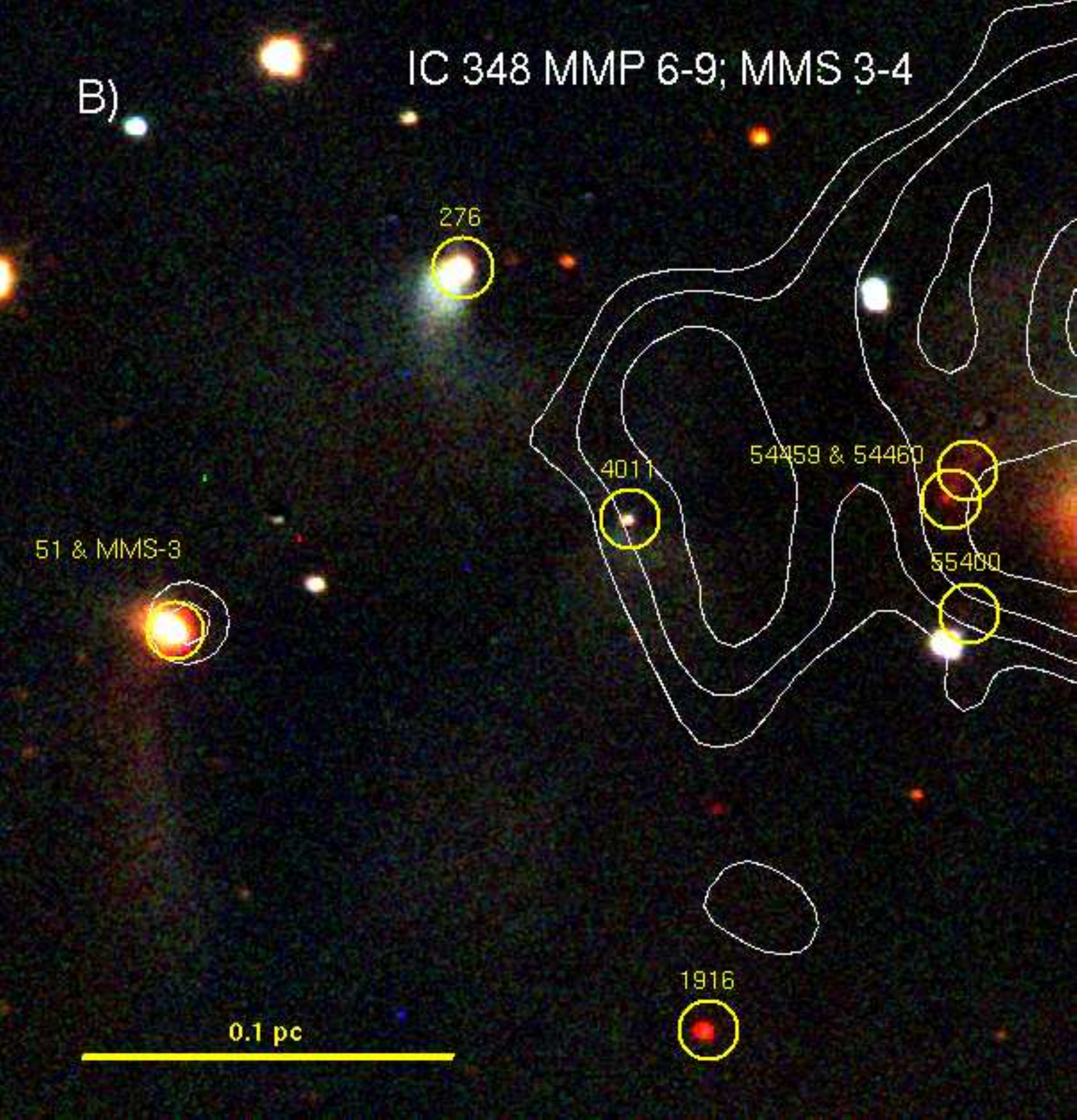}
    \end{minipage}%
    \hspace{0.04\textwidth}%
    \begin{minipage}[c]{0.3\textwidth}
        \centering \includegraphics[angle=0,width=\textwidth]{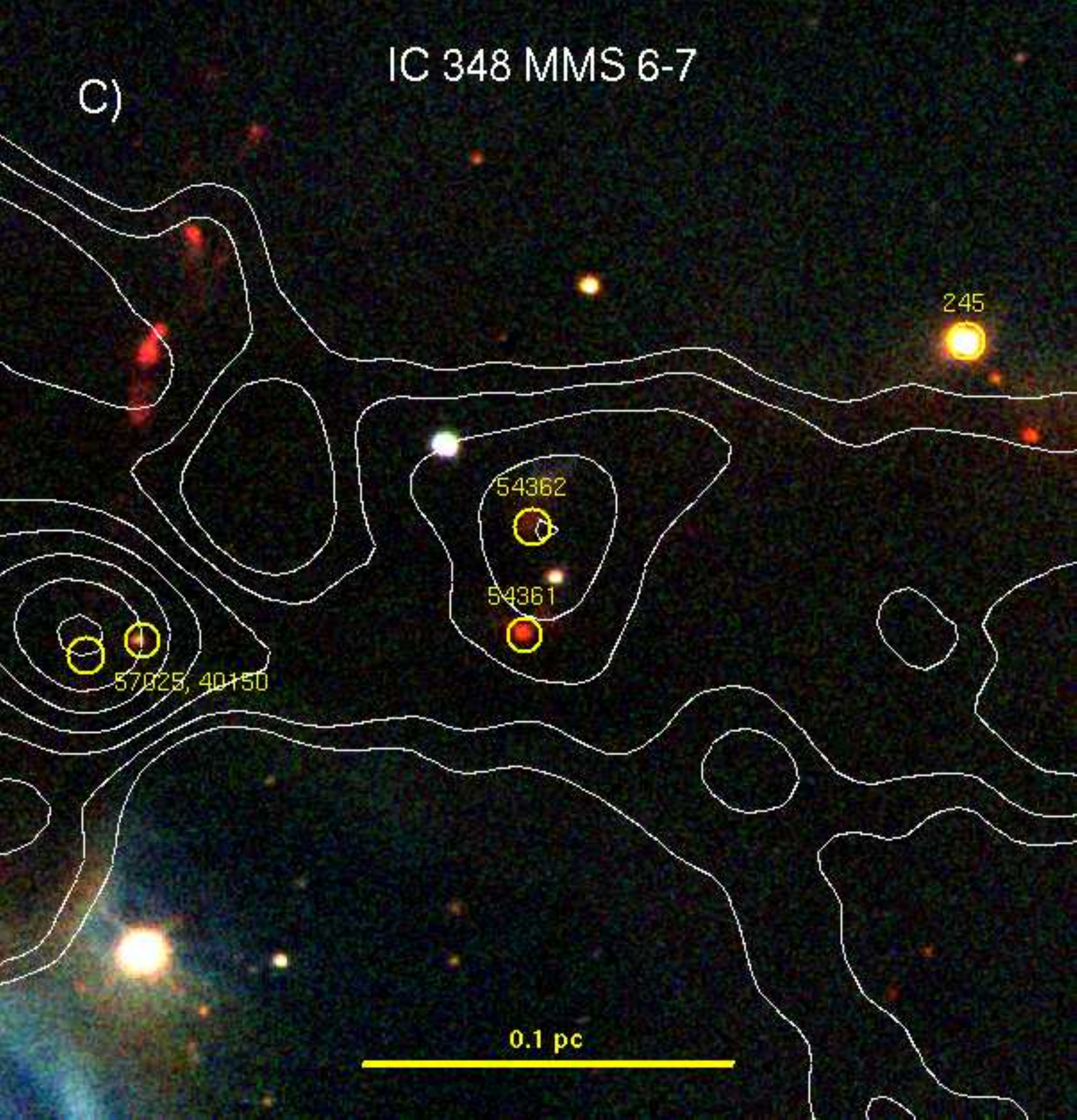}
    \end{minipage}%
\caption{Detailed near-infrared views of protostars in the IC~348 SW molecular ridge.
Images from \protect\citet{2003AJ....125.2029M}(their Figure 1; print edition); 
a 0.1pc yardstick $(D=320\,\mbox{pc})$ is shown in each panel; protostars are circled and labeled.  
SCUBA contours are the same as in Figure \ref{fig:color2}.
Panel A) MMS-1 encloses three protostars but peaks on the red $70\micron$~source~\#1898,  
which is seen as a scattered light cavity in the near-infrared. Source 1898 is separated by $4000$~AU
from the dominate $24\micron$~source~\#1872, which has a featureless spectrum (Figure \ref{fig:spec5}).
Note how the near-IR dust emission traces the SCUBA dust continuum.
Panel B) MMS-3 encloses source \#51, and MMS-4, detected at 1.1mm by \protect\citet{2006ApJ...638..293E}, encloses \#1916,
which has a featureless spectrum (Figure \ref{fig:spec5}).  
Sources \#276, 4011, 54459, 54460 and 55400 cannot be firmly associated with any dust continuum peaks,
though they are all seen in scattered light.
Panel C) SCUBA core MMS-7 peaks on \# 53462, MMS-6 is offset slightly (1500au) to the NNE from the 
class 0 source, \#57025 (IC348-mm) which is thought to drive HH-797 (seen in red continuum arching to the NNW).  
Embedded in scattered light, \#245 also falls outside the dust continuum contours.
\label{fig:color3}}
\end{figure*}%
We also applied a nearest neighbor analysis to these ridge sources.  \citet{2006ApJ...636L..45T} 
examined the nearest neighbor distribution of bright class I sources in the embedded Spokes 
subcluster of NGC~2264, finding a preferred spacing of $27\arcsec$ or 0.1~pc (d=900pc).  
A nearest neighbor analysis for the class II, class I and MM cores in the SW ridge reveals 
no resolved, preferred spacings (Figure \ref{fig:nn}b) but instead all rise to the resolution limit.
Unlike the Spokes, the spacings of protostars are mostly flat except for a peak at or 
below $20\arcsec$ $(<0.03\mbox{pc};\,<6000\mbox{AU})$; this unresolved peak is sharpened by
including starless cores as neighbors to the protostars. Visually inspecting Figures 
\ref{fig:color3}(a-c) reveals the nature of this difference with NGC~2264.  These small spacings 
come from a few protostellar systems of 1-3 members with small $\sim1000-6000$~AU separations, 
although the majority of the class I sources are essentially solitary and widely spaced.
The class II spacing distribution also rises down to the resolution limit.

\subsubsection{Near-Infrared Images}
\label{sec:visual:swridge:3}

In this section we use deep near-infrared images from M03 to illustrate some of these small 1-3 member protostellar subclusters. 
Figure \ref{fig:color3}(a-c) show three closeup views of the protostellar ridge, progressing from east to west.
The eastern MMS-1 core, illustrated in Figure \ref{fig:color3}a, contains three protostars, including the brightest far-IR 
source in the entire cluster. Although the SCUBA core peaks right between protostars \#1898 and \#1872, 
our comparison of the near-IR, \sst\ $24\micron$ and $70\micron$ images leads us to 
conclude that the $70\micron$  source (and thus also the $160\micron$  source) peaks up on the eastern red $K$ band knot 
(source \#1898) rather than on the western protostar, \#1872, which is where the $24\micron$ source peaks (scrutinize 
the color version of Figure \ref{fig:color2}).
Source \#234 could be either the tertiary member of a hierarchal triple or an entirely separate clump fragment.

Unlike MMS-1 in Figure \ref{fig:color3}a, panel (b) shows how most of the protostars are unassociated with 
individual SCUBA cores.  The edge-on source \#4011 and the trio of 54459/54460/55400 are simply adjacent to 
starless SCUBA cores.   All of the protostars in Figure \ref{fig:color3}b are seen in scattered light, including the rather 
solitary flat spectrum protostars \#51 and \#276 as  the flat spectrum protostar \# 245 panel (c).   
On the other hand, two other very good SCUBA/\sst/near-IR~correlations are illustrated in panel (c). 
The class 0 \#57025 and protostar \#54362  both appear almost precisely at their respective SCUBA 
closed contour peaks (within 1000~AU of MMS-6 \& MMS-7, respectively). 
These comparisons reinforce our arguement that most of these SCUBA cores are infact starless.

\subsection{Inferred cluster properties}
\label{sec:visual:ages}

\begin{figure*}
       \centering \includegraphics[angle=90,totalheight=0.60\textwidth]{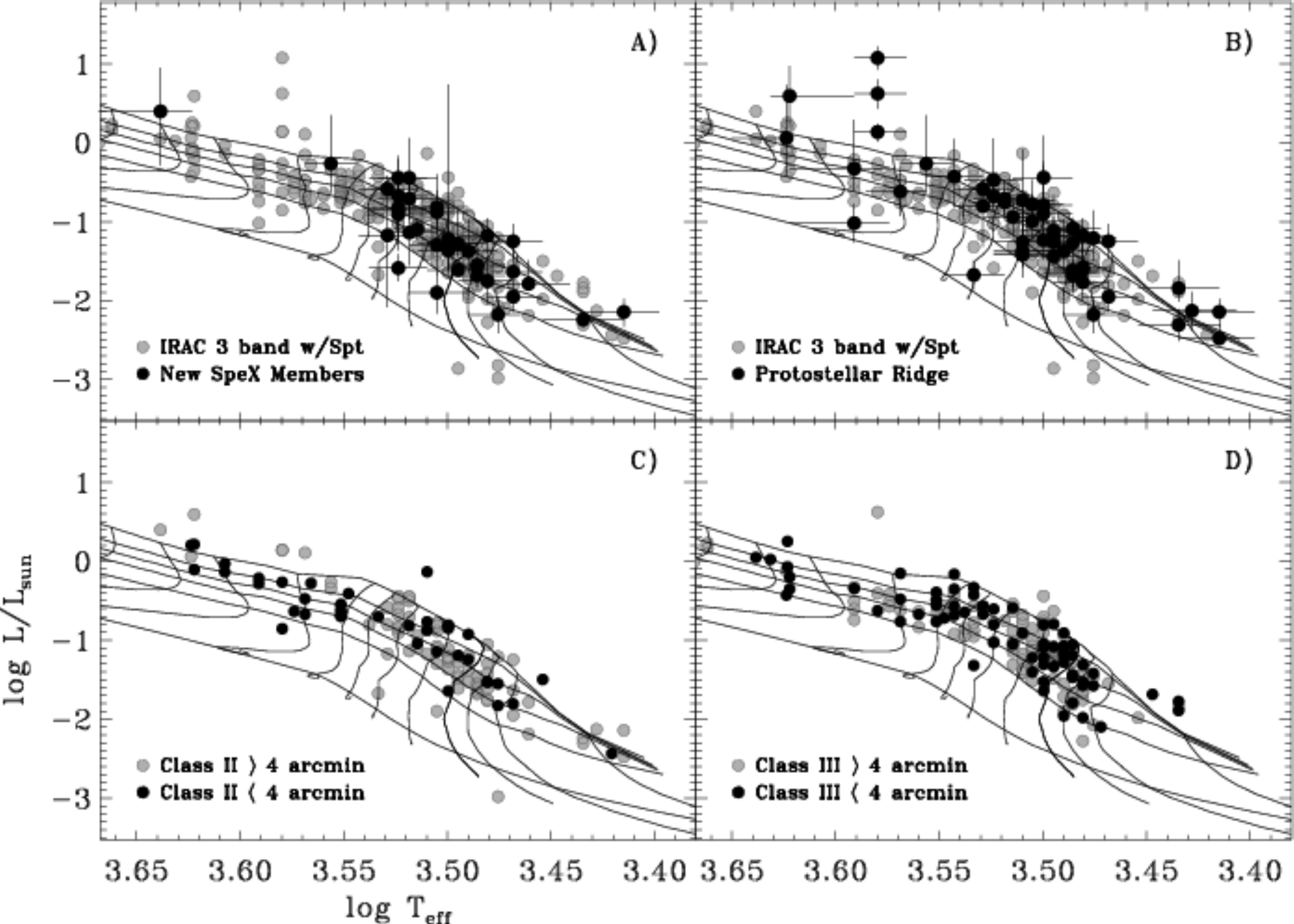}
\caption{Hertzsprung-Russell diagrams. 
Only sources with spectral types and subject to our census constraints (three \irac\ band detections) are included.
A) New \sst\ members compared to the pre-existing IC~348 population.
New members do not differ in their luminosity spread on the HR diagram but are preferentially later types.
B) Those sources on the nebula's southern edge ($\delta<32.07$) and projected toward the protostars 
are compared to the ensemble population.  The sources along the southern protostellar ridge 
appear more luminous on average than the cluster ensemble.
C) Radial dependence of class II sources in the HR diagram. 
D) Radial dependence of class III sources in the HR diagram.  
Isochrones plotted correspond to ages of 1, 2, 3, 5, 10, 100 Myr as ordered by decreasing luminosity.
Evolutionary tracks plotted correspond to stars of 0.03, 0.072, 0.1, 0.13, 0.2, 0.3, 0.4, 0.5, 0.6, 0.8, 
1.0 and 1.3 \solarmass as ordered by increasing T$_{eff}$. 
\label{fig:hr}}
\end{figure*}

\begin{figure*}
    \centering \includegraphics[angle=90,totalheight=0.55\textwidth]{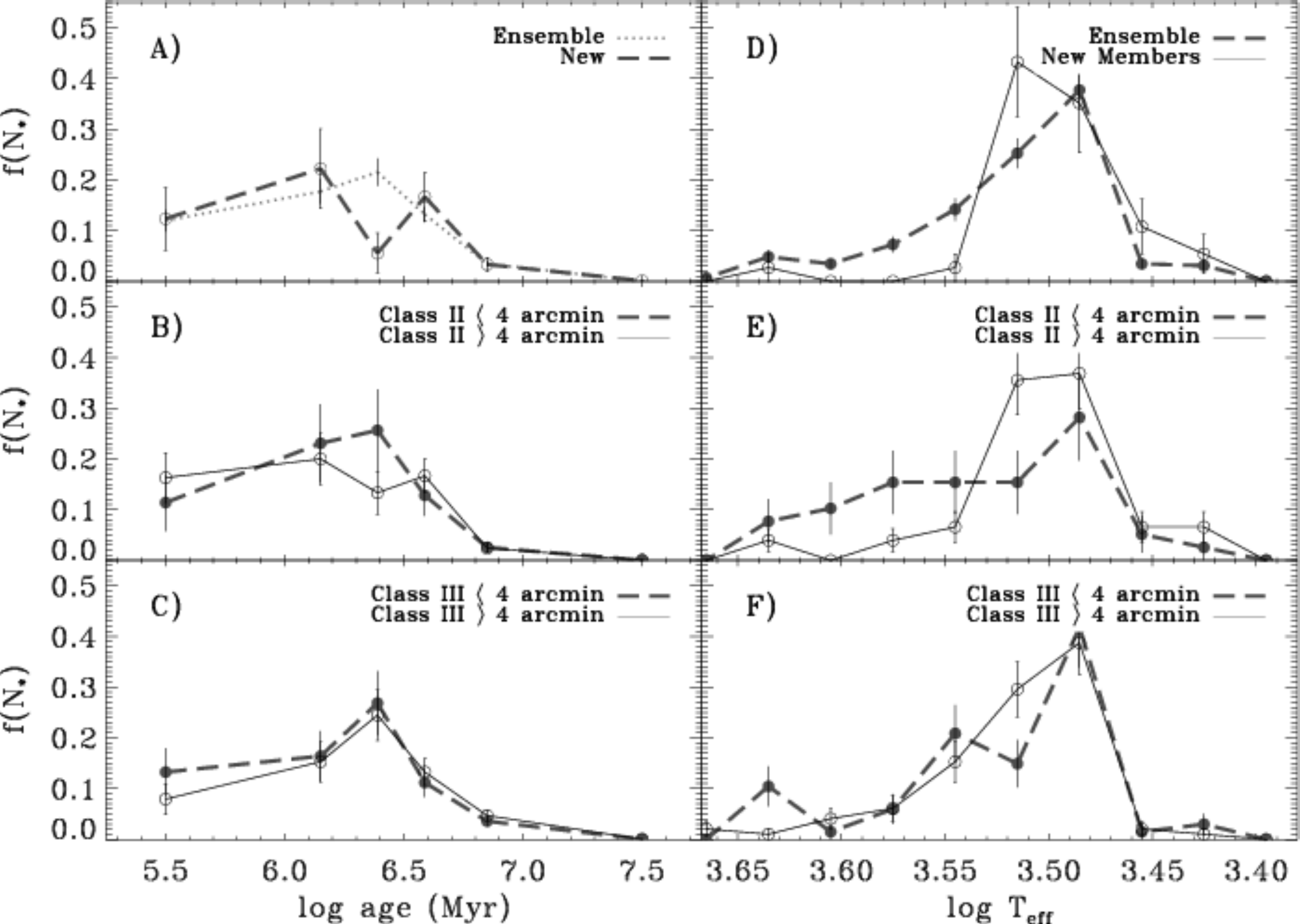}
\caption{Distributions of IC~348 member properties as a function of cluster structure and disk evolutionary phase.   
Panels A-C plot the star forming histories and panels D-F plot the distribution of effective temperatures which 
we use a proxy for the mass function.
All distribution functions were normalized by (individual) population size for these comparisons
and panels A-C were also divided by the bin width in Myr and thus have units of fractional stars / Myr.
Note that any star lying above the 1 Myr isochrone was placed into the first log age bin.
\label{fig:imf}}
\end{figure*}

Considering the expanded borders of the IC~348 cluster traced by our \sst\ census, it is useful 
to ask how the addition of new cluster members over a large physical scale might have modified 
global cluster properties such as the median age or stellar initial mass function (IMF).  
In this section we derived bolometric luminosities for the new and old members and  compared them to 
theoretical isochrones on the Hertzsprung-Russell (HR) diagram to answer this question.
In this exercise all the sources were placed on the HR diagram by dereddening a single passband flux, 
using the $\av$ derived from SED fitting (see Paper I), and applying a bolometric correction (BC), 
which is tabulated as a function of effective temperature and taken from our previous studies. 
Other fixed values or assumptions included a value of $M_{bol,\odot}=4.75$, the use of a subgiant 
spectral type to T$_{eff}$ scale from \citet{1999ApJ...525..466L}, and a distance of 320~pc, which is the value
we have assumed in all of our previous studies of IC~348 members.    One subclass spectral type uncertainties 
were assumed and were propagated into the \lbol\ uncertainty, which was the quadratic sum of the 
$1\sigma$ photometric error,  the $\chi^{2}$ $\av$ fit uncertainty and the variation in BC as a function of \teff.  
The $\chi^{2}$ $\av$ fit uncertainty dominates the error budget of \lbol\ for each star.  
We actually derived \lbol\ at all passbands  from $V$ to $K$
and found that these derivations are extremely self-consistent in the near-IR with essentially
no variation between $\lbol$ derived from the $J$ or $H$ bands though there was some evidence for $K$ band 
excess producing slightly higher bolometric luminosities (typically, however, $<0.2$~dex).  
Figure \ref{fig:hr} presents HR diagrams for sets of members parsed spatially or according to their disk properties.
Isochrones and evolutionary sequences were taken from \citet{1998A&A...337..403B}\footnote{No single 
set of \citeauthor{1998A&A...337..403B} models fit the locations of the GG Tau quadruple or the IC~348 locus 
on the HR diagram \citep[most recently see][and references therein]{2003ApJ...593.1093L}.  
As prescribed previously, we use a mixed set of \citet{1998A&A...337..403B} models with different convective properties for
different mass ranges:  a mixing length parameter 1/Hp=1 for $\mass<0.6\solarmass$ and 1/Hp=1.9 for $\mass>0.6\solarmass$.
Thus, by design, our set of isochrones will yield a constant inferred mean age as a function of \teff\ (\solarmass).}.
Figure \ref{fig:imf} presents the inferred cluster properties based on these HR diagrams and theoretical tracks.

Although they lie preferentially at the edges of previous spectroscopic census, the new, primarily class II sources 
identified in our \sst\ census fall in the same basic locations on the HR diagram as previous members (Figure \ref{fig:hr}a);
specifically, they have a very similar spread in \lbol\ at fixed \teff. 
This spread in \lbol\ at a fixed \teff\ should represent a range of radii for stars of approximately the same stellar mass
and should correspond to the spread in the birth times for contracting pre-main sequence stars.
This \lbol\ spread is, however,  convolved with a distribution of uncertainties, which in this case we find to be 
dominated by uncertainties in extinction estimates, and the  age of a particular star should be viewed with caution.
The ensemble of cluster members ages may yield  some clues about the cluster's star forming history,   
so we quantified this luminosity (age) spread by counting sources  between logarithmically spaced 
isochrones and plotting them in  Figure \ref{fig:imf}a.  
In this way, for example, we can show that the addition 
of new class II sources does not appear to modify the star forming history inferred previously for IC~348.    

A spatially distinct population of protostars spread along the cluster's periphery clearly suggests that star 
formation in IC~348 is not necessarily coeval and that the location of star formation may have varied 
with time across the nebula.  We tested the hypothesis of spatial variations in the SFH for IC~348 members 
by examining radial variations of the cluster loci on the HR diagram (Figures \ref{fig:hr}cd) and the 
inferred SFHs (Figure \ref{fig:imf}bc).   Radial variations of the apparent ages of IC~348 members were reported  
by \citet{1998ApJ...497..736H}  yet our class II \sst\ survey is spatially complete over a much larger area than 
his \halpha\ based survey.  We divided the population into a  $r<4\arcmin$ core and a $r>4\arcmin$ halo, which is 
approximately the same radial distinction used by or discussed in LL95, \citet{1998ApJ...497..736H} and M03;
these two samples correspond to roughly equal proportions of cluster membership ($40\%$ and $60 \%$, specifically). 
We find no significant radial differences in the spread of \lbol\ on the HR diagram or in the extracted SFHs of the 
spatially complete class II populations (Figure \ref{fig:imf}b); although spatially complete only in the core,  
we found no radial variation in the class III SFHs (Figure \ref{fig:imf}c) either.  
Infact, the class II and class III age distributions are essentially indistinguishable, 
displaying a peak at 2.5 My and an age spread of 4~My, which we derived using the half
dpower points of the cluster ensemble age distribution.   
Even if star formation were a function of time {\it and} location in the nebula, the common heritage of stars 
inside and outside the cluster core means that the core is either a distinct and long lived star formation site 
or the merger of many  smaller briefer star formation events whose initial spatial distribution
no longer appears terribly obvious.
\begin{deluxetable*}{lrrrrrrrrrrr}
\tablewidth{0pt}
\tablecaption{Derived properties for IC 348 members. \label{tab:prop}}
\tablehead {
\colhead{   ID}  & 
\multicolumn{2}{c}{\teff} &
\multicolumn{3}{c}{$\av$\tablenotemark{a}} &
\multicolumn{3}{c}{\lbol\tablenotemark{b}} &
\multicolumn{3}{c}{SED Params\tablenotemark{c}} \\
\colhead{   }  & 
\colhead{  Sp.T.} &
\colhead{   (K)} &
\colhead{   Best Fit} &
\colhead{   -$1\sigma$} &
\colhead{   +$1\sigma$} &
\colhead{   Best Fit} &
\colhead{   -$1\sigma$} &
\colhead{   +$1\sigma$} &
\colhead{   $\iracalpha$ } &
\colhead{   $1\sigma$} &
\colhead{   $850\micron$ }
}
\startdata
     1 &     B5 &  15400 &    3.1 &    1.0 &    8.7 &   3.230 &   2.778 &   3.787 &  -2.638 &   0.102   0.005 \\
     2 &     A2 &   8970 &    3.2 &    1.1 &    8.5 &   2.067 &   1.774 &   2.656 &  -1.396 &   0.127   0.009 \\
     3 &     A0 &   9520 &    3.9 &    3.0 &    4.7 &   2.073 &   1.708 &   2.329 &  -2.794 &   0.110   0.000 \\
     4 &     F0 &   7200 &    2.3 &    1.0 &    7.3 &   1.614 &   1.435 &   2.168 &  -2.786 &   0.091   0.004 \\
     5 &     G8 &   5520 &    7.7 &    5.5 &    9.9 &   1.306 &   1.045 &   1.556 &  -1.389 &   0.160   0.014 \\
     6 &     G3 &   5830 &    3.5 &    1.9 &    7.4 &   1.209 &   1.010 &   1.645 &  -1.972 &   0.079   0.011 \\
     7 &     A0 &   9520 &    1.7 &    0.0 &    8.1 &   1.642 &   1.235 &   2.383 &  -2.788 &   0.071   0.009 \\
     8 &     A2 &   8970 &    1.6 &    0.0 &    8.0 &   1.509 &   1.272 &   2.234 &  -2.532 &   0.101   0.008 \\
     9 &     G8 &   5520 &    5.3 &    3.7 &    7.7 &   1.032 &   0.839 &   1.304 &  -2.894 &   0.134   0.012 \\
    10 &     F2 &   6890 &    2.1 &    0.8 &    7.0 &   1.153 &   0.976 &   1.705 &  -2.827 &   0.054   0.057 \\
\enddata
\tablenotetext{a}{The best fit $\av$ from $\chi^2$ fits and the lower and upper $1\sigma$ limits from SED fitting. }
\tablenotetext{b}{The derived log luminosity at $J$ band derived at the best fit $\av$ and the lower and upper 1 sigma $\av$ values; for 
sources without spectral types, $\av$ and \lbol\ estimates were derived assuming a K7 spectral type.}	
\tablenotetext{c}{SED parameters: sources without $1\sigma$~fit uncertainties  in \iracalpha were detected in less than 3 \sst\ \irac\ bands; 
in all cases these values correspond to 95\% upperlimits at $850\micron$ in a $20\arcsec$ beam.  See Table \ref{tab:fproto} for SCUBA
detections of protostars.}
\end{deluxetable*}

Although our new members are preferentially cool stars and thus low mass  $(<0.3\solarmass$; Figures \ref{fig:imf}ad) 
this does not appear to be the result of a bias in our survey.  It is instead a consequence of the IMF,
which peaks in IC~348 for low mass stars \citep{2003ApJ...593.1093L, 2003AJ....125.2029M} coupled with an apparent 
radial variation in this mass function which skews to lower mass stars at large radii where most of our new class II 
sources are found. Besides the unclassifiable protostars we 
found only 1 new early type class II K star (\#1933);  the solar mass members at larger radii are either already 
included in our census and/or perhaps diskless.  We found that the spatially complete class II population of 
IC~348 displays a modest radial variation in the distribution of effective temperatures, which we use as a proxy for mass.  
The hotter, higher mass sources are more concentrated in the cluster's core and cooler lower mass stars prefer the 
cluster halo (Figure \ref{fig:imf}b).  This result using the HR diagram supports the luminosity function analysis of 
M03 which first identified radial MF variations in this cluster.  Further, the M03 IMF analysis is not biased for or against
the presence of disk excess so that despite the inconclusive HR diagram results for the spatially 
incomplete class III population (Figure \ref{fig:imf}c) we conclude that this MF radial skew is real. 

\section{Discussion}
\label{sec:discuss}

\subsection{Total young star population of the IC~348 nebula}
\label{sec:discuss:extent}
 
We have added a substantial contingent of new young stars to the membership of IC~348, 
bringing the total known membership to 363 sources. This is larger than anticipated 
statistically by \citet{2006A&A...445..999C} using 2MASS all sky data.
We now perform an estimate of the total young star population in IC~348, accounting statistically 
for undocumented class III members not identified using our disk based criteria.  
We first use the ratio of class II to class III sources in the \citet{2003ApJ...593.1093L} 
completeness region (70 / 186 = 0.38) to extrapolate from our class II census.
We find 90 class II members within the $10\farcm33$ radius of the \citet{2003AJ....125.2029M} 
survey and, thus, we estimate there should be a population of 227 class III sources or a 
total population estimate of 327 members in a $r\sim1$pc region. 
Although this is consistent with but slightly larger than the $303\:\pm\:28$ members estimated 
by \citeauthor{2003AJ....125.2029M}  using a $2\micron$ luminosity function analysis, 
it suggests that about 30 more class III members remain unconfirmed in this $r\sim1$pc portion
of the IC~348 nebula.

Probably the most efficient way to find the 30 predicted class III sources would be to employ 
deep X-ray surveys; unfortunately, existing X-ray surveys are much smaller than our \sst\ survey, 
and only roughly cover the $20\arcmin$ M03 region. They also miss most of the protostellar ridge.  
A second means to identify young stars is by monitoring for periodic  (variable) stars over large 
cluster areas.  Combining archived X-ray data and recent literature results \citep[e.g.][]{2006astro.ph..6127C}
for these two techniques (see Appendix \ref{app:class3}), we cataloged 27 class III candidates
of which 17 fall within the M03 + Xray portion of our \sst\ survey region.  
Considering that both of these techniques have their own (different) completeness limits 
(less than 1/3rd of the confirmed IC~348 members are periodic while 2/3rds are detected in X-rays) 
these 17 class III candidates confirm our prediction of 30 missing class III members as accurate. 
Wider field X-ray surveys are clearly warranted, especially to elucidate the
radial MF variation we observe for the class II members.

On the larger 2.5pc spatial scale of our \sst\ survey, we conclude that the $118$ known class II  
members suggest a total population size perhaps as large as $\sim420$ IC~348 sources.  
This assumes that the ratio of class II/III members  does not vary much over the survey area.
Thus, we predict approximately 60 class III sources remain either unidentified or lacking
spectroscopic follow up within the immediate vicinity of IC~348.
In total our findings (confirmed or extrapolated) represent a substantial 
(30\%) increase to the traditional population estimate of $\sim300$ sources for IC~348 \citep{1995AJ....109.1682L,2003AJ....125.2029M,2003ARA&A..41...57L}.  
It is, however, unclear where the boundaries of this cluster are and thus where we should stop 
looking for missing members. 
The 2MASS surface density excess identified by \citet{2006A&A...445..999C} extends beyond the 
borders of our \sst\ survey (but appears to underestimate its membership); from a cursory analysis of 
archival \sst\ data\footnote{Wider field \sst\ data of the IC~348 nebula was obtained by an
\sst\ Legacy Science project entitled ``Cores to Disks,'' \citep{2003PASP..115..965E};
the IRAC data was analyzed in \citet{2006ApJ...645.1246J}.  
We downloaded the fluxes from their third incremental release which were posted on a web site 
(\url{http://data.spitzer.caltech.edu/popular/c2d/20051220 enhanced v1/}). We then applied our 
$\iracalpha$ selection criteria to find additional candidates.  Unfortunately, the boundaries 
of the c2d IRAC survey are irregular and one could not simply expand our study to larger cluster radii.} 
it is quite clear that not far from IC~348 there are small aggregates of young stars, including
those around Lk\halpha\ 330~$30\arcmin$ to the NE and around 
MSX6C G160.2784-18.4216 \citep[][]{2003AJ....126.1423K,2006A&A...445..999C} $30\arcmin$ to the SE, 
which may or may not be associated with the star formation we observe within the nebula. 
Even if we were to include all these groups and account for subsequent generations of stars 
yet to form in the protostellar ridge it is clear that in a physically similar volume of space, 
the IC~348 nebula will produce about an order of magnitude fewer stars than the Orion nebula.

\subsection{Physical structure of the IC~348 star cluster}
\label{sec:discuss:structure}

Our \sst\ census of the IC~348 nebula has revealed a couple of new facts about the structure of the 
associated embedded star cluster, which we discuss briefly in this section.   First, our analysis of the 
composite spectral energy distributions of probable cluster members confirms that a population of embedded 
sources along the nebula's southern edge are infact class 0/I protostars, as suggested by previous 
observations of jets and outflows \citep{2006A&A...456..179T,2006AJ....132..467W}.
The protostars follow a ridge of molecular material, are characterized by low spatial  surface densities, and 
are anti-correlated spatially  with the cluster's much more centrally condensed class II and class III population.  
On the other hand these protostars are correlated spatially with a population of millimeter cores, which we find 
however to be mostly starless using our \sst\ data.  

Our analysis of protostars in Trapezium cluster using ground based $3\micron$ data  found somewhat similar results: 
the youngest stars are distributed into an elongated ridge following the densest molecular gas. The youngest Trapezium 
members appear segregated in subclusters with radii of  $\lesssim0.1$~pc,  populations of 10-20+ 
members \citep{2004AJ....128.1254L, 2005ApJS..160..530G}.  We find something rather different in IC~348 
where the protostars are less clustered and have surface densities at least one order of magnitude lower than in Orion
($\mbox{peak}\;\sim 200\,\mbox{stars} \cdot \mbox{pc}^{-2}$~in IC~348 as opposed to the 
$\sim 2000-3000\,\mbox{stars} \cdot \mbox{pc}^{-2}$ we found behind the Trapezium \citep{2004AJ....128.1254L} ).
Moreover, the IC~348 nebula is sufficiently nearby that we can resolve individual protostellar cluster members 
in our \sst\ data and possibly identify the smallest fragmentation scale,  which nonetheless appears unresolved $(<6000\,
\mbox{AU})$.  This is in contrast, for example, to the  Spokes embedded cluster in NGC~2264, where the bright \mips\ sources 
are spaced by 0.1pc yet appear mostly singular in the \sst\ data \citep{2006ApJ...636L..45T}. It is possible that 
higher resolution data will find that the singular Spokes sources will break up into multiples or even small clusters \citep{2006ApJ...642..972Y}, 
but it will be interesting to learn whether their protostellar object densities will approach those we find in 
Orion or are more similar to those we discuss here in IC~348.

We also found that the cluster in the IC~348 nebula is more simply structured than previously thought. 
Using the expanded cluster boundaries provided by our \sst\ census,  we found that the spatial surface density  
of confirmed members is fairly smooth and  that most of the substructure  previously reported in  IC~348 
is not apparently significant.  Only those substructures which appear correlated with molecular gas appear 
to be clusterings of actual members.   To further examine the structure of the IC~348 cluster,  we 
calculated the (spherically symmetric) radial surface density profile of confirmed IC~348 members.
The cluster's surface density drops off smoothly as $r^{-1}$ out to a radius of 1pc, which means the 
space density of stars goes as  $r^{-2}$. This means that the apparent flattening of the radial profile 
for $r>4\arcmin$ seen by \citet{1998ApJ...497..736H} and M03 was the result of variable background 
contamination, which was also the likely cause for the insignificant sub-clusterings found by LL95.  

\subsection{History of star formation in the IC~348 nebula}
\label{sec:discuss:sfh}

Our wide field \sst\ census permits us to reconstruct a more complete history of star formation in the IC~348 nebula.   
Foremost, we found spatially correlated and nearly equal sized populations of class 0/I protostars and starless MM sources 
in a filamentary ridge that is 1pc from the central B star and lying behind the nebula's apparent edge. 
This finding clearly indicates that star formation in the nebula is not finished but is infact ongoing.  
As pointed out by \citet{2005A&A...440..151H}  the large concentration (relative to the entire Perseus cloud)
of MM cores near IC~348 is infact  consistent with a present day star formation rate equivalent to that 
which built the older central cluster,  assuming that each core will eventually produce 1-3 stars.   
That our \sst\ data  indicate that the majority of these cores  appear starless suggests star formation
can continue at this rate into the near future.    

\begin{figure}
   \centering \includegraphics[angle=90,width=0.3\textwidth]{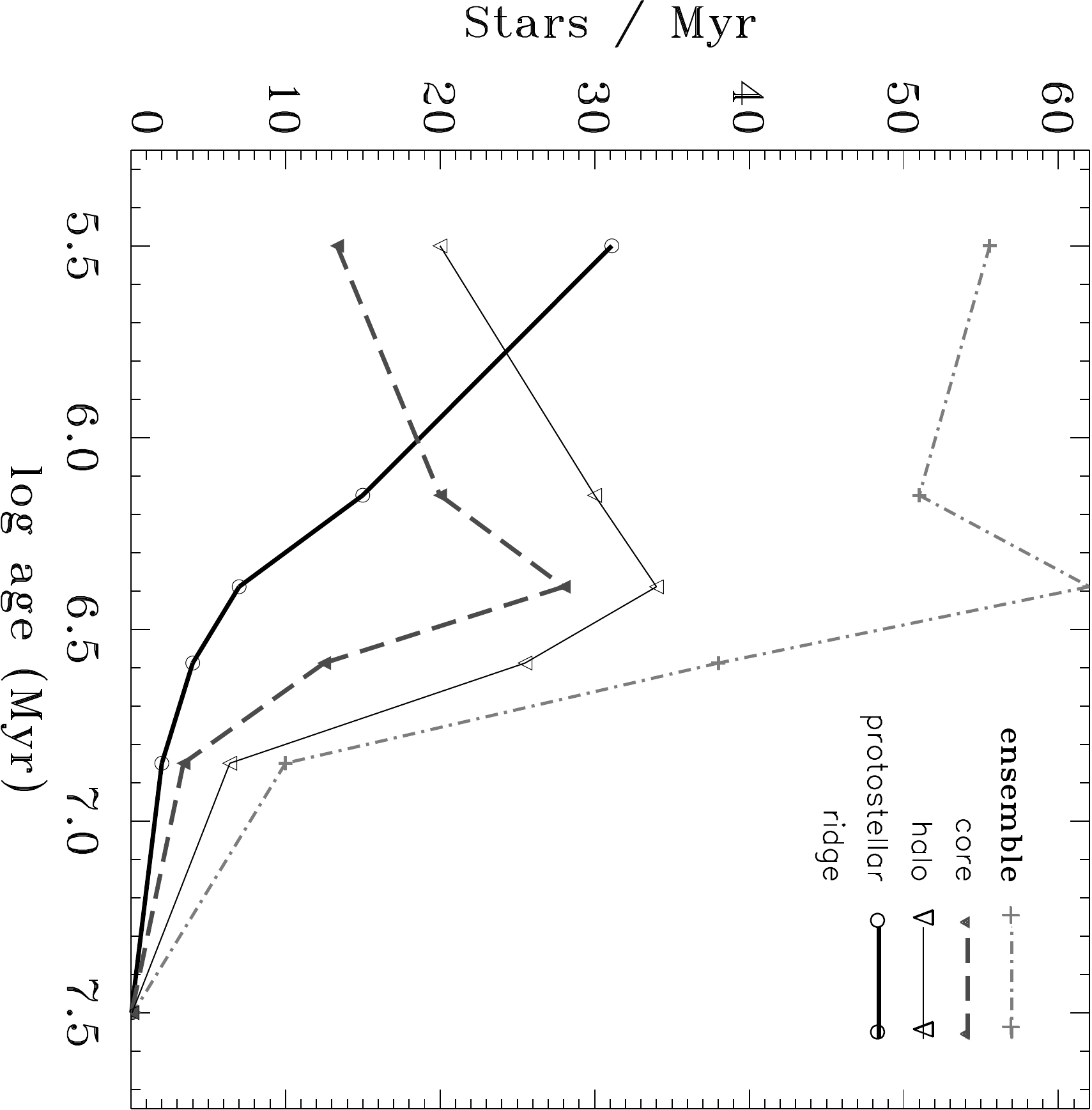}
\caption{History of star formation in the IC~348 nebula.  As in Figure \ref{fig:imf} the 
star formation rate (stars per Myr) is plotted in bins of roughly equal width of logarithmic age 
and normalized by the bin width (in age).   The plot compares the SFH of the IC~348 core 
and the halo as defined in text; they appear to peak at around 2.5 Myr ago. 
The star forming history of the protostellar ridge is a combination of the ages of
members seen in projection toward that ridge  and placed on the HR diagram
(Figure \ref{fig:hr}b) and the ages of the protostars, assuming that
the protostars have formation ages in the past 1 Myr.  It appears to be increasing with time;
regardless, the ensemble star forming history of IC~348 is consistent with roughly
constant star formation over the past 4 Myr. 
\label{fig:sfr}}
\end{figure}

Figure \ref{fig:sfr}a plots the histogram of inferred ages from  the HR diagram, and at first glance the 
SFH for the cluster ensemble is quite broad and suggests a peak at around 2.5 My with perhaps a
decline to the present.  The interpretation of such a ``peak'' depends upon the accurate counting
of the population of embedded protostars, which could not be included into previous studies of 
cluster age spreads \citep[e.g.][]{2000ApJ...540..255P} as they lacked the deep mid-IR data
provided by \sst\ \footnote{\citet{2000AJ....120.3162L} used the statistics of protostellar candidates 
detected at $3.8\micron$ in the Trapezium core of the Orion Nebula Cluster to draw a similar conclusion 
about the quite vigorous present day star formation rate in that nebula.}.  Even if the protostars in the SW ridge were long 
lived \citep[as postulated in][$\tau\,\sim\,1\,\mbox{My}$ for class I]{2006astro.ph..4081W}, the star formation 
rate in the southern molecular ridge is increasing and approaching $\sim50\;\mbox{stars/My}$, 
which already exceeds the average star 
formation rate in the cluster halo (Figure \ref{fig:sfr}).   The sum of the SFH for the cluster ensemble and the protostellar 
ridge confirms essentially a constant star formation rate of $\sim50\;\mbox{stars/My}$ over the past 5~My.
 
Attempts to further quantify of the duration of star formation in IC~348 are very difficult.   
Besides intrinsic differences in birth times, the observed luminosity spread is inflated by the propagated 
uncertainties in the derivation of \lbol\ \citep{1990ApJ...349..197K, 2001AJ....121.1030H}.   
On one hand we have the fact that extremely ``old'' members on the HR diagram could be the result
of gross underestimates of the intervening extinction caused perhaps by the sources being seen edge-on.
On the other hand, the existence of a real luminosity spread on the HR diagram (or color-magnitude diagram) is 
fairly clear evidence for the stars having a range of radii and thus a range of contraction ages independent 
of systematic uncertainties in the theoretical tracks used to interpret them.
From the $V-I_C$ vs $V$ color-magnitude diagram \citet{1998ApJ...497..736H}  argued  that star formation in the IC~348 
nebula was not coeval.  Instead Herbig found the spread of star formation ages in IC~348 was of order 5 My,
which is larger than the members' median age.  Using members drawn from a much larger survey covering
the entire nebula we come the same conclusion: if we conservatively ignore the tails of the observed SFH and use
the half power points in our derived age distribution functions  (Figure \ref{fig:sfr};  cluster ensemble) 
as the age spread we find that non-negligible star formation began at least 4~My ago.   Put another way, if we ignore 
structure in the SFH and infer a constant star formation rate to the present,  the derivation of a median age 
of 2.5~My, implies a star formation duration of $\sim5$~My.   Again, the presence of primarily starless mm 
cores suggests this duration will continue to lengthen.  Note, if we were to assume that the IC~348 nebula 
were closer \citep[250~pc;][]{1999A&AS..137..305S, 2002A&A...387..117B} then the inferred median age and 
duration of star formation (using the half power points of the SFH) would increase by roughly 
$0.5$~My and $2$~My, respectively.

\subsection{The origin \& evolution of the IC~348 star cluster}
\label{sec:discuss:time}

Using the structure and star forming history derived from our \sst\ census,
we can address a few questions about the origin and evolution of the IC~348 star cluster.
Foremost, we observe a difference between the structure of the more populous, centrally condensed 
and somewhat older cluster and the filamentary ridge of likely younger protostars.
As already discussed, there is evidence that the youngest stars in other regions,
such as Orion \citep{2004AJ....128.1254L} and the Spokes cluster in NGC~2264 \citep{2006ApJ...636L..45T},
are also arranged  in small subclusters along a filamentary structure.
\citet{2002MNRAS.334..156S} used numerical simulations of the cluster in 
the Orion Nebula to  show that despite the youth of that cluster $(\tau<1\mbox{ My})$ its 
current structure could be explained by the merger and evaporation of many 
$(N_S\sim100)$ very small subclusters similar perhaps to the protostellar ridge in IC~348. 
Thus, it is possible that the centrally condensed, older cluster 
looks different from the protostellar ridge because of significant dynamical evolution due to 
stellar interactions.  We can examine such a hypothesis by deriving the relevant timescales 
for dynamical evolution to act upon the stars in the 	IC~348 nebula.

Consider the central cluster of  members in the IC~348 nebula: 
within a roughly 1pc radius region there is a total stellar mass of 
$165\;(\frac{N_{\star}}{330}) \times (\frac{\langle\mass\rangle}{0.5})\;\solarmass$.
Were this cluster virialized (by its own stellar mass excluding the natal cloud) 
it would have a 3 dimensional velocity dispersion $(\sigma_{3d})$ of 0.86~km/sec.  
Assuming a star forming efficiency (SFE) less than unity increases this value;
for example, a SFE of $0.3$,  would increase the isotropic virial $\sigma_{3d}$ by a factor of 2.
Rewriting the cluster crossing time, $\tau_{c}=R/v$ \citep{1987gady.book.....B}, as 
\begin{displaymath}
\tau_{c}\: \cong \:1.2 \cdot  \left( \sqrt{ SFE \cdot \left( \frac{330}{N_{\star}} \right) \cdot \left( \frac{0.5}{\langle\mass\rangle} \right) \cdot  \left( \frac{R}{1\mbox{ pc}} \right)^{3} } \right) \mbox{ My},
\end{displaymath}
we find the central cluster has maximum $\tau_{c} \sim 1.2$~My, assuming $SFE=1$.   
The relatively simple radial profile we find for central cluster members and the lack of 
substructure outside of the molecular cloud are consistent with the conclusion that the 
IC~348 cluster is at least one crossing time old \citep{2006ApJ...641L.121T}. 
Indeed, our somewhat conservative estimate  for the duration of star formation in the nebula 
(3-5 My) suggests that the central cluster is at minimum 3-5 crossing times old.  
For systems older than one crossing time, stellar interactions are important and,
subsequently after one relaxation time,
\begin{displaymath}
\tau_{r}\: =\: \frac{0.1 \cdot N}{\ln{N}} \cdot \tau_{c},
\end{displaymath}
they will undergo a change in their velocity of order their velocity; 
this is also the equipartition time for a system \citep{1987gady.book.....B}. 
For the central cluster in the IC~348 nebula the relaxation time corresponds to about five crossing times,
which is of order the duration of star formation in the nebula.  We note (again) that reasonable changes 
to any of these assumptions, e.g. the cluster were initially smaller or had an SFE~$<1$, 
would only increase the dynamical age of IC~348 as expressed in crossing (or relaxation) times. 
Thus, we safely conclude the stars in the nebula have had enough time to undergo an initial relaxation. 
We believe that the mass segregation we observe is thus the product of the equipartition of energy 
during these dynamical encounters and is not primordial.  Put another way which is independent of
whether or not the cluster is relaxed,  if there were primordial mass segregation then its precise
functional form has likely been erased since the cluster is more than a few crossing times old.

Given that sufficient time has passed for the central cluster to undergo dynamical evolution
we find it difficult to differentiate between two viable models for this cluster's origin. 
The current cluster configuration (centrally condensed, smooth radial profile, lack of subclusters)
could be the byproduct of the infall and dissolution of stars or small subclusters that 
formed in filamentary cloud structures, similar to the protocluster ridge.  
The fact that protostellar populations are often observed to be aligned in filamentary 
structures, including, for examples, the Spokes cluster in NGC 2264 and the embedded 
subclusters behind the Trapezium in Orion, lends support to this hypothesis.  
Yet to build up the IC~348 cluster in 3-5 My requires an (constant) infall rate (in stars) of
about $30\solarmass$ per My; there is infact evidence for infall of gas onto the central cluster (see below).
In an alternative model the cluster forms from in single, massive $(>200\solarmass)$~core 
and the protostellar ridge is a subsequent but separate star forming event.  
In this latter case, for example, we could be observing a process of sequential star formation 
in which the nebula's expansion, induced by the presence of the newly formed cluster,
swept up the ridge and triggered a second generation or new burst of star formation within it. 
 
A more detailed comparison of the radial velocities of the stars and gas could provide some clarity.
Stellar radial velocities are, unfortunately, known for only $10\%$ of the cluster members \citep[][
very recently published $v\,\sin{i}$ and heliocentric radial velocities for 27 stars]{2006AJ....132.1555N}.
These measurements, which have a typical uncertainty of 3~km/sec,
yield a median heliocentric radial velocity of 16.5~km/sec for the stars.
This converts to 10~km/sec in the local standard of rest, with a range from 8 to 12 km/sec.
No radial velocities are known for the protostars but the southern molecular filament that 
appears to surround the protostars is blue-shifted relative to the cluster stars 
($v_{r,\star}\,\sim\,-1.5$~km/sec).
On the one hand the blue velocity shift of the southern filament relative to the stars
is consistent with it being swept up (and pushed outward) by the nebula.
Since these relative radial velocities are of order the escape speed at the 
distance of the protostellar ridge ($\sqrt(2)\sigma_{3d}$ or $\sim 1.2$~km/sec 
assuming the star cluster's potential can be treated as that of a 
uniform sphere of mass $165\solarmass$) the protostars may escape.
On the other hand the relative radial velocities of the stars and the gas
provide evidence for continued global infall of gas onto the cluster stars:   
the central filament, which lies behind the cluster stars,
also has a radial velocity of -2~km/sec in the rest frame of the stars 
and this gas is therefore colliding with or falling in toward the cluster.
Perhaps a future study combining additional, higher precision stellar radial velocities and 
a more detailed map of the gravitational potential well created by the star 
cluster and the molecular gas  will provide an origin and fate for the youngest members of the nebula.

There are a few additional conclusions we can draw about the IC~348 nebula and its members.
First, class II and III sources have the same median ``age'' $(\tau\sim2.5\mbox{My})$ and the same 
luminosity spread on the HR diagram.  This means that external to the protostellar ridge, disked and diskless 
stars are in general co-spatial and ``coeval;'' there is absolutely no evidence that the halo represents generations of 
stars which formed before \citep[e.g.][]{1998ApJ...497..736H} or more recently \citep[e.g.][]{2006A&A...456..179T} 
than the cluster core.   Put another way,  we have no information from the spatial distribution of disked and non-disked sources 
(outside the ridge) to indicate when or where they were created.  Interestingly, a uniform spatial and temporal 
distribution of  class II and III  sources suggests that there is a wide dispersion in the timescale 
for (inner) disk evolution,  regardless of the stars' initial configurations.  We can state this same point
another way and  suggest that since  the age spread in IC~348 is of order the disk dissipation timescale 
as derived from young clusters with a range of median ages \citep{2001ApJ...553L.153H}, the dispersion observed in
such a correlation is probably real instead of a byproduct of uncertainties in age or disk excess measurement.  
This uniform spatial and temporal mixing of class II and III members also affirms the notion that accretion 
does not significantly alter the locations of the stars on the HR diagram.

\section{Conclusions}
\label{sec:conc}

Using sensitive \sst\ mid-IR observations we have performed a census of disk-bearing members of the IC~348 
young cluster in Perseus, including class II T-Tauri stars and embedded class 0/I protostars.  
Using spectral indices indicative of excess mid-infrared emission, we identified and scrutinized roughly 200
candidate YSOs about which we can draw the following conclusions:

\begin{description}

\item[1)] There are a total of 118 class II members within a 2.5~pc region in and around the IC~348 nebula.
Using extensive existing and new spectroscopy we determine that 118 of 136 candidate class II sources
are actual members, indicating that the spectral diagnostic, $\iracalpha$, is fairly robust for identifying 
class II stars.

\item[2)] We catalog a population of 31 protostars, of which $\sim20$ are high quality candidates (confirmed via
other source characteristics such as spectra). Three appear likely to be in the youngest class 0 phase.  
The catalog of protostars includes 11 faint class I candidates though this faint sample still appears contaminated by
background sources which are unassociated with the molecular gas cloud.  Some of these $\sim30$ protostars have 
been  previously associated with Herbig-Haro jets and molecular outflows, but lacked an SED analysis appropriate 
to their classification.
Using SED diagnostics to identify class I members was much less effective than for finding  class II YSOs;
more than half of the initial sample of SED selected class I candidates were eliminated as non-member background
contaminants with  strong PAH emission features suggestive of extragalactic sources.  
Reconnaissance spectroscopy of very faint class I candidates  reveal only interlopers which are probably all background galaxies.  

\item[3)]  The size of the class II population suggests a total cluster size of approximately 420 members, 
which includes a prediction of 60 new class III members that were not uncovered by our \sst\ survey. 
This estimate is reinforced by a search of archival X-ray data that covers a much smaller area than 
our \sst\ data but that nonetheless allow us to identify candidates corresponding to about half of these 
predicted members (also, Appendix C).
Comparing various techniques for finding young stars in IC~348, we find that disk excess surveys were 
successful at identifying approximately  1/3rd of the population, which is similar to the fraction of 
members that are periodic photometric variables.  On the other hand,  60-80\% of 
the known population are detected in X-rays.

\end{description}

We further analyzed the properties of the YSOs we identified in the IC~348 nebula, including plotting 
their spatial distributions, deriving their clustering properties and estimating their physical properties 
by placing them on the HR diagram. From this analysis we draw the following conclusions about star 
formation in the IC~348 nebula:

\begin{description}

\item[1)] Protostars and class II/III YSOs are spatially anti-correlated, with protostars restricted to a narrow
filamentary ridge 1pc SW of the exposed cluster's core.  The existence of this protostellar ridge illustrates the need for \sst\ 
surveys to identify securely a cluster's protostellar population before conclusions are drawn about that young 
cluster's structure or star forming history.  

\item[2)] The stars forming in this protostellar ridge are characterized by a lower spatial surface density than
either the central cluster core or those protostellar subclusters found in Orion; they also display no preferred 
{\it resolved} spacings which could  trace the fragmentation scale of the dense molecular gas in the region. 
A few small pairs or triples trace the highest order of multiplicity in the region but most protostars appear 
essentially solitary (down to $400$~AU). 

\item[3)] The structure of the central cluster is much simpler than previously supposed.
Using confirmed cluster members we found that we do not recover most of the small sub-clusterings previously reported 
in the halo of the central cluster.  Instead the central cluster displays a smooth $r^{-1}$ radial surface density profile 
out to a radius of 1~pc.  That the exposed cluster shows little substructure indicates that nebula is more than a 
crossing time old \citep{2006ApJ...641L.121T}.

\item[3)] The star forming history of the IC~348 nebula is consistent with essentially constant star formation
$(\sim50\;\mbox{stars per Myr})$ over the past 2.5- 5 Myr.   The star formation rate in the southern molecular
ridge is roughly the same as that spatially averaged rate which formed the foreground cluster, and an ensemble of 
$\sim15$~starless mm cores mixed with the protostars indicates star formation will continue at a
similar rate in the SW ridge into the near future \citep{2005A&A...440..151H}.

\end{description}

Star formation in the vicinity of the IC~348 nebula has been relatively long lived, corresponding 
to at least a few cluster crossing times.   The cluster is also relaxed as evidenced by the 
segregation of low mass members to the cluster halo, which was reported previously by M03 but is 
confirmed here using the HR diagram. On the one hand this relatively long duration of star formation 
means that we cannot determine a precise origin for the central cluster based simply on its structure; 
such information about its primordial structure appears to have been erased. 
On the other hand, because the youngest protostars in IC~348 have a filamentary distribution 
and this distribution matches what is observed in other embedded clusters, e.g. the Orion and Spokes clusters, 
we tend to favor a model where the central cluster was built from members that formed in filaments or 
perhaps small subclusters and that have since fallen into the central cluster's potential well. 
The relative radial velocities of the stars and gas in IC~348 are infact consistent with global 
infall of molecular gas onto the cluster.
In summary, we believe that what we have observed in the protostellar ridge 1pc SW of the central 
IC~348 cluster represents the primordial building blocks for young embedded clusters.

Finally, the argument that star formation is ``fast,'' i.e., beginning rapidly after parts of an initially turbulent 
cloud passes some critical gravitational threshold, should not preclude the idea that star formation may also be long lived.   
Until either the natal gas reservoir is depleted, resulting in a relatively high star formation efficiency, or the
infall of gas and new stars is disrupted by an ionizing member, star formation continues. Clearly, neither circumstance has yet 
been reached for the IC~348 nebula.  As star formation in the  IC~348 nebula  does not appear destined to soon cease, 
a fairly long period of star formation $(>2.5\mbox{ Myr})$ in a fairly small volume $(R\sim1\mbox{ pc})$ of space 
should be considered when examining numerical renditions of cloud collapse or the dynamics of young stars.
 
\acknowledgments
We thank Alyssa Goodman for discussions regarding the molecular gas in IC~348
and James Di Francesco for the SCUBA $850\micron$ image, which was provided in advance of publication.
We are grateful for comments and questions provided by an anonymous referee.
K. L. was supported by grant NAG5-11627 from the NASA Long-Term Space Astrophysics program.

This work is based [in part] on observations made with the \sst\ Space Telescope, which is
operated by the Jet Propulsion Laboratory, California Institute of Technology under a contract with NASA.
Some of the data presented herein were obtained at  Infrared Telescope Facility, which is operated
by the University of Hawaii under Cooperative Agreement with
the National Aeronautics and Space Administration and at the W.M. Keck Observatory, which is operated 
as a scientific partnership among the California Institute of Technology, the University of California 
and the National Aeronautics and Space Administration. This Observatory was made possible by 
the generous financial support of the W.M. Keck Foundation. The authors wish to recognize and 
acknowledge the very significant cultural role and reverence that the summit of Mauna Kea has 
always had within the indigenous Hawaiian community.  We are most fortunate to have the opportunity 
to conduct observations from this mountain.
Based [in part] on observations obtained with XMM-Newton, an ESA science mission with instruments 
and contributions directly funded by ESA Member States and NASA.

{\it Facilities:} \facility{\sst\ (IRAC, MIPS), IRTF (SpeX), Keck (NIRC), MMT (Blue Channel), Magellan (IMACS), \cxo\ (ACIS), XMM-Newton}


\begin{appendix}

\section{Extinction effects on \iracalpha}
\label{app:av}

\begin{figure*}[!t]
    \begin{minipage}[c]{\textwidth}%
       \centering \includegraphics[angle=0,width=0.35\textwidth]{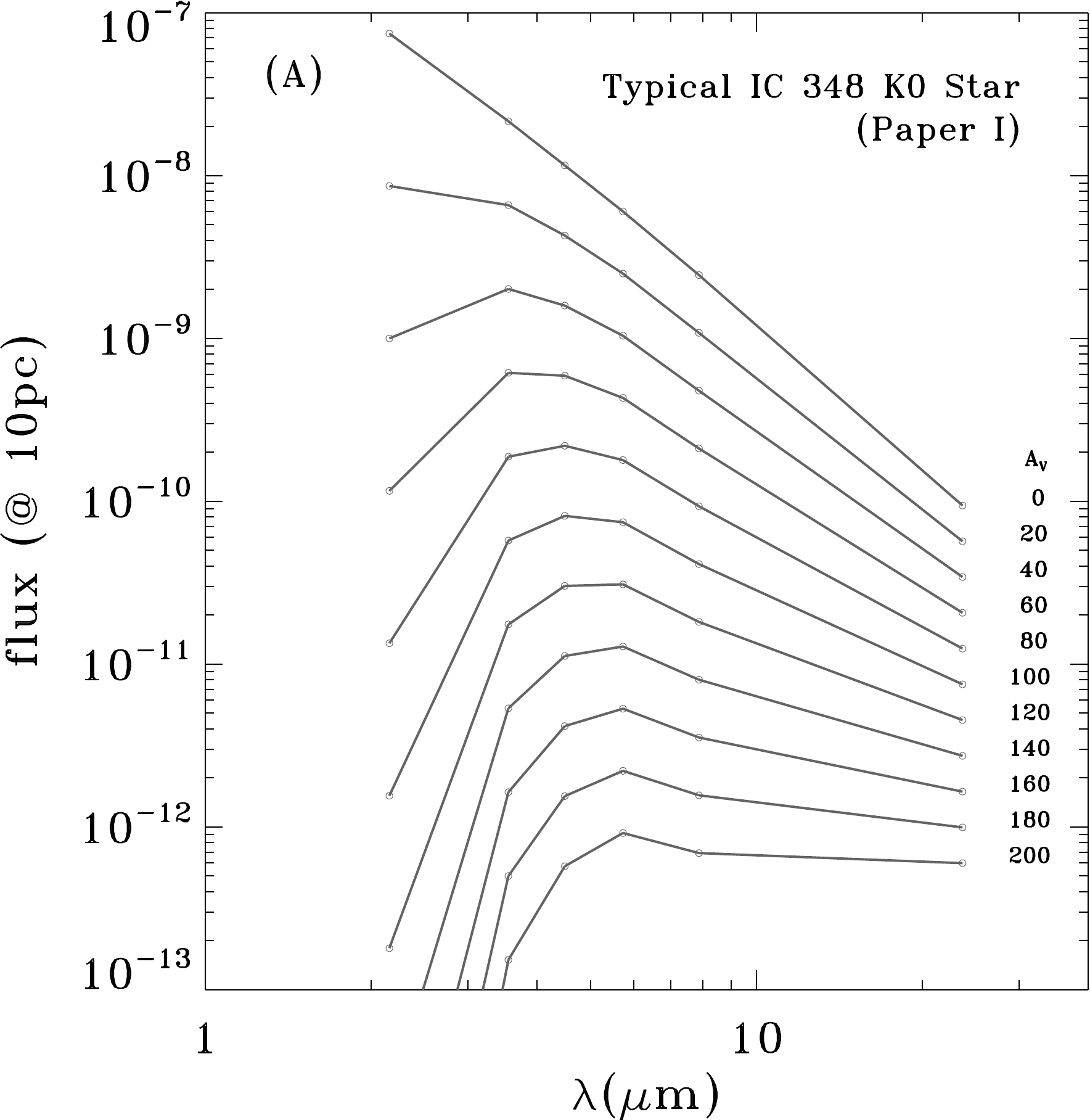}
	\hspace{0.05\textwidth}%
       \centering \includegraphics[angle=0,width=0.35\textwidth]{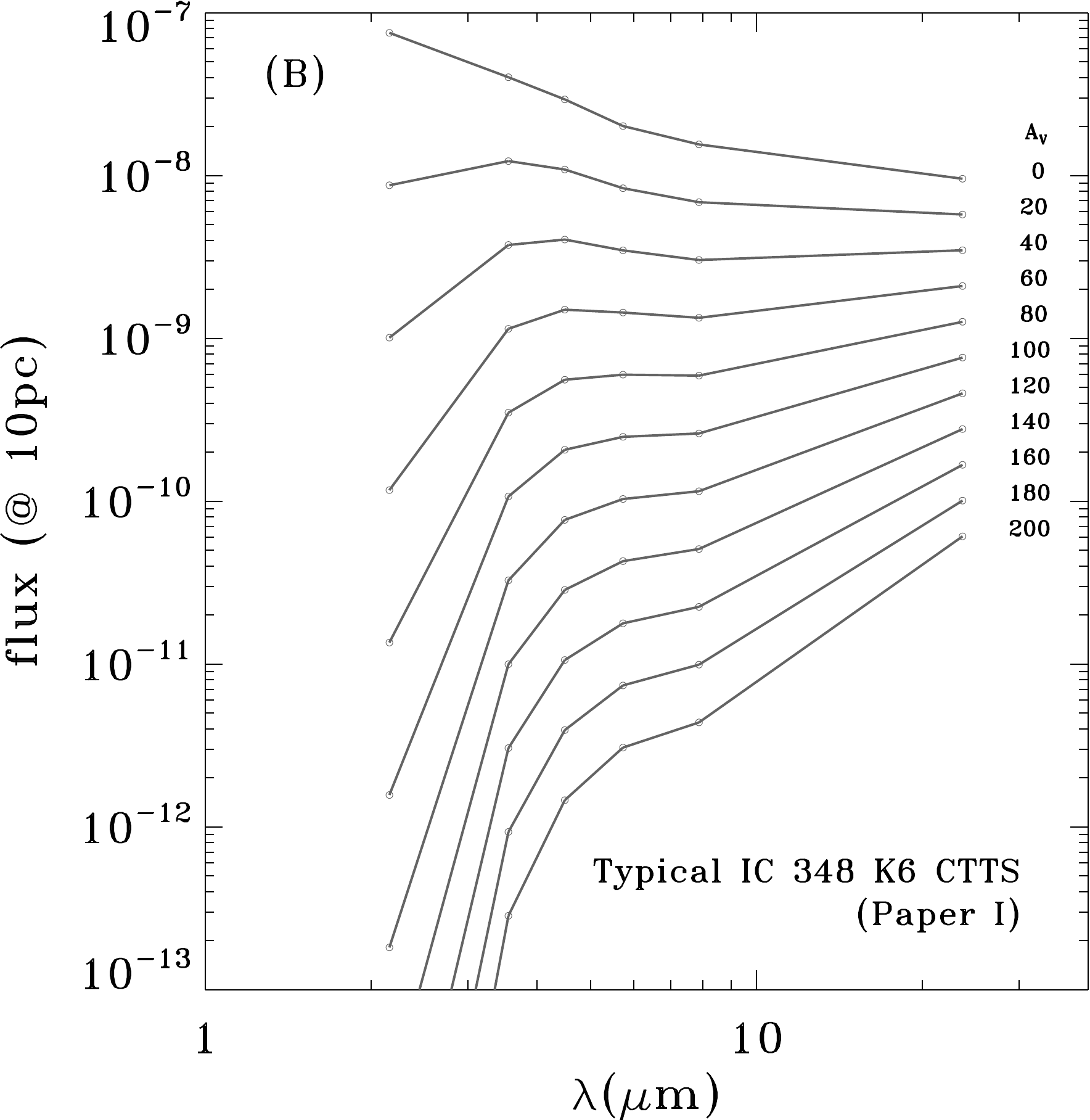}%
    \end{minipage}%
    \hfill \hfill%

    \begin{minipage}[c]{\textwidth}%
       \centering \includegraphics[angle=0,width=0.35\textwidth]{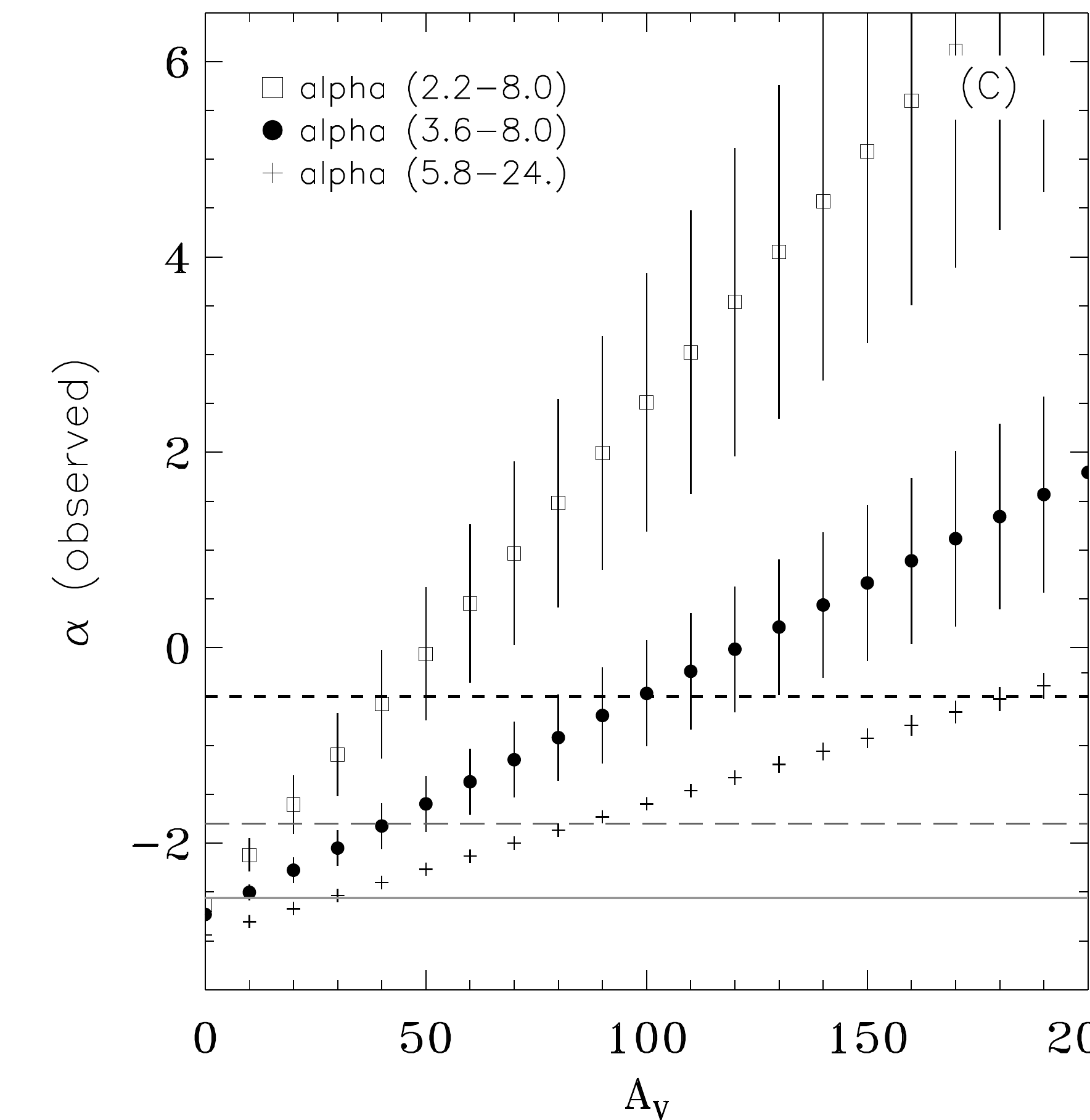}
	\hspace{0.05\textwidth}%
       \centering \includegraphics[angle=0,width=0.35\textwidth]{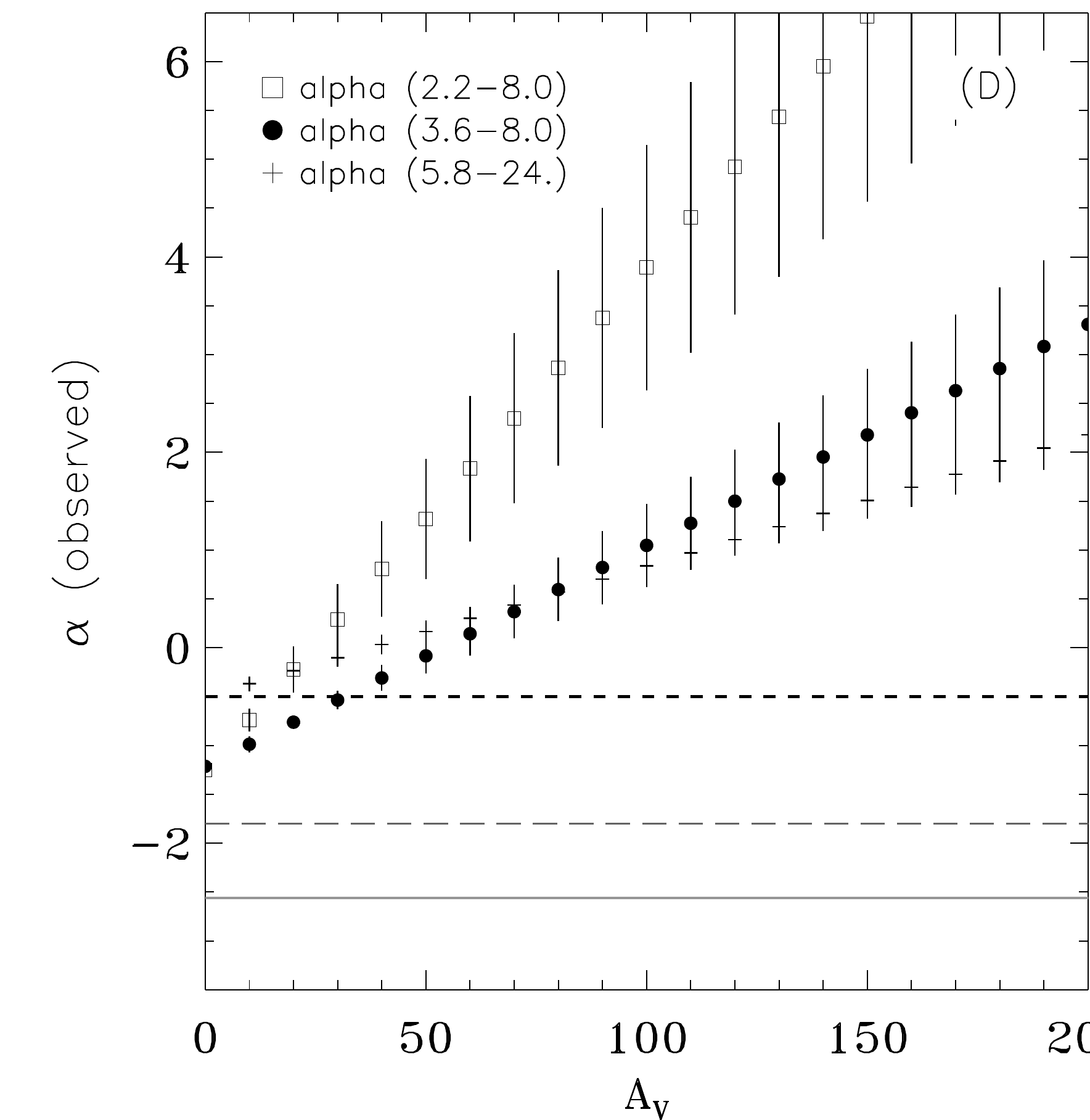}
    \hfill%
    \end{minipage}%
\caption{Effects of reddening on \sst\ spectral energy distributions of T-Tauri stars.  
Panels (A) and (B) show the typical median spectral energy distribution for K0 diskless members and 
K6 class II star+disk members of IC~348 (Paper~I, Table 3) reddened by extinctions up to $\av=200$.  
Panels (C) and (D) plot the change in three spectral indices with increasing extinction; error bars at each
point are the $1\sigma$ fit quality and thus chart the departure of the chosen spectral index from a 
power-law. The slope of the relationship between these spectral indices and $\av$ are given in 
Table \ref{tab:app2}.
\label{fig:app}}
\end{figure*}

We explored the influence of dust extinction on our preferred spectral index, $\iracalpha$. Using a
diskless K0 IC~348 member from Paper~I (Table 3) as a template, we reddened the observed photosphere
by extinctions as large as $\av=200$ using the reddening law from \citet{2005ApJ...619..931I}. Ten
of these reddened SEDs are shown for illustration in Figure \ref{fig:app}a with passbands from $K$
to \mips\ $\SMa$ micron included. %
While the \irac\ slope of the SED requires $\av>100$ to inflect 
to a positive slope, the $K-\SIa\micron$ slope of the SED is inflected by $\av>40$. Note that
even for $\av\sim200$ the $\SIc-24\micron$ SED slope of background stars remains negative. 
Since such large column densities within typical molecular clouds occur only in regions very close to embedded YSOs,
this SED slope proves that background sources with normal Raleigh-jeans SEDs cannot mimic class I
sources except if they were seen through the protostellar envelope of a class I source.

Figure \ref{fig:app}b 
plots the explicit dependence of $\alpha$ on $\av$, from which we can derive
the reddening law for these spectral indices.  We calculate the $\alpha$ index for the \irac,
$K+$\irac\  and $\SIc-24\micron$ portions of the SED, plotting them versus fit quality to demonstrate the
degree of departure from a true power-law as a function of $\av$. These yield the relationships
$\case{A_{\alpha}}{\av}$ given in Table \ref{tab:app2} 
and used in Figure \ref{fig:cmds}.

We repeated this experiment with empirical SEDs for thick disk classical T-Tauri stars in IC~348
(Figure \ref{fig:app}ab). Reddening the median observed SED of K6-M0 IC~348 member (Paper~I, Table
3), we find that $\av>40$ cause both the \sst\ based SED indices to inflect. Actually, for $\av>100$
the $\iracalpha$ index becomes steeper than the $\SIc-24\micron$ slope, a result that would rarely
occur for background field stars.  In principle, background stars could be differentiated from
cluster members by a rising \iracalpha~slope coupled with a negative $\SIc-24\micron$ slope. 
Again, indices using $K$, which include those indices calculated by \citet{2006ApJ...645.1246J},
are very sensitive to extinction causing source classifications including
that band to become degenerate for $\av>20$.  
A fixed value of $\iracalpha=-1$, for example, could correspond either to a 
typical class II YSO or to an extremely heavily reddened $(\av\sim75)$ diskless star.
However, the \irac\ SEDs of heavily reddened diskless stars are distinct from those of typical Class 
II star+disk sources:  the power-law \irac\ SEDs of class II objects are intrinsically shallow but 
heavily reddened diskless stars are bent downward at 2-5 micron; their poor power-law SED
fits should distinguish them as being diskless.   Finally, we see that characteristic dip at $\SId\micron$,
often seen in the SEDs of embedded YSOs (see Figure \ref{fig:bclass1}) can be produced by large $(\av>100)$ 
reddenings of normal cTTs.  Such a dip does not necessitate particular envelope geometries, 
although the observation of such large column densities may only be possible through an envelope 
\citep{1987ApJ...319..340M}.

\begin{deluxetable}{cc}
\tablecaption{$\frac{A_{\alpha,SED}}{\av}$ \label{tab:app2}}
\tablehead{
\colhead{SED Range} &
\colhead{$\frac{A_{\alpha,SED}}{\av}$}
}
\startdata
$K - \SId$    & 0.0514 \\
$\SIa - \SId$ & 0.0226 \\
$\SIc - \SMa$ & 0.0134
\enddata
\end{deluxetable}%
\begin{figure*}
   \includegraphics[angle=0,totalheight=0.9\textwidth]{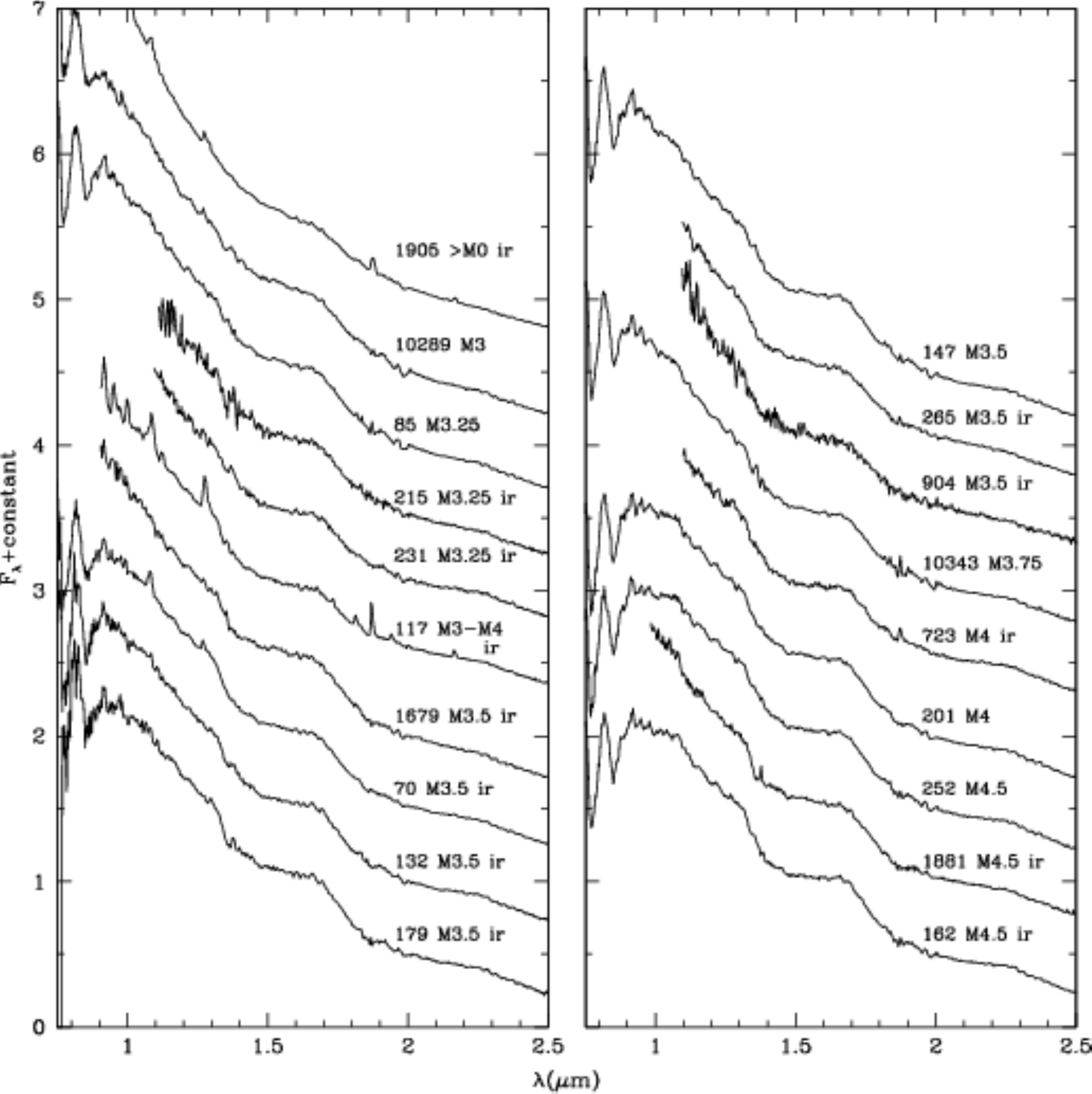}
\caption{
SpeX near-IR spectra of  34 candidate class II and protostellar members of IC~348 
and a sample of optically-classified members of IC~348 and Taurus. 
The candidates are labeled with the types  derived from a comparison to the 
IR spectra of the optically-classified  objects (``ir"). The spectra are ordered according 
to the spectral features in these data. They have a resolution of $R=100$, are 
normalized at $1.68\,\micron$, and are dereddened (\S\ref{sec:spexclass}).
Object names containing five digits or less apply to IC~348, while all other
names refer to Taurus members.
Protostar \#904 (M3.5 ir; $K\sim14.3$) is shown here.
\label{fig:spec1}}
\end{figure*}
\begin{figure*}
   \includegraphics[angle=0,totalheight=0.9\textwidth]{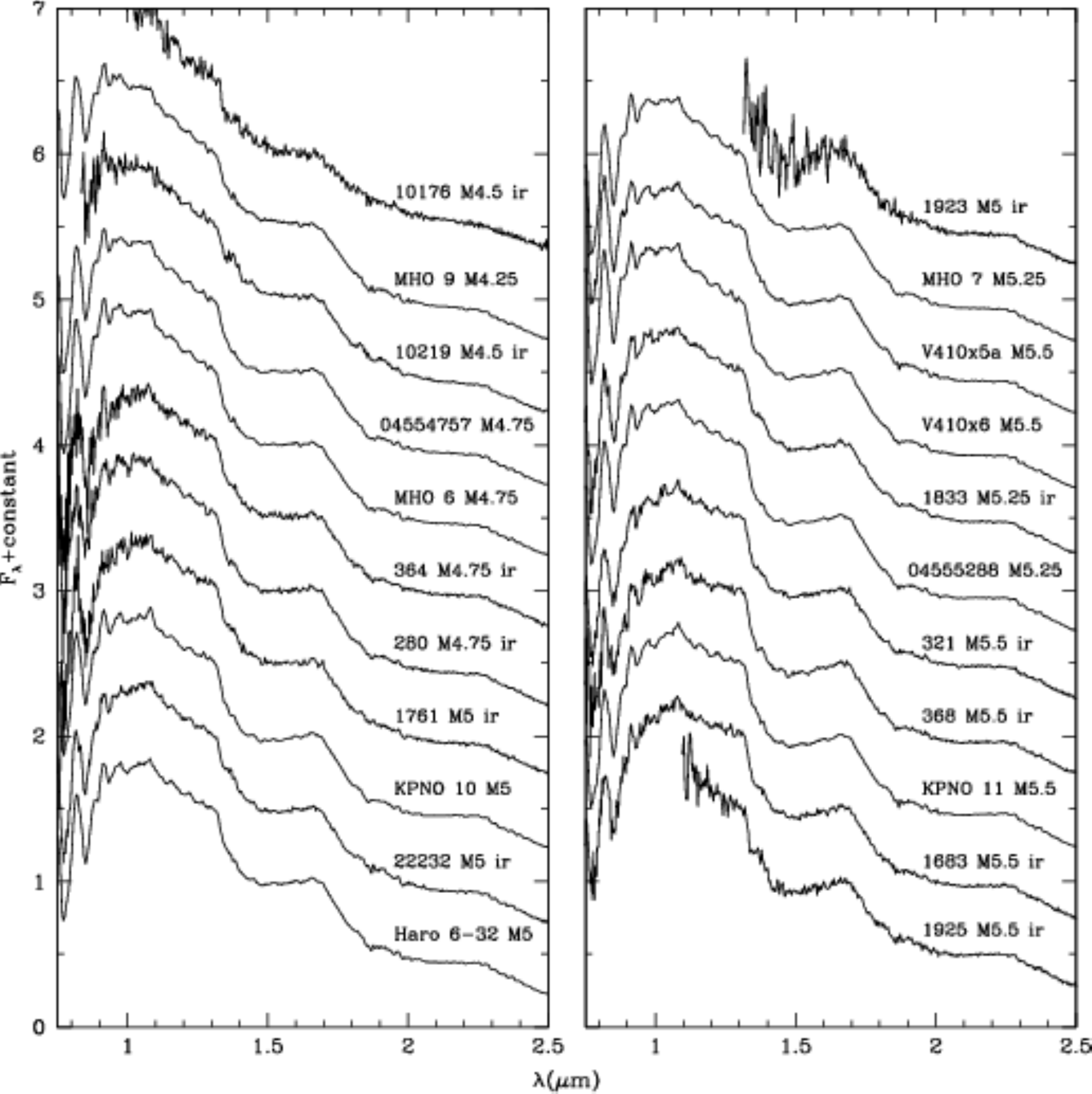}
\caption{
Dereddened SpeX near-IR spectra of candidate IC~348 YSOs.
Same as in Figure~\ref{fig:spec1}.
\label{fig:spec2}}
\end{figure*}
\begin{figure*}
   \includegraphics[angle=0,totalheight=0.9\textwidth]{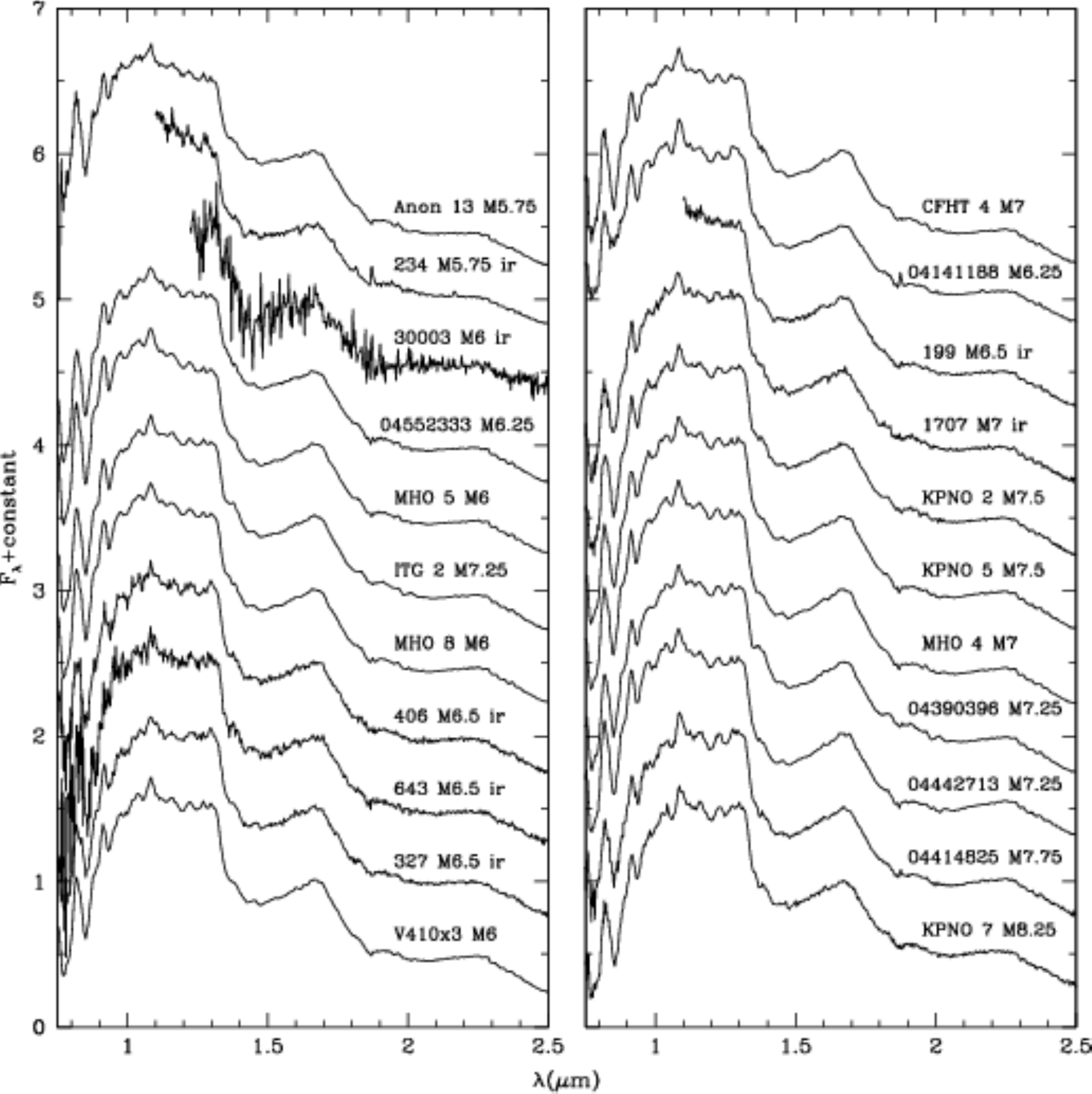}
\caption{
Dereddened SpeX near-IR spectra of candidate IC~348 YSOs.
Same as in Figure~\ref{fig:spec1}. Note the $\mbox{H}_2$ emission
of protostellar candidate \#234.  Another protostellar candidate,
\#30003, is faint $(K\sim15.2)$ and embedded in a scattered light cavity.
\label{fig:spec3}}
\end{figure*}
\begin{figure*}
   \includegraphics[angle=0,width=0.9\textwidth]{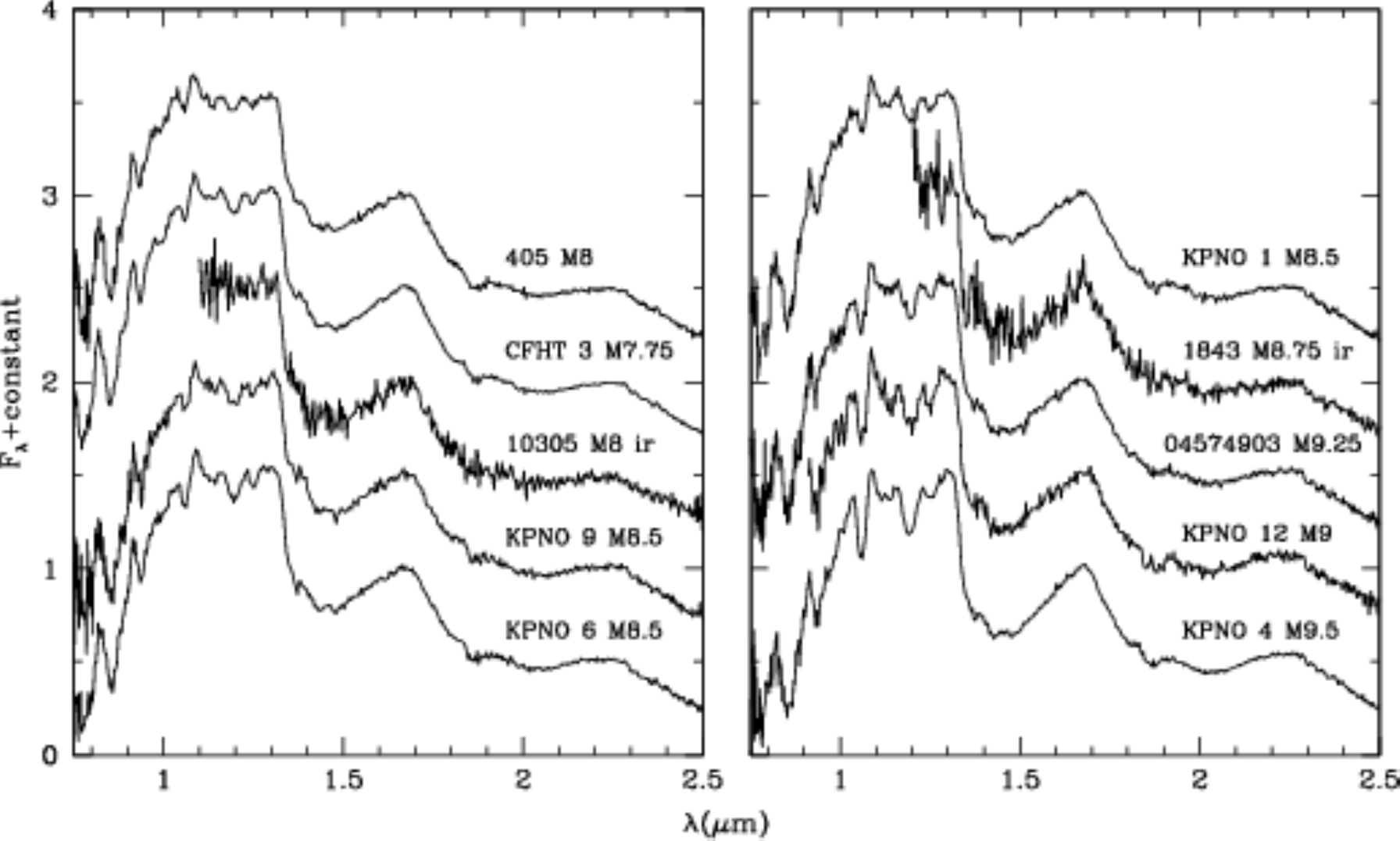}
\caption{
Dereddened SpeX near-IR spectra of candidate IC~348 YSOs.
Same as in Figure~\ref{fig:spec1}.
\label{fig:spec4}}
\end{figure*}
\begin{figure*}
    \begin{minipage}[c]{\textwidth}
    \hspace{0.01\textwidth}%
      \begin{minipage}[c]{0.6\textwidth}
       \centering \includegraphics[angle=0,width=\textwidth]{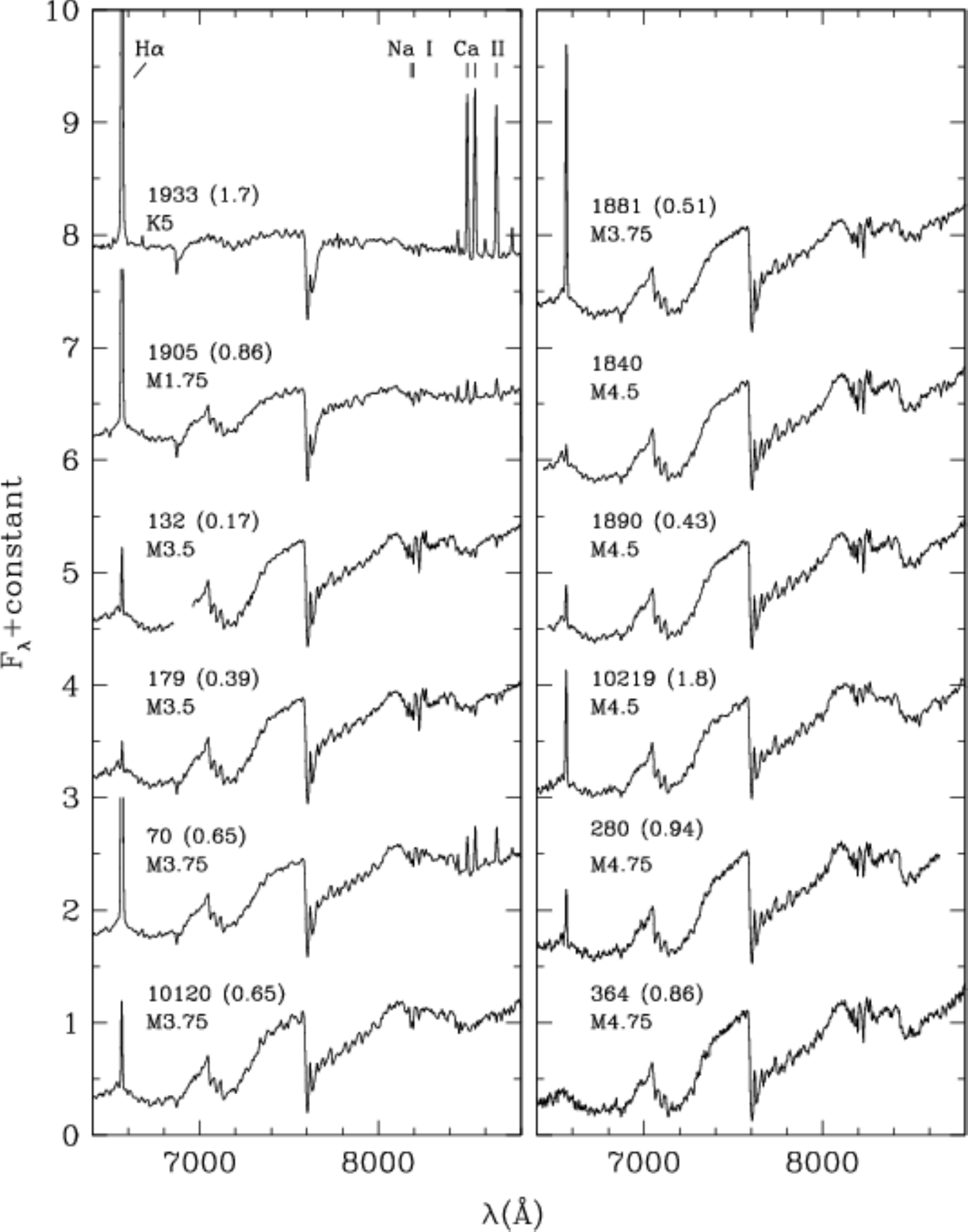}
      \end{minipage}%
      \hspace{0.03\textwidth}%
      \begin{minipage}[c]{0.3\textwidth}
       \centering \includegraphics[angle=0,width=\textwidth]{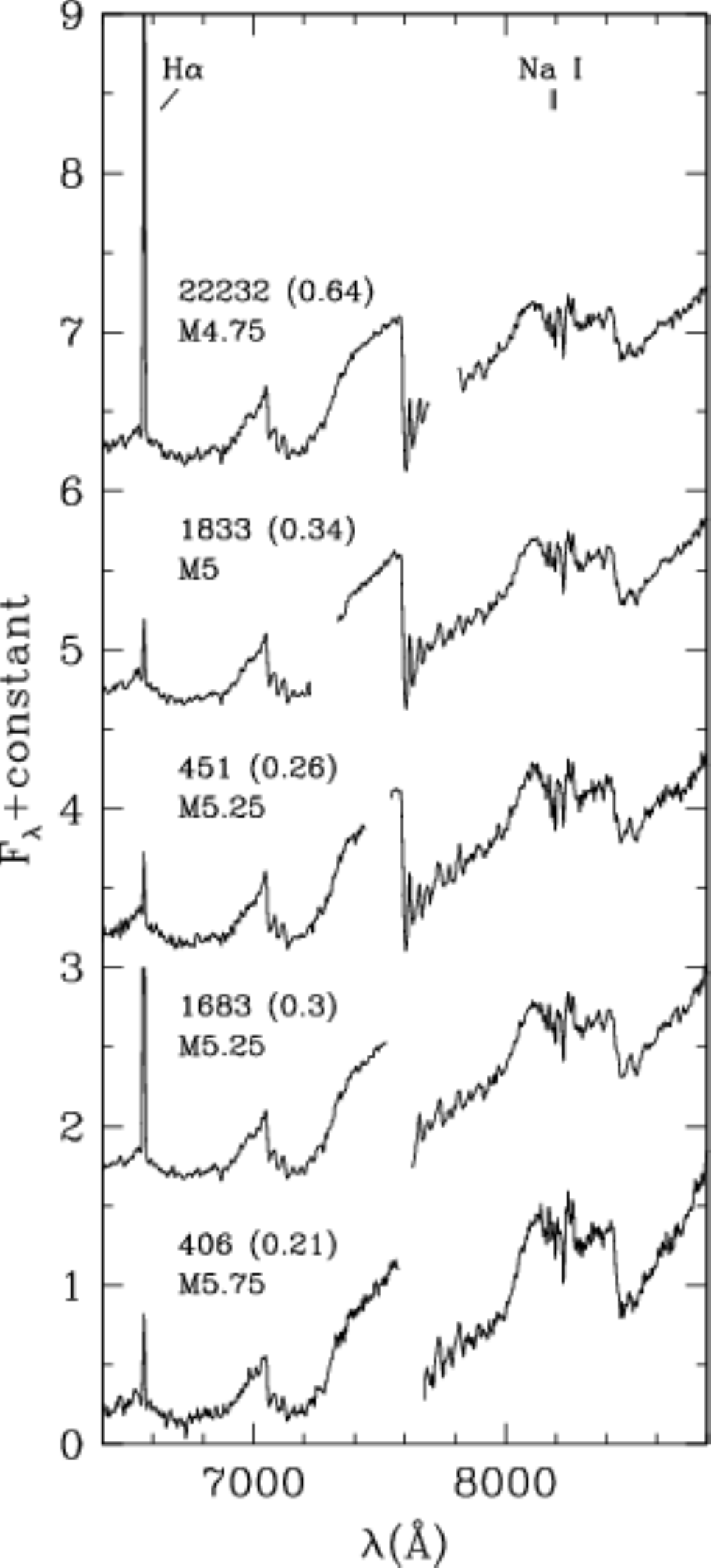}
      \end{minipage}%
    \end{minipage}%
\caption{
Optical spectra of 17 new IC~348 members identified in this work.
Spectra were obtained with the IMACS instrument on the Magellan~I telescope 
and the Blue Channel spectrograph on the MMT. The IMACS spectra were obtained 
in multi-slit mode and some of the spectra fell across two CCDs, 
resulting in gaps in the spectra.  The spectra have been corrected for 
extinction, which is quantified in parentheses by the magnitude 
difference of the reddening between 0.6 and 0.9~\micron\ ($E(0.6-0.9)$).
The data are displayed at a resolution of 8~\AA\ and are normalized at
7500~\AA.   
\label{fig:op-spec1}}
\end{figure*}
\begin{figure*}
    \centering
    \begin{minipage}[c]{\textwidth}
       \centering  \includegraphics[angle=90,totalheight=0.6\textwidth]{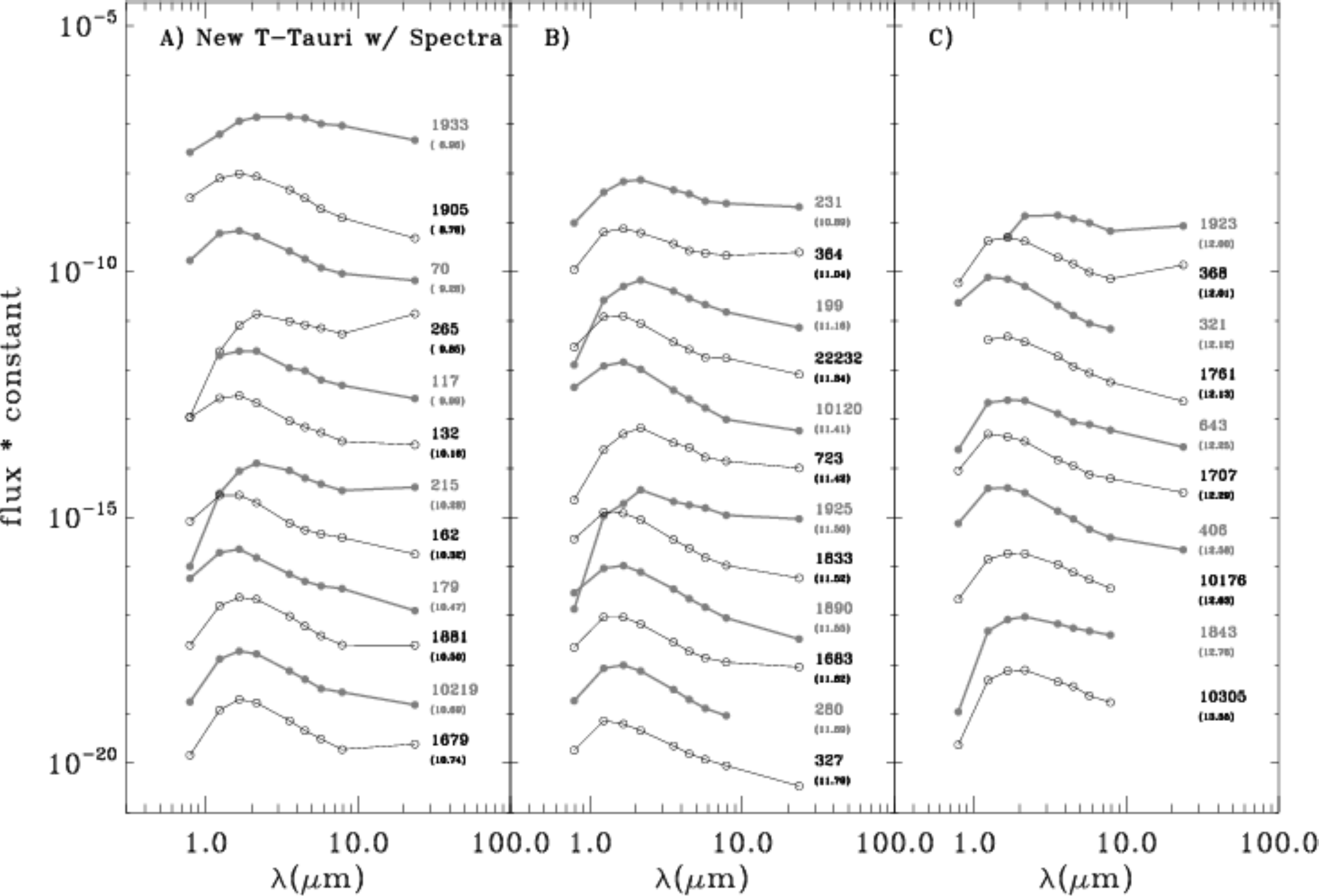}
    \end{minipage}%
\caption{Observed spectral energy distributions of 31 new class II IC~348 members with 
SpeX spectra (see  Figures \ref{fig:spec1} -- \ref{fig:spec4} for
{\it dereddened} optical or near-infrared spectra of these objects; also Table \ref{tab:spex}).
Sources are sorted in order of decreasing $\SIc\micron$ brightness,
which is given in parenthesis beneath each source's identification number.
Plotting symbols, line thickness and line color alternate from SED to SED for clarity.
\label{fig:class2s}}
\end{figure*}
\section{Spectroscopy of new members}
\label{app:spex}

\subsection{Infrared Spectra}
\label{sec:spexobs}

We selected for spectroscopy objects in the \irac\ images that display \irac\ SEDs indicative of disk
excess, are sufficiently bright for the spectrometer employed ($K\lesssim15$), and have not been
previously classified as field stars or cluster members.  A resulting sample of 39
candidate cluster members with spectroscopic confirmation is provided in
Table~\ref{tab:spex}. We also observed a sample of 36 known late-type members of IC~348 and Taurus
\citep{1998AJ....115.2074B,2002ApJ...580..317B,1998ApJ...493..909L,2003ApJ...590..348L,2003ApJ...593.1093L,1999ApJ...525..466L,2004ApJ...617.1216L}, which are listed in
Figures~\ref{fig:spec1}-\ref{fig:spec4}. These optically-classified objects will be used as the
standards during the classification of the candidates in \S\ref{sec:spexclass}. These data were
collected with the spectrometer SpeX \citep{2003PASP..115..362R} at the NASA Infrared Telescope Facility (IRTF) on
the nights of 2004 November 11-13 and 2005 December 12-14. The instrument was operated in the prism
mode with a $0\farcs8$ slit, producing a wavelength coverage of 0.8-2.5~\micron\ and a resolution of
$R\sim100$. The spectra were reduced with the Spextool package \citep{2004PASP..116..362C}, which included a
correction for telluric absorption \citep{2003PASP..115..389V}.

\subsection{Classification}
\label{sec:spexclass}

To measure spectral types for the candidate members of IC~348 that we observed spectroscopically in
\S\ref{sec:spexobs}, we used the absorption bands of VO and TiO ($\lambda<1.3$~\micron) and H$_2$O
($\lambda>1$~\micron). These bands are the primary spectral classification diagnostics for late-type
objects \citep{1991ApJS...77..417K,2001ApJ...548..908L,2001AJ....121.1710R} and are broad enough to be easily detected at the low resolution
of our data. Because near-IR H$_2$O absorption bands are stronger in young objects than in field
dwarfs at a given optical spectral type \citep{1999ApJ...525..440L,2001MNRAS.326..695L,2004ApJ...600.1020M}, spectral types of young objects
derived from H$_2$O with dwarf standards will be systematically too late. Instead, to arrive at
accurate spectral types, optically-classified young objects rather than dwarfs should be used when
measuring spectral types of young sources from steam \citep{1999ApJ...525..440L,2003ApJ...593.1093L}, which is the
approach we adopted in our classification of the candidates in IC~348.

To facilitate the comparison of the band depths between the candidates and the optically-classified
known members, we have dereddened the spectra to the same slope as measured by the ratios of fluxes
at 1.32 and 1.68~\micron. These dereddened spectra are not meant to be precise estimates of the
intrinsic, unreddened appearance of these stars since the slopes likely vary with spectral type. As
shown in Figures~\ref{fig:spec1}-\ref{fig:spec4}, we first arranged the dereddened spectra of the
previously known, optically-classified members of IC~348 and Taurus in order of the strengths of
their molecular absorption features. With a few minor exceptions, the IR features change
monotonically with optical type. We then measured a spectral type for each candidate by visually
comparing the absorption features in its spectrum to those in the data of the optically-classified
objects. Through this analysis, we found that 34 of the 39 candidates in our sample exhibit M types.
We have inserted these 34 sources in the sequence of optically-classified objects in
Figures~\ref{fig:spec1}-\ref{fig:spec4} and have labeled them with the types derived from these IR
spectra, which have uncertainties of $\pm\:0.5$~subclass unless noted otherwise. 

The detection of late-type, stellar photospheric features now demonstrates that these
objects indeed are young stars, and thus members of IC~348. Other available evidence of youth and
membership for these objects is listed in Table~\ref{tab:spex}, which is based on the diagnostics
described by \citet{2003ApJ...593.1093L} and \citet{2005ApJ...623.1141L}.   Of the 35 sources with
new spectral types,  32 are classified as class II T-Tauri stars, while 3 have flat or rising
mid-IR SEDs and are classified as protostellar.  The composite SEDs of these 32 class II 
sources are shown in Figure \ref{fig:class2s}.

\subsection{Optical Spectra}
\label{sec:imacsobs}

We obtained optical spectra of 20 \sst\ selected sources in IC~348 using the 
Blue Channel spectrograph on the MMT during the nights of 2004 December 10 and 11 and 
with the Inamori Magellan Areal Camera and Spectrograph (IMACS) on the Magellan~I telescope at Las
Campanas Observatory during the night of 2005 January 4.  The resulting spectra have a 
wavelength coverage of 6300-8900~\AA\ and a resolution of 3~\AA.  The procedures for the
collection and reduction of these data were similar to those described by \citet{2004ApJ...617.1216L}.

The sources were classified in the same manner as data for Taurus taken on the same nights
\citep{2006ApJ...645..676L}. Of the 20 targets, 17 were members and 3 were determined
to be non-members (the infrared excess of which were very weak; Table \ref{tab:nonmembers}).
Of the members, 13 sources had both optical and SpeX IR spectra and the
spectral types derived from them in general agreed very well, except for two class II
members whose infrared spectra were indeterminate (1905, 1933).
Four sources have only optical  spectral types, including two new class II members 
(1890 and 10120) and two class III members (Appendix \ref{app:class3}).
The reduced, dereddened, optical spectra of these 17 members are displayed in Figure \ref{fig:op-spec1}.
In addition to spectral types, we measured particularly useful optical spectral features (e.g. $\halpha$) 
\begin{deluxetable}{lll}
\tablewidth{0pt}
\tablecaption{Spectral features of new members\label{tab:app1}}
\tablehead{
\colhead{Source} &
\colhead{EW(H$\alpha$,unc)} &
\colhead{Other spectral information\tablenotemark{a} }
}
\startdata
70      & 145\phd\phn\phn\phn(10)      & HeI 6678; \\
        &                              & Ca II: [3.1(0.3),4.1(0.3),3.1(0.4]  \\
132     &   \phn11.5\phn\phn(0.5)      &   \\
179     &\phn\phn4.8\phn\phn(0.5)      &   \\
273     &                              &   \\
280     & \phn18\phd\phn\phn\phn(5)    &   \\
364     &$<1$                          &   \\
401     &                              &   \\
406     & \phn22\phd\phn\phn\phn(2)    &   \\
451     & \phn10\phd\phn\phn\phn(1)    & NaK inconclusive. \\
1683    & \phn62\phd\phn\phn\phn(4)    &   \\
1833    & \phn12.5\phn\phn(1)          &   \\
1840    & \phn\phn4.5\phn\phn(0.5)     &   \\
1881    & \phn45\phd\phn\phn\phn(4)    &   \\
1890    & \phn\phn7.5\phn\phn(0.5)     &   \\
1905    & \phn45\phd\phn\phn\phn(3)    & [OI]6300;, HeI 6678; \\
        &                              & Ca II: [1.7(0.2),1.6(0.2),1.6(0.2]  \\
1933    & \phn55\phd\phn\phn\phn(4)    & HeI 6678; \\
        &                              & Ca II: [17.5(1),18.0(1),18.3(1)]  \\ 
10120   & \phn17.5\phn\phn(1)          &   \\
10219   & \phn23\phd\phn\phn\phn(3)    &   \\
22232   & \phn85\phd\phn\phn\phn(10)   &   \\
\enddata
\tablenotetext{a}{The equivalent widths of Ca II are given in order of
8499,8543,8664\AA.}
\end{deluxetable}%
for these members (\ref{tab:app1}).

\section{Class III membership}
\label{app:class3}

Our \sst\ census cannot uniquely identify diskless cluster members and we did not attempt an exhaustive
search for anemic disk candidates.  In this appendix we describe how we used archival X-ray and recent 
optical monitoring results to tabulate candidate members lacking strong disk signatures (class III), 
which we used to justify our extrapolated population estimate given in Section \ref{sec:discuss:extent}.

Matching the 220 \cxo\ X-ray sources identified in the uniformly processed 
ANCHORS data\footnote{ANCHORS:  an Archive of \cxo\ Observations of Regions of Star Formation;
\cxo\ Archival Proposal 06200277; S. Wolk, PI. See \url{http://hea-www.harvard.edu/~swolk/ANCHORS/}.  
Data (53ksec ACIS;  \dataset[ADS/Sa.CXO#obs/00606]{Chandra ObsId 606}) originally observed (2000-09-21) 
and published by \citet{2001AJ....122..866P}.}  to our source catalog provides the following statistics:  
$15\%$ (31) of these sources have no match to our \sst\ catalog or are lacking near-IR photometry -- 
these are all likely from background galaxies; 12 X-ray sources with $\iracalpha>-0.5$, 
consisting of 4 flat spectrum protostars,  4 low luminosity candidate class I and 
4 rejected low luminosity class I sources; 2 known foreground stars and 162 known members.  
We inspected the composite SEDs of the remaining 13 sources; 
on the $I-J$ vs $I$ color-magnitude diagram three of them fall below the main sequence at the distance 
of IC~348 and were rejected; the remaining 10 fall into the locus of X-ray detected known members.
Similarly, a cross match of our master catalog to the 71 unique X-ray sources in wider-field XMM data  
\citep{2004A&A...422.1001P}\footnote{\dataset[ADS/Sa.XMM#obs/0110880101]{XMM ObsId 0110880101; ObsDate: 2003-02-02}} 
yielded a further 11 candidate X-ray members along with 39 probable extragalactic sources, 17 confirmed members 
and 3 known non-members.  Lastly, we searched a recent catalog of IC~348 periodic sources \citep{2006astro.ph..6127C} and 
found 5 periodic unconfirmed members that fell inside our \sst\ survey region but outside of pre-existing X-ray surveys.  
Of these five, one falls below the main sequence and we did not consider it a member 
(\citeauthor[][source \#140]{2006astro.ph..6127C}); 
thus, within the boundaries of our \sst\ census the total number of candidate IC~348 
member identified by these two techniques is 25.
Only two of these 25 candidate members (\#104 and 185) were detected at $\SMa\micron$; 
both are anemic disk members with $\iracalpha=-2.39\mbox{ \& }-2.48$, respectively. 
The SEDs of the remaining 23 candidate members are consistent with stellar photospheres.  
Two X-ray selected  candidates  were fortuitously assigned random slits in our 
IMACS observations:  \#~273 is an M4.25 type member and \#~401 is an M5.25 type member.
Two other anemic disk sources (\# 451 and 1840), which are neither X-ray sources nor periodic,
have optical spectral types and luminosities  that suggest they are infact members. 
Note that excess of \#451 is very weak while the gravity sensitive NaK  features are indeterminate; 
thus, its membership is poorly defined.  As discussed above, these 27 sources correspond to about
half the predicted number of class III members based on the statistics of class II members
identified in our \sst\ census.
Source names, cross-references, positions and photometry of these class III source are given in Table \ref{tab:fclass3}.

\end{appendix}


%

\clearpage
\begin{landscape}
	\begin{deluxetable}{llrrrrrrrrrrrrrrrrr}
	\tablewidth{0pt}
	\tablecaption{IC~348 class III members and candidates \label{tab:fclass3}}
	\tablehead{
	\colhead{ID} &
	\colhead{Cross-reference} &
	\multicolumn{2}{c}{Position (J2000)} &
	\multicolumn{15}{c}{Photometry\tablenotemark{(2)}} \\
	\colhead{} &
	\colhead{\tablenotemark{(1)}} &
	\colhead{RA} &
	\colhead{DEC} &
	\colhead{$I_c$} &
	\colhead{$z$} &
	\colhead{$J$} &
	\colhead{$H$} &
	\colhead{$K_s$} &
	\colhead{$3.6$} &
	\colhead{unc} &
	\colhead{$4.5$} &
	\colhead{unc} &
	\colhead{$5.8$} &
	\colhead{unc} &
	\colhead{$8.0$} &
	\colhead{unc} &
	\colhead{$24$} &
	\colhead{unc}  
	}
	\startdata
	\multicolumn{19}{c}{Members} \\ \hline
	     273 & XMMU J034352.1+320343   & 03:43:52.09 & 32:03:40.0 & 16.02  & 15.20  & 13.77 & 13.01  &  12.66 & 12.31 &  0.01 & 12.01 &  0.02 & 12.22 &  0.07 & 12.14 &  0.06 &  7.91 & -9.00 \\
	     401 & CXOANC J034431.3+321448 & 03:44:31.20  & 32:14:47.2 & 18.72 & 17.60 & 15.35 & 14.20 & 13.62 & 13.10 &  0.02 & 12.90 &  0.02 & 13.03 &  0.08 & 13.35 &  0.04 &  6.27 & -9.00 \\ 
	     451 & 2MASS J03434521+3205247 & 03:43:45.23  & 32:05:24.8 & 17.62 & 16.73 & 15.30 & 14.54 & 14.02 & 13.72 &  0.02 & 13.62  &  0.04 & 13.30 &  0.09 & 13.43 & -9.00 &  8.66 & -9.00 \\
	   1840 & 2MASS J03431992+3202412 & 03:43:19.93  & 32:02:41.4 & 15.73  & 15.00 & 13.81 & 13.16 & 12.84 & 12.81 &  0.03 & 12.39  &  0.02 & 12.51 &  0.06 & 12.33 &  0.10 &  8.14  & -9.00  \\ \hline
	\multicolumn{19}{c}{Candidates} \\ \hline
	     102 & CXOANC J034435.8+321503 & 03:44:35.898 &  32:15:02.35 & 16.05 & 15.17 & 12.86 & 11.59 & 11.13 & 10.80 &  0.01 & 10.74 &  0.03 & 10.67 &  0.04 & 10.65 &  0.03 &  7.92 & -9.00 \\
	     104 & XMMU   J034429.9+321921 & 03:44:29.98 &  32:19:22.7 & 14.71 & 13.89 & 12.49 & 11.50 & 11.14 & 10.84 &  0.00 & 10.79 &  0.08 & 10.64 &  0.04 & 10.43 &  0.04 &  8.01 &  0.07\\
	     118 & XMMU   J034402.1+321940 & 03:44:02.19 &  32:19:40.1 & 14.11 & 13.56 & 12.52 & 11.60 & 11.33 & 11.12 &  0.01 & 11.09 &  0.02 & 11.04 &  0.05 & 11.06 &  0.04 &  8.49 & -9.00\\
	     126 & CB2006 024                     & 03:43:47.89 &  32:17:56.9 & 14.52 & 13.89 & 12.53 & 11.67 & 11.38 & 11.25 &  0.03 & 11.17 &  0.02 & 11.06 &  0.03 & 11.11 &  0.02 &  8.65 & -9.00\\
	     148 & CXOANC J034435.9+321554 & 03:44:34.71 &  32:15:54.4 & 16.30 & 15.28 & 13.30 & 12.07 & 11.58 & 11.09 &  0.01 & 11.06 &  0.03 & 10.95 &  0.06 & 10.93 &  0.05 &  3.81 & -9.00\\
	     155 & XMMU   J034359.5+321551 & 03:43:59.55 &  32:15:55.4 & 15.23 & \nodata & 13.05 & 12.00 & 11.68 & 11.39 &  0.01 & 11.45 &  0.03 & 11.40 &  0.04 & 11.32 &  0.06 &  6.46 & -9.00\\
	     185 & CXOANC J034421.6+321511 & 03:44:21.56 &  32:15:09.8 & 14.93 & 14.23 & 12.98 & 12.18 & 11.91 & 11.72 &  0.01 & 11.71 &  0.03 & 11.67 &  0.06 & 11.38 &  0.03 &  8.88 &  0.28\\
	     196 & CXOANC J034440.9+321721 & 03:44:40.90 &  32:17:19.1 & 17.01 & 15.98 & 13.94 & 12.67 & 12.17 & 11.73 &  0.01 & 11.64 &  0.04 & 11.60 &  0.03 & 11.61 &  0.04 &  6.35 & -9.00\\
	     204 & CXOANC J034439.9+321601 & 03:44:39.86 &  32:15:58.1 & 15.71 & 14.94 & 13.49 & 12.53 & 12.20 & 11.83 &  0.01 & 11.78 &  0.02 & 11.86 &  0.04 & 11.81 &  0.13 &  6.43 & -9.00\\
	     208 & XMMU   J034438.7+321905 & 03:44:38.79 &  32:19:05.6 & 15.19 & 14.51 & 13.26 & 12.36 & 12.10 & 11.82 &  0.02 & 11.77 &  0.02 & 11.77 &  0.12 & 12.51 &  1.15 &  4.91 & -9.00\\
	     283 & XMMU   J034434.1+321957 & 03:44:34.38 &  32:19:56.9 & 15.49 & 14.89 & 13.84 & 13.07 & 12.84 & 12.56 &  0.01 & 12.45 &  0.05 & 12.45 &  0.07 & 12.37 &  0.12 &  6.19 & -9.00\\
	     288 & CXOANC J034347.1+321320 & 03:43:47.12 &  32:13:21.4 & 15.52 & 15.05 & 13.85 & 13.20 & 12.91 & 12.77 &  0.01 & 12.70 &  0.02 & 12.67 &  0.09 & 12.66 &  0.13 &  7.65 & -9.00\\
	     323 & CXOANC J034358.1+321356 & 03:43:58.12 &  32:13:57.1 & 15.97 & 15.46 & 14.23 & 13.41 & 13.22 & 13.06 &  0.01 & 13.00 &  0.02 & 12.92 &  0.05 & 13.16 &  0.17 &  6.35 & -9.00\\
	    1680 & XMMU   J034505.1+315752 & 03:45:05.07 &  31:57:54.3 & 17.80 & 16.89 & 15.31 & 14.68 & 14.18 & 13.84 &  0.02 & \nodata & \nodata & 13.61 &  0.11 & \nodata & \nodata &  9.80 & -9.00\\
	    1688 & CXOANC J034510.2+320451 & 03:45:10.02 &  32:04:48.8 & 16.99 & 16.23 & 14.76 & 13.97 & 13.62 & 13.29 &  0.01 & 13.20 &  0.03 & 13.14 &  0.07 & 13.29 &  0.11 &  9.80 & -9.00\\
	    1888 & CXOANC J034419.8+315919 & 03:44:19.78 &  31:59:19.0 & \nodata & \nodata & \nodata & \nodata & 15.12 & 12.07 &  0.01 & 11.07 &  0.02 & 10.57 &  0.04 & 10.41 &  0.03 &  7.74 & -9.00\\
	    1931 & XMMU   J034452.9+320005 & 03:44:53.19 &  32:00:06.7 & \nodata & \nodata & 18.61 & 15.15 & 13.21 & 12.20 &  0.01 & 11.91 &  0.03 & 11.59 &  0.06 & 11.72 &  0.04 & 10.22 & -9.00\\
	   10019 & CB2006 138                     & 03:45:18.00 &  32:19:33.0 & \nodata & \nodata & 13.46 & 12.47 & 12.14 & 11.79 &  0.01 & 11.68 &  0.02 & 11.78 &  0.04 & 11.52 &  0.04 &  8.12 & -9.00\\
	   10138 & CB2006 141                     & 03:45:21.06 &  32:18:17.8 & \nodata & \nodata & 13.01 & 11.90 & 11.49 & 11.05 &  0.01 & 10.89 &  0.01 & 10.77 &  0.03 & 10.88 &  0.04 &  7.49 & -9.00\\
	   10284 & XMMU   J034535.3+320328 & 03:45:35.45 &  32:03:25.8 & 14.94 & 14.30 & 13.17 & 12.48 & 12.14 & \nodata & \nodata & \nodata & \nodata & \nodata & \nodata & \nodata & \nodata &  9.59 & -9.00\\
	   10373 & XMMU   J034522.0+320203 & 03:45:22.15 & 32:02:04.1 & 15.12 & 14.48 & 13.23 & 12.58 & \nodata & 12.06 &  0.00 & 11.95 &  0.01 & 11.88 &  0.03 & 11.94 &  0.06 &  8.85 & -9.00\\
	   54519 & XMMU   J034413.8+315534  & 03:44:13.77 &  31:55:34.4 & \nodata & \nodata & 13.64 & 12.19 & 11.59 & 11.02 &  0.01 & \nodata & \nodata & 10.88 &  0.05 & \nodata & \nodata &  8.39 & -9.00\\
	   54916 & CB2006 116                      & 03:44:40.63 &  32:23:10.7 & \nodata & \nodata & 12.74 & 11.85 & 11.50 & 11.28 &  0.01 & 11.28 &  0.01 & 11.25 &  0.03 & 11.21 &  0.07 &  8.72 & -9.00\\
	\enddata
	\tablenotetext{(1)}{CXOANC: ANCHORS catalog; XMMU: \protect\citet{2004A&A...422.1001P}; CB2006: \protect\citet{2006astro.ph..6127C}; 2MASS: \protect\citet{2006AJ....131.1163S}.}
	\tablenotetext{(2)}{Origin of photometry are as follows: ${I_c}z$: \protect\citet{2003ApJ...593.1093L}; $JH{K_s}$: \protect\citet{2003AJ....125.2029M}; \sst\ $3.6 - 24\micron$: this paper. 
	The listed magnitude is an upper limit if the listed uncertainty is given as -9. }
	\end{deluxetable}
\clearpage
\end{landscape}

\end{document}